\documentclass[a4paper]{statsoc}
\usepackage{amsmath,verbatim,amssymb,epsfig,enumitem,url,amsbsy,enumitem,color,natbib,bm,bbm}
\usepackage{geometry}
\newcommand{\change}[1]{{\leavevmode\color{black}{#1}}}
\renewcommand\refname{ \noindent \bf References \vspace{0.0in}}
  \renewenvironment{thebibliography}[1]{
    \begin{oldthebibliography}{#1}
      \setlength{\parskip}{0ex}
      \setlength{\itemsep}{0.ex}}
  {\end{oldthebibliography}}

\usepackage{etoolbox}
\usepackage{ulem}

\makeatletter
\patchcmd{\@makecaption}
  {\parbox}
  {\advance\@tempdima-\fontdimen2} % decrease the width!
  {}{}
\makeatother  

\newcommand{\bea}{\begin{eqnarray*}}
\newcommand{\eea}{\end{eqnarray*}}
\newcommand{\be}{\begin{eqnarray}}
\newcommand{\ee}{\end{eqnarray}}
\newcommand{\bay}{\begin{array}}
\newcommand{\eay}{\end{array}}
\newcommand{\bi}{\begin{itemize}}
\newcommand{\ei}{\end{itemize}}
\newcommand{\ben}{\begin{enumerate}}
\newcommand{\een}{\end{enumerate}}
\newcommand{\bcen}{\begin{center}}
\newcommand{\ecen}{\end{center}}

\newcommand{\tp}{{\tau}}
\newcommand{\Mean}{{\mathbb{E}}}
\newcommand{\Var}{{\mbox{Var}}}
\newcommand{\Cov}{{\mbox{Cov}}}

\newcommand{\diag}{{\mbox{diag}}}
\newcommand{\prob}{{\mathbb{P}}}

\newcommand{\vc}{\textrm{vec}}

\newcommand{\DE}{\textrm{DE}}
\newcommand{\IE}{\textrm{IE}}
\newcommand{\R}{\mathbb{R}}
\newcommand{\M}{\mathbb{M}}
\newcommand{\RNum}[1]{\uppercase\expandafter{\romannumeral #1\relax}}
\newtheorem{thm}{Theorem}
\newtheorem{exmp}{Model}
\newtheorem{lem}{Lemma}
\newtheorem{defs}{Definition}
\newtheorem{asmp}{Assumption}
\newtheorem{cor}{Corollary}
\newtheorem{prop}{Proposition}
\newenvironment{pf}{{\noindent\it Proof}\ }{\hfill $\square$\par}
\usepackage{pythonhighlight}
\usepackage{multirow} 
\usepackage{algorithm}
\usepackage{algcompatible}

\makeatletter
\newenvironment{breakablealgorithm}
  {% \begin{breakablealgorithm}
   \begin{center}
     \refstepcounter{algorithm}% New algorithm
     \hrule height.8pt depth0pt \kern2pt% \@fs@pre for \@fs@ruled 画线
     \renewcommand{\caption}[2][\relax]{% Make a new \caption
       {\raggedright\textbf{\ALG@name~\thealgorithm} ##2\par}%
       \ifx\relax##1\relax % #1 is \relax
         \addcontentsline{loa}{algorithm}{\protect\numberline{\thealgorithm}##2}%
       \else % #1 is not \relax
         \addcontentsline{loa}{algorithm}{\protect\numberline{\thealgorithm}##1}%
       \fi
       \kern2pt\hrule\kern2pt
     }
  }{% \end{breakablealgorithm}
     \kern2pt\hrule\relax% \@fs@post for \@fs@ruled 画线
   \end{center}
  }
\makeatother
\usepackage{subfigure}
\graphicspath{{fig/}}
\makeatother

\usepackage{xr-hyper}
\usepackage{hyperref}
\makeatletter
\newcommand*{\addFileDependency}[1]{
\typeout{(#1)}
\@addtofilelist{#1}
\IfFileExists{#1}{}{\typeout{No file #1.}}
}
\makeatother

\title{Policy Evaluation for Temporal and/or Spatial Dependent Experiments} %in Ride-sharing Platforms}

\author[Shikai Luo$^{a*}$, Ying Yang$^{b*}$, Chengchun Shi$^{c*}$]{Shikai Luo$^{a*}$, Ying Yang$^{b*}$, Chengchun Shi$^{c*}$, Fang Yao$^d$,\\
 Jieping Ye$^{e}$,   and Hongtu Zhu$^{f}$\thanks{The first three authors contributed equally to this paper.  
         Address for correspondence:
        Hongtu Zhu, Ph.D., E-mail: 
        htzhu@email.unc.edu.    
         }\hspace{.2cm}\\}
\address{$^a$AI-Lab, Didi Chuxing, Beijing, P.R. China \\
        $^b$Academy of Mathematics and Systems Science, Chinese Academy of Sciences, Beijing, P.R. China\\
                $^c$Department of Statistics at London School of Economics and Political Science, London, UK \\
                 $^d$School of Mathematics, Peking University, Beijing, P.R. China\\
                 $^e$Alibaba Damou Academy, Hangzhou,  P.R. China\\
                $^f$University of North Carolina at Chapel Hill, North Carolina, USA}
\begin{document}

\begin{abstract}
The aim of this paper is to establish a causal link between the policies implemented by technology companies and the outcomes they yield within intricate temporal and/or spatial dependent experiments. We propose a novel temporal/spatio-temporal Varying Coefficient Decision Process (VCDP) model, capable of effectively capturing the evolving treatment effects in situations characterized by temporal and/or spatial dependence. Our methodology encompasses the decomposition of the Average Treatment Effect (ATE) into   the Direct Effect (DE) and the Indirect Effect (IE). We subsequently devise comprehensive procedures for estimating and making inferences about both DE and IE. Additionally, we provide a rigorous analysis of the statistical properties of these procedures, such as   asymptotic power. To substantiate the effectiveness of our approach, we carry out extensive simulations and real data analyses.


\end{abstract}

\section{Introduction}\label{sec:introduction}
 \change{The utilization of A/B testing, or randomized controlled experiments, has rapidly expanded across various technology companies, including   Google  and Twitter. This practice is employed to inform data-driven decisions regarding new policies, such as services,   or products, effectively establishing itself as the gold standard for product development \citep[see][for an overview]{larsen2023statistical}. 
For instance, in the context of ride-sharing platforms such as Uber, prior to implementing new policies related to order dispatch or subsidies, they frequently undertake a series of online experiments for policy evaluation. These platforms have significantly reshaped human transportation dynamics through the widespread adoption of smartphones and the Internet of Things \citep{Alonso-Mora2017,Hagiu2019,qin2022reinforcement}. These technology-driven companies strive to create efficient spatio-temporal systems incorporating various policies, all aimed at enhancing key platform metrics such as supply-demand equilibrium and total driver income \citep{zhou2021graph,qin2022reinforcement}. 
The switchback design stands out as a widely adopted experimental approach within the domain of online experimentation. This design involves dividing an experimental day into distinct non-overlapping time intervals, alternating between treatment and control policies across several cities for a specified duration, often spanning an even number of days, such as $n=14$ \footnote{https://eng.lyft.com/experimentation-in-a-ridesharing-marketplace-b39db027a66e}.

In the realm of policy evaluation within these technology companies, several significant statistical challenges arise. 
Firstly, the data generating process is often non-stationary. Consider the context of ridesharing platforms as an example. At specific time intervals, metrics like online driver numbers (supply) and call order numbers (demand) can be visualized as spatio-temporal networks that exhibit substantial variation throughout a day, peaking during rush hours. These metrics interact across time and locations in intricate ways. 
Secondly, the market features typically exhibit daily trends, manifesting as spatio-temporal random effects. This trend violates the assumption of conditional independence between market outcomes and past data history. For more in-depth discussions, refer to Section \ref{sec:T method}. 
Thirdly, complex spatio-temporal interference effects add further intricacy to the estimation and inference of treatment effects. 
Lastly, the sample size is often limited, while effect sizes tend to be small. In ridesharing applications, for instance, most AB test experiment durations do not exceed 20 days \citep{shi2023dynamic}, and the size of treatment effects typically ranges between 0.5\% and 2\% \citep{Tang2019}.

  The primary objective of this paper is to develop a robust statistical framework for analyzing the causal connections between the policies implemented by these companies and their corresponding outcomes, even in the presence of the aforementioned challenges. Our four major contributions can be summarized as follows. Firstly, we address the challenges by introducing linear and neural network-based Varying Coefficient Decision Process (VCDP) models. These models accommodate dynamic treatment effects over time and/or space, even in the presence of non-stationarity, random effects, interference, and spatial spillovers. These models account for market features as mediators to incorporate historical policy carryover effects. Furthermore, by assuming network interference and employing mean field approximation (as detailed in Section \ref{sec:STVCDP}), we effectively operate an  ``effective treatment" \citep{manski2013identification} or ``exposure mapping'' \citep{aronow2017estimating}  in the spatio-temporal system. Our approach extends beyond the switchback design to any dynamic treatment allocation setup.

   Secondly, we develop estimation methods for our VCDPs. For linear VCDPs, we propose a two-step process involving the calculation of least squares estimates and kernel smoothing to refine the estimates. Kernel smoothing leverages neighboring observations across time and/or space, enhancing estimation efficiency and overcoming the challenge of weak signals and small sample sizes. Additionally, we decompose average treatment effects (ATEs) into Direct Effects (DE) and Indirect Effects (IE). Similar decompositions have been considered in the literature of causal inference in time series \citep[see e.g.,][]{boruvka2018assessing,bojinov2019time}.  We introduce a Wald test for DE detection and a parametric bootstrap for IE inference, enhancing the detectability of ATE in cases where IE's variance significantly exceeds that of DE. This decomposition also aids decision-makers in understanding policy mechanisms and devising more effective strategies (refer to Section \ref{sec:discuss}).

Thirdly, we rigorously study the asymptotic properties of our test procedure under the setting where the number of treatment decision stages per day ($m$) diverges with the sample size ($n$). Although this aligns with ride-sharing platforms, it poses theoretical complexities as the continuous mapping theorem \citep{1996Weak} is inapplicable when $m\to\infty$. Details are provided in Section \ref{sec:theoretical analysis}. Importantly, our analysis reveals that the switchback design is likely to yield more efficient estimators compared to a simple alternating-day design that randomly assigns treatment throughout each day. 
  
Fourthly, we evaluate the finite sample performance of our parameter estimators and test statistics using extensive simulations and real datasets from Didi. Our empirical findings validate our theoretical assertions. Notably, the empirical power of our test increases with the frequency of switchbacks, further affirming the benefits of the switchback design.

%In essence, our work equips these companies with a potent framework to analyze the causal relationships between their policies and outcomes, addressing multifaceted statistical challenges and providing a foundation for effective decision-making. 
   }
  
\subsection{Related works}\label{sec:relatedwork} 
  The key idea of A/B testing is to apply causal inference methods to estimating the treatment effect of a new change under the assumption of ``no interference'' as a part of the {\it stable unit treatment value assumption} \citep[SUTVA,][]{rubin1980discussion}. 
  Despite of its ubiquitousness, however,  the standard A/B testing is not directly applicable for causal inference under interference, which frequently occurs  in many complex systems, particularly for spatio-temporal systems. 
  For instance,  researchers from Google and eBay have observed that advertisers (or users) interact within online auctions.

  There has been substantial interest in the development of causal inference under interference. See the comprehensive reviews in  \cite{Halloran2016}, \cite{Reich2020}, and \cite{savje2021average} and references therein. 
  Since there is   a consensus that causal inferences are impossible without  any assumptions 
  on the interference structure, 
  capturing  interference effects requires new definitions of the estimands of interest
  and new  models for causal effects. {For instance, \cite{bojinov2019time} considered the $p$ lag causal effect, whereas \cite{aronow2020design} introduced a spatial ``average marginalized response". In contrast, our target parameter is the global average treatment effect,  which is the expected return difference under  the new policy  against   the control policy  in the entire market. }
  In addition, there are four major types of models for the interference processes.   
  Firstly,  early methods assumed specific structural models to restrict the interference process  \citep{Lee2007}. Secondly,  
  the partial interference assumption  has been widely used to   restrict interference  only  in known and disjoint groups of units \citep{sobel_what_2006,Tchetgen2012,zigler_estimating_2012,Halloran2016,pollmann2020causal}. Thirdly, the local or  network-based interference  assumption was introduced to deal with  interference  between %connected 
  local units   in a geographic space or connected nodes in an exposure graph 
  \citep{Bakshy2014,perez-heydrich_assessing_2014,Verbitsky-Savitz2012,Puelz2019,aronow2020design}. {Our VCDPs are closely related to the second and third types of models, but they focus on interference across time \textit{and} space. Most aforementioned works studied the interference effect across time \textit{or} space and  were  motivated by research questions in environmental and epidemiological studies. It remains unknown about   their  generalization to ride-sharing markets. Fourthly, recent models capture the interference effect via congestion or price effects in a marketplace \citep{munro2021treatment,wager2021experimenting,johari2022experimental}.  \change{These solutions rely on an assumption of Markovanity or stationarity and are design-dependent. In contrast, our approach accommodates non-stationarity and is capable of managing non-Markovianity in scenarios where outcome errors exhibit time-correlated patterns.} %our proposal allows non-stationarity, can be extended to non-Markov settings, and remains valid under a wide range of designs.

  Our proposal is closely related to a growing literature on off-policy evaluation (OPE) methods in sequential decision making \citep[see][for a review]{uehara2022review}. In the literature, augmented inverse propensity score weighting methods \citep[see e.g.,][]{zhang2013robust,luedtke2016statistical,jiang2016doubly,thomas2016data} %are applicable for OPE in general non-Markovian environments. 
  have been proposed for valid OPE. Nonetheless, these methods suffer from the curse of horizon \citep{liu2018breaking} in that the variance of the resulting estimator grows exponentially fast with respect to $m$, leading to  inefficient  estimates in the large $m$ setting. 
  Efficient model-free OPE methods have been proposed by \citet{kallus2020double,kallus2022efficiently,liao2020batch,liao2021off,luckett2020estimating,shi2021deeply,shi2022statistical} under the Markov decision process \citep[MDP, see e.g.,][]{puterman2014markov} model assumption. 
  %,  which requires the immediate reward and future observation to be conditionally independent of the past data history given the current action and observation.  
  \change{Recently, \cite{hu2021pomdp} proposed a model-free OPE method in partially observed MDPs (POMDPs) that avoids the curse of horizon.} %as a modeling framework for off-policy evaluation of dynamic treatment rules.  However, such MDP model assumption excludes the existence of random effects and is typically violated in our application.}   
  Our proposal is model-based and is ultimately different from most existing model-free OPE methods that did not consider the random effects, spatial interference effects, and the decomposition into DE and IE.     In addition, little  has been done on OPE for spatio-temporal dependent experiments.

  Finally, our paper is related to a line of works on quantitative approaches to ride-sharing platforms. %at the intersection of operations research and statistics. 
  In particular, \cite{bimpikis2019spatial} proposed supply-and-demand models and investigated the impact of the demand pattern on the platform's prices and profits. %Nonetheless, estimation and inference of the model parameters were not covered. 
  \cite{castillo2017surge} studied how the surging prices can prevent wild goose chase (e.g., drivers pick up distant customers) and conducted regression analysis to verify the  nonmonotonicity of supply on pickup times. %wasting time and reducing earnings.
  However, estimation and inference of  target policy's treatment effect have not been considered in these papers. \cite{cohen2021frustration} employed the difference in differences methods to estimate the treatment effects of %dynamic pricing can improve the drivers' earnings. 
  different types of compensation on the engagement of riders who experienced a frustration. Their analysis is limited to   staggered designs. 
  \cite{garg2021driver} studied the theoretical properties of driver-side payment mechanisms and compared additive surge against multiplicative surge numerically. However, they did not consider the spatial spillover effects of these policies. Our paper complements the existing literature by developing a general framework to efficiently infer a target policy's direct and indirect effects based on data collected from spatio-temporal dependent experiments and analyzing the advantage of switchback designs in the presence of spatio-temporal random effects.}

\subsection{Paper outline}

  The rest of the paper is organized as follows. 
  In Section \ref{sec:po_mdp}, we introduce a potential outcome framework for problem formulation, % under the temporal alternation design and  
  propose two novel temporal VCDP models under temporal dependent experiments, %while 
  and develop estimation  and testing procedures for both DE and IE. %in order to test the dynamic treatment effect.  
  In Section 
  \ref{sec:ST method}, we further propose %a potential outcome framework and 
  two spatio-temporal VCDP models %for the 
  under spatio-temporal dependent experiments
  %alternation design, while 
  and develop the associated estimation and testing procedures.    
  In Section \ref{sec:theoretical analysis}, we systematically investigate the theoretical properties of estimation and testing procedures (e.g., consistency and power) developed in Sections \ref{sec:po_mdp} and \ref{sec:ST method}. \change{We also illustrate the benefits of employing the switchback design in theory.}   In  Section \ref{sec:simulation studies}, 
  we use numerical simulations to examine the finite sample performance of our estimation and testing procedures. \change{Furthermore, we numerically explore the benefits of the switchback design}. In Section \ref{sec:real data analysis}, we apply the proposed procedures to evaluating different  policies in Didi Chuxing.

\section{Policy evaluation for temporal dependent experiments}
\label{sec:po_mdp}
In this section, we present the proposed methodology for policy evaluation in temporal dependent experiments for  one experimental region. 

\subsection{A potential outcome framework}
 We use the potential outcome framework to present our %off-policy evaluation 
 model in non-stationary environments. %MDPs.  
 We divide each day into $m$ equally spaced nonoverlapping intervals. At each time interval, the platform can implement either the new or old policy. We use $A_{\tp}$  to denote the policy implemented at the $\tp$th interval for any integer $\tp \ge 1$. Let $S_{\tp}$  be some state variables measured at the $(\tp-1)$-th interval in a given day. All the states share the same support, which is assumed to be a compact subset of $\R^d$,  where $d$ denotes the dimension of the state. Let $Y_{\tp}\in \mathbb{R}$ be the outcome of interest measured at time $\tp$.    

  Firstly, we define the average treatment effect (ATE) as the difference between the new and old policies.  
  Let $\bar{a}_{\tp}=(a_1,\ldots,a_{\tp})^\top\in \{0,1\}^{\tp}$ denote a treatment history vector up to time $\tp$, where 
  $1$ and $0$   denote the new policy and the old one, respectively. 
  We define $S_{\tp}^*(\bar{a}_{\tp-1})$ and $Y_{\tp}^*(\bar{a}_{\tp})$ as the counterfactual state and the counterfactual outcome, respectively.  Then ATE can be defined   as follows. 
  \begin{defs}
  %For a given reference distribution function $\mathbb{G}$, ATE %of temporal-alternating design 
  ATE is   the difference between two value functions  given by 
  \begin{eqnarray*}
  \textrm{ATE}=\sum_{\tp =1}^{m} \Mean \{Y_{\tp}^*(\bm{1}_{\tp})-Y_{\tp}^*(\bm{0}_{\tp})\},
  \end{eqnarray*}
  where $\bm{1}_{\tp}$ and $\bm{0}_{\tp}$ denote vectors of 1s and 0s of length $\tp$, respectively. 
  \end{defs}

  Secondly, we can decompose ATE as the  sum of direct effects (DE) and indirect effects (IE). Let $R_\tp$ denote the conditional mean function of the outcome given the data history,
  \begin{eqnarray*}
 \Mean \{Y_\tp^*(\bar{a}_\tp)|S_\tp^*(\bar{a}_{\tp-1}),Y_{\tp-1}^*(\bar{a}_{\tp-1}),S_{\tp-1}^*(\bar{a}_{\tp-2}),Y_{\tp-2}^*(\bar{a}_{\tp-2}),\ldots, S_1\}=R_\tp(a_\tp,S_\tp^*(\bar{a}_{\tp-1}),a_{\tp-1},S_\tp^*(\bar{a}_{\tp-2}),\ldots,S_1).
  \end{eqnarray*}
  %Specifically, it follows from TCMIA 
  It follows that $\textrm{ATE}$ can be rewritten as  
  \begin{eqnarray}\label{eq:decomp ATE}
  \begin{split}
  &~&\sum_{\tp=1}^m \Mean \{R_{\tp}(1,S_\tp^*(\bm{1}_{\tp-1}),1, S_{\tp-1}^*(\bm{1}_{\tp-2}),\ldots,S_1)-R_\tp(0,S_\tp^*(\bm{0}_{\tp-1}),0, S_{\tp-1}^*(\bm{0}_{\tp-2}),\ldots,S_1)\} \\
  &=& \underbrace{\sum_{\tp=1}^m \Mean \{R_{\tp}(1,S_\tp^*(\bm{0}_{\tp-1}),0, S_{\tp-1}^*(\bm{0}_{\tp-2}),\ldots,S_1)-R_\tp(0,S_\tp^*(\bm{0}_{\tp-1}),0, S_{\tp-1}^*(\bm{0}_{\tp-2}),\ldots,S_1)\}}_{\DE}\\
  &+&\underbrace{\sum_{\tp=1}^m \Mean  \{R_{\tp}(1,S_\tp^*(\bm{1}_{\tp-1}),1, S_{\tp-1}^*(\bm{1}_{\tp-2}),\ldots,S_1)-R_\tp(1,S_\tp^*(\bm{0}_{\tp-1}),0, S_{\tp-1}^*(\bm{0}_{\tp-2}),\ldots,S_1)\}}_{\IE}. 
  \end{split}
  \end{eqnarray}

  The DE represents the sum of the short-term treatment effects on the immediate outcome over time assuming that the baseline policy is being employed in the past.  In contrast,  IE characterizes the carryover effects of past policies. %that work through the state vector.  
  Our problems of interest are to estimate DE, IE and test the following hypotheses:
  \begin{eqnarray}
  &H_0^{DE}: \DE\le 0\,\,\,\,\textrm{versus}\,\,\,\, H_1^{DE}:\DE>0.\label{hypo:de t}\\
  &H_0^{IE}: \IE\le 0\,\,\,\,\textrm{versus}\,\,\,\,H_1^{IE}:\IE>0.\label{hypo:ie t}
  \end{eqnarray}
  If both  $H_1^{DE}$ and $H_1^{IE}$  hold,  then the new policy is better than the baseline one. 

  Thirdly,  since  all other potential variables except $S_1$  cannot be observed,  we follow    the causal inference literature  and assume the consistency assumption (CA), the sequential randomization assumption (SRA) and the positivity assumption (PA) as follows: 
  \begin{itemize} 
  \item{\bf CA}. $S_{\tp}^*(\bar{A}_{\tp-1})=S_{\tp}$ and $Y^*(\bar{A}_{\tp})=Y_{\tp}$ for any $\tp\ge 1$, where $\bar{A}_{\tp}$ denotes the observed policy history up to time $\tp$. 
  \item {\bf SRA}. $A_{\tp}$ is conditionally independent of all potential variables given $S_{\tp}$ and $\{(S_{j},A_{j},Y_j)\}_{j<\tp}$. 
  \item {\bf PA}. For any $\tp\ge 1$, the probability\footnote{When data are not identically distributed, the observed data distribution corresponds to a mixture of individual trajectory distributions with equal weights.} that the observed action at time $\tp$ equals one given the observed data history is strictly bounded between zero and one. 
  \end{itemize}
  The SRA allows the policy to be adaptively assigned based on the observed data history (e.g., via the $\epsilon$-greedy algorithm). It is automatically satisfied under the temporal switchback design, in which the policy assignment mechanism is independent of the data. The PA is also automatically satisfied under this design, in which at each time, half actions equal zero whereas the other half equal one. Moreover, CA, SRA and PA ensure that DE and IE are estimable from the observed data, as shown below.

  \begin{lem}\label{lemma:identification}
  Under CA, SRA and PA, we have
  \begin{eqnarray}\label{eqn:reward}
  R_{\tp}(a_{\tp},s_{\tp},\ldots,s_1)&=&\Mean (Y_{\tp}|A_{\tp}=a_{\tp},S_{\tp}=s_{\tp},\ldots,S_1=s_1), \\ 
  \label{eqn:rtau}
  \Mean \{R_{\tp}(a,S_{\tp}^*(\bar{a}_{\tp-1}),\ldots,S_1)\}&=&\Mean [\Mean [R_{\tp}(a,S_{\tp},\
  \ldots,S_1)| \{A_j=a_j\}_{1\le j<\tp}, \{S_j,Y_j\}_{1\le j<\tp} ]].
  \end{eqnarray}
  \end{lem}

  Lemma  \ref{lemma:identification} implies that the causal estimand can be represented as a function of the observed data.

\subsection{TVCDP model}
\label{sec:T method}

  We introduce  two  TVCDP regression models to model $Y_{i, \tp}$ and the conditional distribution of $S_{i, \tp}$ given the data history, forming  the basis of our estimation and testing procedures. 
  Suppose that the  experiment is conducted over $n$ days. Let  $(S_{i,\tp},A_{i,\tp},Y_{i,\tp})$ be the state-policy-outcome triplet measured at the $\tp$th time interval of the $i$th day for $i=1, \ldots, n$ and $\tp=1, \ldots, m$. {The proposed TVCDP model is composed of the following set of additive noise models, %as follows,
  \begin{eqnarray}\label{model:TVCM DE}
  \begin{split}
  Y_{i,\tp} &= f_{1,\tp}(S_{i,\tp},A_{i,\tp}) + e_{i,\tp}, \\
  % \label{model:TVCM IE}
  S_{i,\tp+1} & =  f_{2,\tp}(S_{i,\tp}, A_{i,\tp}) + \varepsilon_{i,\tp S},
  \end{split}
  \end{eqnarray}
  where $f_{1,\tp}(\cdot)$ and $f_{2,\tp}(\cdot)$ are the regression functions. %and $Z_{i,\tp} = (1,S_{i,\tp}^\top, A_{i,\tp})^\top$ corresponds to the current state-action pair. 
  
  \change{We would like to highlight several key points. Firstly, in addition to defining the standard outcome regression model $f_{1,\tau}$ as described in equation \eqref{model:TVCM DE}, it is crucial to specify how past actions influence future states. This is accomplished through the inclusion of $f_{2,\tau}$, which plays a pivotal role in quantifying temporal interference effects.

Secondly, we introduce a specific assumption related to the error structure. This assumption is fundamental as it allows us to incorporate temporal random effects effectively.%}
 % The proposed time-varying model can capture the variation of the data across different time interval and quantify the temporal interference effects by $f_{2,\tp}$. 
  %Moreover, to include  the temporal random effects, we assume  $e_{i,\tp}=\eta_{i,\tp} + \varepsilon_{i,\tp}$ and $\varepsilon_{i,\tp S}$ are the noise components with  $\eta_{i,\tp}$ characterizing the day-specific temporal variation across different days and $\varepsilon_{i,\tp}$ and $\varepsilon_{i,\tp S}$ being  measurement errors.} 
 % \change{%We assume the following noise structures for model \eqref{model:TVCM DE}.
  \begin{asmp}\label{assump:eps t}
  (i) The outcome noise $e_{i,\tp}=\eta_{i,\tp}+\varepsilon_{i,\tp}$ is a combination of two mutually independent stochastic processes: day-specific temporal variation  $\eta_{i,\tp}$  and measurement error $\varepsilon_{i,\tp}$.
  %and state noises $\varepsilon_{i,\tp S}$ satisfy:  $\eta_{i,\tp},\varepsilon_{i,\tp}$ and $\varepsilon_{i,\tp S}$ are mutually independent; 
  (ii) The processes $\{\eta_{i,\tp}\}_{i,\tp}$ are identical realizations of a zero-mean stochastic process with covariance function $\{\Sigma_{\eta}(\tp_1,\tp_2)\}_{\tp_1,\tp_2}$. Additionally, all components of $\Sigma_{\eta}(t_1,t_2)$ have bounded and continuous second derivatives with respect to $t_1$ and $t_2$.  (iii) %$\{\varepsilon_{i,\tp}\}_{i,\tp}$ and $\{\varepsilon_{i,\tp S}\}_{i,\tp}$ are independent measurement errors with zero means,   $\Var(\varepsilon_{i,\tp})=\sigma_{\varepsilon,\tp}^2$, and $\mbox{Cov}(\varepsilon_{i,\tp S})=\Sigma_{\varepsilon,\tp S}$.  
 The measurement errors $\{\varepsilon_{i,\tp}\}_{i,\tp}$ and $\{\varepsilon_{i,\tp S}\}_{i,\tp}$ are   independent over time. 
 They have zero mean values  and exhibit  $\Var(\varepsilon_{i,\tp})=\sigma_{\varepsilon,\tp}^2$  and $\mbox{Cov}(\varepsilon_{i,\tp S})=\Sigma_{\varepsilon,\tp S}$.
\end{asmp}

  % We assume that $\eta_{i,\tp},\varepsilon_{i,\tp},$ and $\varepsilon_{i,\tp S}$ are mutually independent;  $\{\varepsilon_{i,\tp}\}_{i,\tp}$ and $\{\varepsilon_{i,\tp S}\}_{i,\tp}$ are independent measurement errors with zero means,   $\Var(\varepsilon_{i,\tp})=\sigma_{\varepsilon,\tp}^2$, and $\mbox{Cov}(\varepsilon_{i,\tp S})=\Sigma_{\varepsilon,\tp S}$; and $\{\eta_{i,\tp}\}_{i,\tp}$ are identical copies of a mean-zero stochastic process with covariance function $\{\Sigma_{\eta}(\tp_1,\tp_2)\}_{\tp_1,\tp_2}$. 
  \noindent   It's important to note that the day-specific random effects are present only in the outcome regression model. However, our approach can be extended to scenarios where these random effects also exist in the state regression model. We provide a detailed discussion of this extension in Section \ref{sec:discuss}. Additionally, it's worth mentioning that both the conditional mean and covariance functions, namely $f_{1,\tp}$, $f_{2,\tp}$, $\sigma^2_{\varepsilon,\tp}$, and $\Sigma_{\varepsilon,\tp S}$, are time-dependent. This captures the nonstationarity inherent in the data generating process. 
  }

\change{Our TVCDP models \eqref{model:TVCM DE} have strong connections with the MDP model that is commonly used in reinforcement learning. %in the reinforcement learning literature \citep[see e.g.,][]{puterman2014markov}. 
Specifically, models \eqref{model:TVCM DE} reduce to   non-stationary (or time-varying) MDP models \citep{kallus2022efficiently}  when there are no day-specific random effects in $\{e_{i,\tp}\}_{i,\tp}$. }
%However, there are a few differences between the two models. First, the state variables are unobserved in LQG and need to be inferred based on the observation $Y_{i,\tp}$. Second,  
%when only the coefficients of the treatment variables are time-varying, models \eqref{model:TVCM DE} reduce to linear quadratic Gaussian model \citep{Lewis2020}.}
However,  the proposed time varying models are no longer MDPs due to the existence of the day-specific random effects.} In particular, $Y_{i,\tp}$ in \eqref{model:TVCM DE} is dependent upon past responses given \change{$Z_{i,\tp}=(1,S_{i,\tp}^\top,A_{i,\tp})^\top$}, leading to the violation of the conditional independence assumption. %Existing OPE solutions ignore the existence of the random effects and are less efficient than our proposal 
%Meanwhile, the regression functions depend on the past history only through the current state-action pair. Such a simplification effectively reduces the dimensionality, allowing us to break the curse of horizon in presence of partial observability. 
In addition,   the market features at each time serve as mediators that mediate the effects of past actions on the current outcome.
  
  Next, we consider two specific function approximations for $f_1$ and $f_2$ and derive their related IE and DE as follows. 

  \begin{exmp}\label{exmp:linear}
Linear temporal varying coefficient decision process (L-TVCDP) assumes
  \begin{align*}
  &Y_{i,\tp} = \beta_0(\tp) + S_{i,\tp}^\top\beta(\tp) + A_{i,\tp}\gamma(\tp) + e_{i,\tp}
  = Z_{i,\tp}^\top \theta(\tp) + e_{i,\tp}, \\
  & S_{i,\tp+1} =\phi_0(\tp)+\Phi(\tp) S_{i,\tau}+A_{i,\tp} \Gamma(\tp) + \varepsilon_{i,\tp S}
  =  \Theta(\tp)Z_{i,\tp} + \varepsilon_{i,\tp S},
  \end{align*}
where $\theta(\tp) = (\beta_0(\tp), \beta(\tp)^\top,  \gamma(\tp))^\top$ is a $(d+2)\times 1$ vector of time-varying coefficients, $\Theta(\tp) = [\phi_0(\tp) ~~ \Phi(\tp) ~~ \Gamma(\tp)]$ is a $d\times (d+2)$ coefficient matrix and  {$Z_{i,\tp}=(1,S_{i,\tp}^\top,A_{i,\tp})^\top$}. 
\end{exmp}
\change{Model \ref{exmp:linear} shares a close connection with the linear quadratic Gaussian model (LQG), well studied in the fields of RL and control theory \citep[see, for example,][]{lale2021adaptive}. To be more specific, Model \ref{exmp:linear} can be seen as a simplified, one-dimensional observation variant of LQG under certain conditions. This happens when the outcome regression model doesn't incorporate $A_{i,\tp}$ and the autocorrelated noise $\eta_{i,\tp}$. However, there's a crucial distinction between LQG and our proposed model. In LQG, the state variables are hidden and must be deduced from the observed $Y_{i,\tp}$ values. This contrasts with similar models used in literature for estimating dynamic treatment effects \citep{Lewis2020}. 
} 

When $\{\eta_{i,\tp}\}_{i,\tp}$ become the fixed effects and satisfy $\eta_{i,\tp}=\eta_{i}$ for any $i$ and $\tp$,  the outcome regression model of L-TVCDP includes both the day-specific fixed effects $\{\eta_i\}_i$ and the time-specific fixed effects $\{\beta_0(\tp)\}_{\tp}$. It is  similar to the two-way fixed effects model  in the panel data literature \citep{de2020two,wooldridge2021two,arkhangelsky2021double,imai2021use}. Furthermore, we  derive the closed-form expressions for DE and IE under L-TVCDP, whose proof can be found in Section \ref{pf:prop1} of  the supplementary document.

\begin{prop}\label{concl:linear}
Under the L-TVCDP model, we have $\DE=\sum_{\tp=1}^m\gamma(\tp)$ and
  \begin{eqnarray} \label{eq:DE IE T}
  \IE=\sum_{\tp=2}^m \beta(\tp)^\top \left\{ \sum_{k=1}^{\tp-1} \left(\Phi(\tp-1)\Phi(\tp-2)\ldots\Phi(k+1)%\prod_{l=k+1}^{\tp-1} \Phi(l)
  \right) \Gamma(k) \right\},
  \end{eqnarray} 
  where by convention, the product $\Phi(\tp-1)\Phi(\tp-2)\ldots\Phi(k+1)=1$ when $\tau-1<k+1$. 
\end{prop}

\begin{exmp}\label{exmp:neural}
Neural networks temporal varying  decision process (NN-TVCDP) assumes
  \begin{align*}
  &Y_{i,\tp} 
  =g_{0}(\tp,S_{i,\tp})\cdot \mathbb{I}(A_{i,\tp}=0)+g_{1}(\tp,S_{i,\tp})\cdot \mathbb{I}(A_{i,\tp}=1)+ e_{i,\tp}, \\
  & S_{i,\tp+1}
  =G_{0}(\tp,S_{i,\tp})\cdot \mathbb{I}(A_{i,\tp}=0)+G_{1}(\tp,S_{i,\tp})\cdot \mathbb{I}(A_{i,\tp}=1) + \varepsilon_{i,\tp S},
  \end{align*}
 where  $\mathbb{I}(\cdot)$ denotes the indicator function of an event and $g_{0}(\cdot, \cdot)$, $g_{1}(\cdot, \cdot)$, $G_{0}(\cdot, \cdot)$, and $G_{1}(\cdot, \cdot)$ are parametrized via some (deep) neural networks. %and $\varepsilon_{i,\tp S}$ are some i.i.d. Gaussian random variables.
\end{exmp}

Under NN-TVCDP,   DE and IE are, respectively, given by  
\begin{equation}\label{eq:nn de ie}
    \DE=\sum_{\tp=1}^m\Mean \left\{g_{1}\left(\tp, {S}_{\tp}^0\right)-g_{0}\left(\tp, {S}_{\tp}^0\right)\right\}\quad \mbox{and} \quad \IE=\sum_{\tp=1}^m\Mean \left\{g_{1}\left(\tp, {S}_{\tp}^1\right)-g_{1}\left(\tp, {S}_{\tp}^0\right)\right\},
\end{equation}
where $S_{\tp}^0$ and  $S_{\tp}^1$ are defined recursively by $S_{\tp}^0=G_0(\tp-1,S_{\tp-1}^0)$ and $S_{\tp}^1=G_1(\tp-1,S_{\tp-1}^1)$.

\subsection{Estimation and testing procedures for DE in the L-TVCDP model}
  \label{sec:DE T}
  We describe our estimation and testing procedures for DE in the L-TVCDP model and present their pseudocode in Algorithm  \ref{alg:T DE}   as follows.

{
  \begin{breakablealgorithm}
  \renewcommand{\algorithmicrequire}{\textbf{Input:}}
  \renewcommand{\algorithmicensure}{\textbf{Output:}}
  \caption{Inference of DE in the L-TVCDP model\label{alg:T DE}}
  \begin{algorithmic}[1]
  \STATE Compute the OLS estimator $\widehat{\bm{\theta}}$ according to \eqref{eqn:OLS}.
  \STATE Employ kernel smoothing to compute a refined estimator $\widetilde{\bm{\theta}}$ according to \eqref{eqn:kernelEst} and calculate the estimate $\widehat{\DE}$ by \eqref{eqn:DE}.   
  \STATE Estimate the variance of $\widehat{\bm{\theta}}$ as follows:\\
  \quad (3.1).  Estimate the conditional variance of $\bm{Y}_i$ given $\{Z_{i,\tp}\}_{\tp}$ using \eqref{eqn:sigmay}; \\
  \quad (3.2). Estimate the variance of $\widehat{\bm{\theta}}$ by the sandwich estimator \eqref{eqn:sandwich}. 
  \STATE Estimate the variance of $\widetilde{\bm{\theta}}$ by $\widetilde{\bm{V}}_{\theta}=\bm{\Omega} \widehat{\bm{V}}_{\theta} \bm{\Omega}^\top$ and compute the standard error of $\widehat{\DE}$, denoted by $\widehat{se}(\widehat{\DE})$.
  \STATE Reject $H_0^{DE}$ if $\widehat{\DE}/\widehat{se}(\widehat{\DE})$ exceeds the upper $\alpha$th quantile of a standard normal distribution.
  \end{algorithmic}  
  \end{breakablealgorithm}
}

  Step 1 of Algorithm  \ref{alg:T DE}  is to obtain an initial estimator of   $\theta(\tp)$ by computing its   ordinary least squares (OLS)  estimator, defined as
  \begin{eqnarray}\label{eqn:OLS}
  \widehat{\theta}(\tp)=(\sum_{i=1}^n Z_{i,\tp} Z_{i,\tp}^\top )^{-1} 
  (\sum_{i=1}^n Z_{i,\tp} Y_{i,\tp})~~~\mbox{for}~~ 1\le \tp\le m. 
  \end{eqnarray}

  Step 2 of Algorithm  \ref{alg:T DE}  is to employ kernel smoothing to refine the initial estimator. Specifically, for a given kernel function $K(\cdot)$, we introduce the refined estimator
  \begin{eqnarray}\label{eqn:kernelEst}
  \widetilde{\theta}(\tp) = (\widetilde{\beta}_0(\tp), \widetilde{\beta}(\tp)^\top,  \widetilde{\gamma}(\tp))^\top=\sum_{\tp=1}^m \omega_{\tp,h}(t) \widehat{\theta}(\tp),
  \end{eqnarray}
  for any $t\in [0,m]$ and a bandwidth parameter $h$, where  $\omega_{\tp,h}(t)=K((t-\tp)/(mh))/\sum_{j=1}^m K((t-j)/(mh))$ is 
  the weight function.    Our DE estimator   is given by
  \begin{eqnarray}\label{eqn:DE}
   \widehat{\DE} = \sum_{\tp=1}^m \widetilde{\gamma}(\tp).
  \end{eqnarray}
  We will show in Section \ref{sec:theoretical analysis} that as $\min(n,m)\rightarrow\infty$, $\widehat{\DE}$ is asymptotically normal.
  To derive a Wald test for \eqref{hypo:de t}, it remains to estimate its variance $\Var(\widehat{\DE})$.

  There are two major advantages of using the smoothing step here. 
  First, it allows us to estimate the time-varying coefficient curve $\theta(t)$ without restricting $t$ to the class of integers. 
  Second, the smoothed estimator has smaller variance, leading to a more powerful test statistics. To elaborate, according to model \eqref{model:TVCM DE} for L-TVCDP, the variation of the OLS estimator comes from two sources, the day-specific random effect and the measurement error. The use of smoothing  removes the random fluctuations due to the measurement error. See 
  Theorem \ref{thm:t theta asymp} in Section \ref{sec:theoretical analysis} for a formal statement.  This smoothing technique has been widely applied in the analysis of varying-coefficient models \citep[see e.g.,][]{zhu2014spatially}.

  Step 3 of Algorithm  \ref{alg:T DE} is to estimate the covariance matrix of the initial estimator $\widehat{\bm{\theta}}=(\widehat{\theta}^\top(1),\ldots,\widehat{\theta}^\top(m))^\top$. We first estimate the residual $e_{i,\tp}$ by $\widehat{e}_{i,\tp}=Y_{i,\tp}-Z_{i,\tp}^\top \widetilde{\theta}(\tp)$. It allows us to estimate the day-specific random effect via smoothing, i.e., 
  $\widehat{\eta}_{i}(t)=\sum_{j=1}^m \omega_{j,h}(t)\widehat{e}_{i,\tp}.  $ 
  Second,   the measurement error can be estimated by  
  $ \widehat{\varepsilon}_{i,\tp}=\widehat{e}_{i,\tp}-\widehat{\eta}_{i,\tp}$ 
  for any $i$ and $\tp$, where $\widehat{\eta}_{i,\tp}=\widehat{\eta}_{i}(\tp)$.  Third, we  estimate the conditional covariance matrix of $\bm{Y}_i=(Y_{i,1},\ldots,Y_{i,m})^\top$ given $\{Z_{i,\tp}\}_{\tp}$ based on these estimated residuals.  Under model \eqref{model:TVCM DE} for  L-TVCDP, the covariance between $Y_{i,\tp_1}$ and $Y_{i,\tp_2}$ conditional on $\{Z_{i,\tp}\}_{\tp}$ is given by $\Sigma_y(\tp_1,\tp_2)=\sigma_{\varepsilon,\tp_1}^2\mathbb{I}(\tp_1=\tp_2)+\Sigma_{\eta}(\tp_1,\tp_2),$  which can be consistently estimated by 
  \begin{eqnarray}\label{eqn:sigmay}
  \widehat{\Sigma}_y(\tp_1,\tp_2)\equiv \frac{1}{n}\sum_{i=1}^n \widehat{\varepsilon}_{i,\tp_1}^2 \mathbb{I}(\tp_1=\tp_2)+\frac{1}{n}\sum_{i=1}^n\widehat{\eta}_{i,\tp_1}\widehat{\eta}_{i,\tp_2}.
  \end{eqnarray}
  This allows us to estimate $\Var(\bm{Y}_i|\{Z_{i,\tp}\}_{\tp})$ by $\widehat{\bm{\Sigma}}=\{\widehat{\Sigma}_y(\tp_1,\tp_2)\}_{\tp_1,\tp_2}$. Finally, the covariance matrix of $\widehat{\bm{\theta}}$ can be consistently estimated by the sandwich estimator,
  \begin{eqnarray}\label{eqn:sandwich}
  \widehat{\bm{V}}_{\theta}=(\sum_{i=1}^n \bm{Z}_i^\top \bm{Z}_i)^{-1}  (\sum_{i=1}^n \bm{Z}_i^\top \widehat{\bm{\Sigma}} \bm{Z}_i)  (\sum_{i=1}^n \bm{Z}_i^\top \bm{Z}_i )^{-1},
  \end{eqnarray}
  where $\bm{Z}_i$ is a block-diagonal matrix computed by aligning $Z_{i,1}^\top$, $\ldots$, $Z_{i,m}^\top$ along its diagonal.

  Step 4 of Algorithm  \ref{alg:T DE} is to estimate the covariance matrix of the refined estimator $\widetilde{\bm{\theta}}=(\widetilde{\theta}^\top(1),\ldots,\widetilde{\theta}^\top(m))^\top$. A key observation is that each $\widetilde{\theta}(\tp)$ is essentially a weighted average of $\{\widehat{{\theta}}(\tp)\}_{\tp}$. Writing in matrix form, we have $\widetilde{\bm{\theta}}=\bm{\Omega} \widehat{\bm{\theta}}$, where $\bm{\Omega}$ is a block-diagonal matrix computed by aligning $\omega_{1, h}(\tp)\bm{J}_{p}$, $\ldots$, $\omega_{m, h}(\tp)\bm{J}_{p}$ along its diagonal and $\bm{J}_p$ is a $p\times p$ matrix of ones. 
  As such, we estimate the covariance matrix of $\widetilde{\bm{\theta}}$ by
  $
  \bm{\widetilde{V}}_{\theta}=\bm{\Omega} \bm{\widehat{V}}_{\theta} \bm{\Omega}^\top.
  $ 
  This in turn yields a consistent estimator for the variance of $\widehat{\DE}$, as $\widehat{\DE}$ is a linear combination of $\widetilde{\bm{\theta}}$.

  Step 5 of Algorithm  \ref{alg:T DE} is to  construct a Wald-type test statistic based on $\widehat{\DE}$ and its standard error $\widehat{se}(\widehat{\DE})$. We reject the null hypothesis in \eqref{hypo:de t} if $\widehat{\DE}/\widehat{se}(\widehat{\DE})$ exceeds the upper $\alpha$th quantile of a standard normal distribution. Size and power properties of the proposed test are investigated in Section \ref{sec:theoretical analysis}. 

\subsection{Estimation and testing procedures for IE  in the L-TVCDP model }
\label{sec:IE T}
  We describe our estimation and testing procedures  for IE  in the L-TVCDP model and present their pseudocode in Algorithm  \ref{alg:T IE}   as follows. %As commented before, the 

  \begin{breakablealgorithm}
  \renewcommand{\algorithmicrequire}{\textbf{Input:}}
  \renewcommand{\algorithmicensure}{\textbf{Output:}}
  \caption{Inference of IE in the L-TVCDP model\label{alg:T IE}}
  \begin{algorithmic}[1]
  \STATE Compute the OLS estimator 
  \begin{equation*}
   \widehat{\bm{\Theta}} =\{\widehat{\Theta}(1),\ldots,\widehat{\Theta}(m-1)\}^\top  =  \{\sum_{i=1}^n \bm{Z}_{i,(-m)} \bm{Z}_{i,(-m)}^\top \}^{-1}
    \{\sum_{i=1}^n \bm{Z}_{i,(-m)} \bm{S}_{i,(-1)}^\top \},
  \end{equation*}
  where $\bm{S}_{i,(-1)}$ and $\bm{Z}_{i,(-m)}$ are block-diagonal matrices computed by aligning $S_{i,2}^\top$, $\ldots$, $S_{i,m}^\top$ and $Z_{i,1}^\top$, $\ldots$, $Z_{i,m-1}^\top$ along their diagonals, respectively. 
  \STATE Compute the refined estimator $\widetilde{\bm{\Theta}}=\{\widetilde{\Theta}(1),\ldots,\widetilde{\Theta}({m-1})\}^\top= \bm{\Omega}  \widehat{\bm{\Theta}}$. 
  \STATE Construct the plug-in estimator $\widehat{\IE}$ according to \eqref{eqn:IE}. 
  \STATE Compute the estimated residual $\widehat{\varepsilon}_{i,\tp S}=S_{i,\tp+1}-Z_{i,\tp} \widetilde{\Theta}(\tp)$ for any $i$ and $\tp$.
  \FOR{$b=1,\ldots,B$}
  \STATEx Generate i.i.d. standard normal random variables $\{\xi_i^b\}_{i=1}^n$;
  \STATEx Generate pseudo outcomes $\{\widehat{S}_{i,\tp}^b\}_{i,\tp}$ and $\{\widehat{Y}_{i,\tp}^b\}_{i,\tp}$ according to \eqref{eqn:pseudooutcome};
  \STATEx Repeat Steps 1-2 in Algorithm \ref{alg:T DE} and Steps 1-3 in Algorithm \ref{alg:T IE} to compute $\widehat{\IE}^b$. 
  \ENDFOR
  \STATE Reject $H_0^{IE}$ if $\widehat{\IE}$ exceeds the upper $\alpha$th empirical quantile of $\{\widehat{\IE}^b-\widehat{\IE}\}_b$. 
  \end{algorithmic}  
  \end{breakablealgorithm}

  Steps 1-3 of    Algorithm  \ref{alg:T IE} are to compute a consistent estimator $\widehat{\IE}$ for IE. Specifically, in Step 1 of    Algorithm  \ref{alg:T IE}, we apply OLS regression to derive an initial estimator $\widehat{\bm{\Theta}}$ for $\bm{\Theta}=\{{\Theta}(1),\ldots,  {\Theta}({m-1})\}^\top$. In Step 2 of    Algorithm  \ref{alg:T IE}, we employ kernel smoothing to compute a refined estimator $\widetilde{\bm{\Theta}}=\bm{\Omega} \widehat{\bm{\Theta}}$ to improve its statistical efficiency, as in Algorithm \ref{alg:T DE}. In Step 3 of    Algorithm  \ref{alg:T IE}, we  plug  in $\widetilde{\bm{\Theta}}$ and $\widetilde{\bm{\theta}}$ for $\bm{\Theta}$ and $\bm{\theta}$ in model \ref{exmp:linear}, leading to   
  \begin{eqnarray}\label{eqn:IE}
  \widehat{\IE}=\sum_{\tp=2}^m \widetilde{\beta}(\tp)^\top \left\{ \sum_{k=1}^{\tp-1} \left( \widetilde{\Phi}(\tp-1)\widetilde{\Phi}(\tp-2)\ldots\widetilde{\Phi}(k+1) \right) \widetilde{\Gamma}(k) \right\},
  \end{eqnarray}
  where $\widetilde{\beta}(\tau)$, $\widetilde{\Phi}(\tau)$ and $\widetilde{\Gamma}(\tau)$ are the corresponding estimators for $\beta(\tau)$, $\Phi(\tau)$ and $\Gamma(\tau)$, respectively. 

  Step 4 of    Algorithm  \ref{alg:T IE} is to compute the estimated residuals $\widehat{E}_{i,\tp}=S_{i,\tp+1}-Z_{i,\tp} \widetilde{\Theta}(\tp)$ for all $i$ and $\tp$, which are used to generate pseudo outcomes in the subsequent bootstrap step. 

  Step 5 of    Algorithm  \ref{alg:T IE} is to use bootstrap to simulate the distribution of $\widehat{\IE}$ under the null hypothesis.  
  The key idea is to compute the bootstrap samples for  $\widetilde{\bm{\theta}}$ and $\widetilde{\bm{\Theta}}$ and use the plug-in principle to construct the bootstrap samples for $\widehat{\IE}$.  A key observation is that $\widetilde{\bm{\theta}}$ and $\widetilde{\bm{\Theta}}$ depend linearly on the random errors, so the wild bootstrap method \citep{wu1986jackknife} is applicable. 
  We begin by generating i.i.d. standard normal random variables $\{\xi_i\}_{i=1}^n$. We next generate pseudo-outcomes given by 
  \begin{eqnarray}\label{eqn:pseudooutcome}
  \widehat{S}_{i,\tp+1}=\widetilde{\Theta}(\tp)\widehat{Z}_{i,\tp}+\xi_i \widehat{\varepsilon}_{i,\tp S}  \,\,\hbox{and}\,\,\widehat{Y}_{i,\tp}=\widehat{Z}_{i,\tp}^\top \widetilde{\theta}(\tp)+\xi_i \widehat{e}_{i,\tp},
  \end{eqnarray}
  where $\widehat{Z}_{i,\tp}$ is a version of $Z_{i,\tp}$ with $S_{i,\tp}$ replaced by $\widehat{S}_{i,\tp}$.
  Furthermore,  we apply Steps 1-2 of Algorithm \ref{alg:T DE} and Steps 1-3 of Algorithm \ref{alg:T IE} to compute the bootstrap version of $\widehat{\IE}$ based on these pseudo outcomes in  (\ref{eqn:pseudooutcome}). The above procedures are repeatedly applied to simulate a sequence of bootstrap estimators $\{\widehat{\IE}^b\}_{b=1}^B$ based on which the decision region can be derived. 

\subsection{Estimation procedure in NN-TVCDP model}

{
We first introduce how to estimate the regression functions $g_{0}$, $g_{1}$, $G_{0}$ and $G_{1}$. 
Take $g_{0}$ as an instance, we consider minimizing the following empirical objective function
$$\sum_{i=1}^n\sum_{\tp=1}^m%\mathcal{D}_0}
(1-A_{i,\tp})\left\{Y_{i,\tp}-g_{0}(\tp,S_{i,\tp})\right\}^2.$$
% to obtain estimates of $a_{k\ell}$, which we denote by $\widehat{a}_{k\ell}$. 
Instead of separately estimating $g_0(\tau,\bullet)$ for each $\tau$, we treat $\tau$ as part of the features and jointly estimate $\{g_0(\tau,\bullet)\}_{\tau}$ by solving the above optimization. It allows us to borrow information across different time points to improve the estimation accuracy.

Next, we introduce the estimation procedures for DE and IE. We impose a parametric model (e.g., Gaussian) for the density function $f_{\varepsilon_{\tau S}}$ of the measurement error $\varepsilon_{i,\tau S}$ and summarize the steps below.

\begin{enumerate}
\item[1.] Use neural networks to estimate $g_0$, $g_1$, $G_0$ and $G_1$ by solving their corresponding least square objective functions. Denote the corresponding estimators by $\widehat g_{0}$, $\widehat g_{1}$, $\widehat G_{0}$,  and $\widehat G_{1}$, respectively. 
\item[2.] Compute the residual
$\widehat\varepsilon_{i,\tp S}=
S_{i,\tp+1}-\left\{\widehat G_{0}(\tp, S_{i,\tp})\cdot \mathbb{I}(A_{i,\tp}=0)+\widehat G_{1}(\tp,S_{i,\tp})\cdot \mathbb{I}(A_{i,\tp}=1)\right\}$ 
and use $\widehat\varepsilon_{i,\tp S}$ to compute the density function estimator $\widehat{f}_{\varepsilon_{\tau S}}$.
\item[3.] Use Monte Carlo to estimate the distributions of the potential states $S_{i,\tp}^*(\bm{1}_{\tp-1})$ and $S_{i,\tp}^*(\bm{0}_{\tp-1})$ conditional on $S_{i,1}$. Specifically, for $\tau=1,\ldots,m$, $i=1,\ldots,n$, and $k=1,\ldots,M$,  we use $\widehat{f}_{\varepsilon_{\tau S}}$ to generate error residuals $\{\widehat{\varepsilon}_{i,\tau S, k}\}_{k=1}^M$,  where $M$ denotes the number of Monte Carlo replications. Next, we set $\widehat{S}_{i,1,k}^1=\widehat{S}_{i,1,k}^0=S_{i,1}$ for any $i$ and $k$, and sequentially construct Monte Carlo samples $\{\widehat{S}_{i,\tp,k}^1\}_{k=1}^{M},\{\widehat{S}_{i,\tp,k}^0\}_{k=1}^{M}$ by setting
$ \widehat{S}_{i,\tau+1,k}^1=\widehat{G}_1(\tau,\widehat{S}_{i,\tau,k}^1)+\widehat{\varepsilon}_{i,\tau S, k} $ and $ \widehat{S}_{i,\tau+1,k}^0=\widehat{G}_0(\tau,\widehat{S}_{i,\tau,k}^0)+\widehat{\varepsilon}_{i,\tau S, k}
$.
\item[4.] Based on \eqref{eq:nn de ie}, we estimate DE and IE by using 
\begin{eqnarray*} 
\widehat{DE}&=&\frac{1}{nM}\sum_{i=1}^n\sum_{k=1}^M\sum_{\tp=1}^m\left\{ \widehat g_{1}(\tp, \widehat{S}_{i,k,\tp}^0)- \widehat g_{0}(\tp, \widehat{S}_{i,k,\tp}^0)\right\}~~\mbox{and}~~ \\
\widehat{IE}&=&\frac{1}{nM}\sum_{i=1}^n\sum_{k=1}^M\sum_{\tp=2}^m\left\{ \widehat g_{1}(\tp, \widehat{S}_{i,k,\tp}^1)-\widehat g_{1}(\tp, \widehat{S}_{i,k,\tp}^0)\right\}.
\end{eqnarray*}
\end{enumerate}
}

\section{Policy evaluation for spatio-temporal dependent experiments}
\label{sec:ST method}

  In this section, we present the proposed methodology for policy evaluation in spatio-temproal dependent experiments by extending our proposal in temporal dependent experiments. We highlight 
  %under the spatio-temporal alternation design for  multiple experimental regions by extending that for the temporal alternation design in  Section \ref{sec:po_mdp}. For simplicity, we describe 
  several key differences between the spatio-temporal dependent experiment and the temporal dependent one. 
  %alternation design and the temporal one.      

\subsection{A potential outcome framework}  \label{sec:STPOF}
 
  Firstly, we introduce the spatio-temporal dependent experiments %alternation design 
  as follows. 
  Specifically, a city is split into $r$ non-overlapping regions. Each region receives a sequence of policies over time and different regions may receive different policies at the same time. In our  application, we employ the spatio-temporal dependent alternation design to randomize these policies. In each region, we independently randomize the initial policy (either A or B) and then apply the temporal alternation design. %Consequently, different regions may receive different policies at the same time. 
  %with the criteria to make orders of each region answered by drivers of the same region. Then the experiment is conducted simultaneously on these regions with different align strategies.
  As discussed in the introduction, one major challenge for policy evaluation is that the spatial proximities will induce spatio-temporal interference among locations across time. In the example of ride-sharing platforms, for many call orders, their pickup locations and destinations belong to different regions. Therefore, applying an order dispatch policy at one region will change the distribution of drivers of its neighbouring areas as well, so the order dispatch policy at one location could influence  outcomes of those neighbouring areas, inducing interference among spatial units. 

  Secondly, to quantify the spatio-temporal interference, we allow the potential outcome of each region to depend on polices applied to its neighbouring areas as well. 
  %The system is rather complex and the interference (also called spillover) happens. First, the division criteria is satisfied among controlled groups, however, in treatment groups, orders around the edge of a region may be answered by drivers in the neighboring regions. Second, the destinations of orders are usually cross regional, and the treatments of one region can change the distribution of drivers on the whole domain and influence the future supply and demand relationship of all regions.  To handle this, denote 
  Specifically, for the $\iota$th region, let $\bar{a}_{\tp,\iota}=(\bar{a}_{1,\iota},\ldots,\bar{a}_{\tp,\iota})^\top$ denote its treatment history up to time $\tp$ and   $\mathcal{N}_\iota$ denote   the neighbouring regions of $\iota$.  Let $\bar{a}_{\tp,[1:r]}=(\bar{a}_{\tp,1},\ldots,\bar{a}_{\tp,r})^\top$ denote the treatment history associated with all regions. Similarly, let $S_{\tp,\iota}^*(\bar{a}_{\tp-1,[1:r]})$ and $Y_{\tp,[1:r]}^*(\bar{a}_{\tp,[1:r]})$ denote the potential state and outcome associated with the $\iota$th region, respectively. Let $S_{\tp,[1:r]}^*(\bar{a}_{\tp-1,[1:r]})$ denote the set of potential states at time $\tp$. 

Similarly, we introduce CA and SRA in the spatio-temporal case as follows. 
  \begin{itemize} 
  \item{\bf CA}. $S_{\tp+1,\iota}^*(\bar{A}_{\tp,[1:r]})=S_{\tp+1,\iota}$ and $Y_{\tp,\iota}^*(\bar{A}_{\tp,[1:r]})=Y_{\tp,\iota}$ for any $\tp\ge 1$ and $1\le\iota\le r$,  where $\bar{A}_{\tp,[1:r]}$ denotes the set of observed treatment history up to time $\tp$. 
  \item {\bf SRA}. $A_{\tp,[1:r]}$, the set of observed policies at time $\tp$, is conditionally independent of all potential variables given $S_{\tp,[1:r]}$ and $\{(S_{j,[1:r]},A_{j,[1:r]},Y_{j,[1:r]})\}_{j<\tp}$. 
  \end{itemize}
  SRA automatically holds under the spatio-temporal alternation design, in which the policy assignment mechanism is conditionally independent of the data given the policies assigned at the initial time point. 

  Thirdly, we are interested in the overall treatment effects. Define ATE as the difference between the new and old policies aggregated over different regions.
  \begin{defs}
  %For a given reference distribution function $\mathbb{G}$, ATE %of temporal-alternating design 
  ATE is defined as  the difference between two value functions  given by 
  \begin{align*}
    \textrm{ATE}_{st}=\sum_{\iota=1}^r\sum_{\tp =1}^{m} \Mean \{Y_{\tp,\iota}^*(\bm{1}_{\tp,[1:r]})-Y_{\tp,\iota}^*(\bm{0}_{\tp,[1:r]})\}.
  \end{align*}
  % where $\bm{1}_{\tp}$ and $\bm{0}_{\tp}$ denote vectors of 1s and 0s of length $\tp$, respectively. 
  \end{defs}
  Let $R_{\tp,\iota}$ denote the conditional mean function of $Y_{\tp,\iota}^*(\bar{a}_{\tp,[1:r]})$ given the past policies and potential states. 
  Similarly, we can decompose ATE as the sum of DE and IE, which are, respectively,  given by
  %the treatment effects on the whole domain. Define  
  \begin{align*}
  \DE_{st}&=\sum_{\iota=1}^r\sum_{\tp=1}^m \Mean \{R_{\tp,\iota}\big(\bm{1}_{\tp,[1:r]},S_{\tp,\iota}^*(\bm{0}_{\tp-1,[1:r]}),\bm{0}_{\tp-1,[1:r]},\ldots,S_1\big)-R_{\tp,\iota}\big(\bm{0}_{\tp,[1:r]},S_{\tp,\iota}^*(\bm{0}_{\tp-1,[1:r]}),\bm{0}_{\tp-1,[1:r]},\ldots,S_1\big)\},\\
  \IE_{st}&=\sum_{\iota=1}^r\sum_{\tp=1}^m \Mean \{R_{\tp,\iota}\big(\bm{1}_{\tp,[1:r]},S_{\tp,\iota}^*(\bm{1}_{\tp-1,[1:r]}),\bm{1}_{\tp-1,[1:r]},\ldots,S_1\big)-R_{\tp,\iota}\big(\bm{1}_{\tp,[1:r]},S_{\tp,\iota}^*(\bm{0}_{\tp-1,[1:r]}),\bm{0}_{\tp-1,[1:r]},\ldots,S_1\big)\}.
  \end{align*}
  %and similar to (\ref{hypo:de t})--(\ref{hypo:ie t}), we want to test:
  We aim to test the following hypotheses: 
  \begin{eqnarray}
  &H_0^{DE}: {\DE}_{st}\le 0\,\,\,\,{v.s}\,\,\,\,H_1^{DE}:{\DE}_{st}>0,\label{hypo de st}\\
  &H_0^{IE}: {\IE}_{st}\le 0\,\,\,\,{v.s}\,\,\,\,H_1^{IE}:{\IE}_{st}>0.\label{hypo ie st}
  \end{eqnarray}

\subsection{Spatio-temporal VCDP models}\label{sec:STVCDP}

  We introduce the spatio-temporal VCDP (STVCDP) models to model $Y_{\tp,\iota}$ and $S_{\tp,\iota}$, respectively. Suppose that the experiment is conducted across $r$ regions over $n$ days. Let 
  $(S_{i,\tp,\iota}, A_{i,\tp,\iota}, Y_{i,\tp,\iota})$ denote the state-policy-outcome triplet measured from  the $\iota$th region at the $\tp$th time interval of  the $i$th day for $i=1, \ldots, n$, $\tp=1, \ldots, m$, and $\iota=1, \ldots, r$. 
%  \change }
  The STVCDP model is given as follows,
  \begin{align*}
    % \label{model:STVCM DE}
    Y_{i,\tp,\iota} &= f_{1,\tp,\iota}\big(S_{i,\tp,\iota},A_{i,\tp,\iota},\bar{A}_{i,\tp,\mathcal{N}_\iota}\big) + e_{i,\tp,\iota},\\
    S_{i,\tp+1,\iota} &= f_{2,\tp,\iota}\big(S_{i,\tp,\iota},A_{i,\tp,\iota},\bar{A}_{i,\tp,\mathcal{N}_\iota}\big) + \epsilon_{i,\tp,\iota},\nonumber
  \end{align*}
  where %$Z_{i,\tp,\iota} = (1, S_{i,\tp,\iota}^\top, A_{i,\tp,\iota}, \bar{A}_{i,\tp,\mathcal{N}_\iota})^\top$, 
  $\bar{A}_{i,\tp,\mathcal{N}_\iota}$ denotes the average of $\{A_{i,\tp,k}\}_{k\in \mathcal{N}_\iota}$,  
  % $\theta(\tp,\iota),\Theta(\tp,\iota)\in\R^{d\times (d+2)}$ are the sets of spatio-temporal varying coefficients, 
  and $\{e_{i,\tp,\iota},\epsilon_{i,\tp,\iota}\}$ are the random noises. 
  % We assume $e_{i,\tp,\iota}=\eta_{i,\tp,\iota}+\varepsilon_{i,\tp,\iota}=\eta_{i,\tp,\iota}^I+\eta_{i,\tp,\iota}^{II}+\eta_{i,\tp,\iota}^{III}+\varepsilon_{i,\tp,\iota}$, where the random errors $\{\eta_{i,\tp,\iota}^I\}$, $\{\eta_{i,\tp,\iota}^{II}\}$, $\{\eta_{i,\tp,\iota}^{III}\}$, and $\{\varepsilon_{i,\tp,\iota}\}$ for $i=1,\ldots,n$ are independent, and are i.i.d. copies of  zero mean random processes with covariance structures $\Sigma_{\eta^I}(\tp_1,\iota_1,\tp_2,\iota_2)$,  $\Sigma_{\eta^{II}}(\tp_1,\iota_1,\tp_2)\mathbb{I}(\iota_1=\iota_2)$, $\Sigma_{\eta^{III}}(\tp_1,\iota_1,\iota_2)\mathbb{I}(\tp_1=\tp_2)$, and $\sigma_{\varepsilon}^2(\tp_1,\iota_1)\mathbb{I}((\tp_1,\iota_1)=(\tp_2,\iota_2))$, respectively, and the measurement errors $\{\varepsilon_{i, \tp,\iota}\}_{i, \tp, \iota}$ and $\{\epsilon_{i, \tp,\iota}\}_{i, \tp, \iota}$ are independent across different location/time combinations,  $\{\eta_{i, \tp,\iota}^{II}\}_{i, \tp,\iota}$ are independent across different regions, and $\{\eta_{i, \tp,\iota}^{III}\}_{i, \tp,\iota}$ are independent over time. 
  \change{In parallel to Assumption \ref{assump:eps t}, we impose the following noise assumption for the STVCDP model.
  \begin{asmp}\label{assump:eps st}
 (i) The outcome noise $e_{i,\tp,\iota}%=\eta_{i,\tp,\iota}+\varepsilon_{i,\tp,\iota}=$$
  	=\eta_{i,\tp,\iota}^I+\eta_{i,\tp,\iota}^{II}+\eta_{i,\tp,\iota}^{III}+\varepsilon_{i,\tp,\iota}$ can be decomposed into four mutually independent processes: %and state noises $\epsilon_{i,\tp,\iota}$ satisfy: 
  	$\{\eta_{i,\tp,\iota}^I\}$, $\{\eta_{i,\tp,\iota}^{II}\}$, $\{\eta_{i,\tp,\iota}^{III}\}$, and $\{\varepsilon_{i,\tp,\iota}\}$.  %, and $\{\epsilon_{i,\tp,\iota}\}$ for $i=1,\ldots,n$ are independent; 
  	(ii) The $\{\eta_{i,\tp,\iota}^I\}$, $\{\eta_{i,\tp,\iota}^{II}\}$ and $\{\eta_{i,\tp,\iota}^{III}\}$ are i.i.d. copies of some zero-mean random processes with covariance functions $\Sigma_{\eta^I}(\tp_1,\iota_1,\tp_2,\iota_2)$,  $\Sigma_{\eta^{II}}(\tp_1,\iota_1,\tp_2)\mathbb{I}(\iota_1=\iota_2)$, and $\Sigma_{\eta^{III}}(\tp_1,\iota_1,\iota_2)\mathbb{I}(\tp_1=\tp_2)$, respectively.  These covariance functions have bounded and continuously differentiable second-order derivatives. 
   (iii) %the measurement errors $\{\varepsilon_{i, \tp,\iota}\}_{i, \tp, \iota}$ and $\{\epsilon_{i, \tp,\iota}\}_{i, \tp, \iota}$ are independent across different location/time combinations with zero means,
The measurement errors $\{\varepsilon_{i,\tp,\iota}\}_{i,\tp,\iota}$ and the state noises $\{\epsilon_{i,\tp,\iota}\}_{i,\tp,\iota}$ are independent over different location/time combinations, have zero means, and satisfy $\Var(\varepsilon_{i,\tp,\iota})=\sigma_{\varepsilon}^2(\tp,\iota)$ and $\Cov(\epsilon_{i,\tp,\iota})=\Sigma_{\epsilon,\tp,\iota}$. 
      %The outcome noise $e_{i,\tp,\iota}=\eta_{i,\tp,\iota}+\varepsilon_{i,\tp,\iota}=$$\eta_{i,\tp,\iota}^I+\eta_{i,\tp,\iota}^{II}+\eta_{i,\tp,\iota}^{III}+\varepsilon_{i,\tp,\iota}$ and state noises $\epsilon_{i,\tp,\iota}$ satisfy: $\{\eta_{i,\tp,\iota}^I\}$, $\{\eta_{i,\tp,\iota}^{II}\}$, $\{\eta_{i,\tp,\iota}^{III}\}$, $\{\varepsilon_{i,\tp,\iota}\}$, and $\{\epsilon_{i,\tp,\iota}\}$ for $i=1,\ldots,n$ are independent; $\{\eta_{i,\tp,\iota}^I\}$, $\{\eta_{i,\tp,\iota}^{II}\}$ and $\{\eta_{i,\tp,\iota}^{III}\}$ are i.i.d. copies of  zero mean random processes with covariances $\Sigma_{\eta^I}(\tp_1,\iota_1,\tp_2,\iota_2)$,  $\Sigma_{\eta^{II}}(\tp_1,\iota_1,\tp_2)\mathbb{I}(\iota_1=\iota_2)$, $\Sigma_{\eta^{III}}(\tp_1,\iota_1,\iota_2)\mathbb{I}(\tp_1=\tp_2)$, respectively, where $\Sigma_{\eta^{I}}(\tp_1, \tp_2,\iota_1,\iota_2),\Sigma_{\eta^{II}}(\tp_1, \iota_1, \tp_2),$ and $\Sigma_{\eta^{III}}(\tp_1, \iota_1, \iota_2)$ have bounded and continuous second-order derivatives; the measurement errors $\{\varepsilon_{i, \tp,\iota}\}_{i, \tp, \iota}$ and $\{\epsilon_{i, \tp,\iota}\}_{i, \tp, \iota}$ are independent across different location/time combinations with zero means, $\Var(\varepsilon_{i,\tp,\iota})=\sigma_{\varepsilon}^2(\tp,\iota)$ and $\Cov(\epsilon_{i,\tp,\iota})=\Sigma_{\epsilon,\tp,\iota}$.
  \end{asmp}}

  \change{We make  three remarks.   Firstly, as per the STVCDP model, the outcome in the $\iota$th region is influenced solely by the current actions $A_{i,\tp,\iota}$ and those from its neighboring areas. This assumption is often valid in various applications, such as ride-sharing platforms. For instance, the policy in one location may impact other locations only through its effect on the distribution of drivers. Within each time unit, a driver can travel at most from one location to its neighboring ones. Consequently, outcomes in one location are independent of policies applied to non-adjacent locations.

   Secondly, in our spatial interference model, we adopt the mean field approximation. Under this approach, the outcome $Y_{\tp,\iota}$ and next state $S_{\tp+1,\iota}$ in a given region depend on the treatments of neighboring regions  $\{A_{\tp,k}\}_{k\in \mathcal{N}{\iota}}$ only through their average  $\bar{A}_{\tp,\mathcal{N}{\iota}}$. The mean field approximation is a commonly used technique in multi-agent reinforcement learning for policy learning and evaluation. It's worth noting that studies, such as \citet{shi2022multi}, have shown that the average effect $\bar{A}_{\tp,\mathcal{N}{\iota}}$ effectively summarizes the impact of $\{A_{\tp,k}\}_{k\in \mathcal{N}{\iota}}$. This approach aligns with assumptions frequently made in the causal inference literature dealing with spatial interference \citep{sobel_what_2006,hudgens2008toward,zigler_estimating_2012,perez-heydrich_assessing_2014,sobel_causal_2014,liu_inverse_2016,savje2021average}.

  	%Third, in addition to the average effect, any low dimensional key statistics of $\{A_{ij}:j\in\mathcal{N}_\iota\}$ could be an alternative, such as $\sum_{j\in\mathcal{N}_\iota} \theta_{\iota j}A_{ij}$ and $\theta_\iota\mathbb{I}_{\{\sum_{j\in\mathcal{N}_\iota}A_{ij}>0\}}$. The estimation and inference results can be derived similarly. 

Thirdly, besides the average effect, alternative low-dimensional summary statistics of $\{A_{ij}:j\in\mathcal{N}_\iota\}$ can be considered, such as $\sum_{j\in\mathcal{N}_\iota} \theta_{\iota j}A_{ij}$ and $\theta_\iota\mathbb{I}_{\{\sum_{j\in\mathcal{N}_\iota}A_{ij}>0\}}$ \citep{hu2022average}. The resulting estimation and inference procedures can be similarly derived.

}

   Similar to model \eqref{model:TVCM DE}, we allow general function approximation for $f_1$ and $f_2$. %in \eqref{model:STVCM DE} can take a general class of functions. 
  %For space economy, we delineated linear STVCDP model (L-STVCDP), but the estimation and inference procedure can be carried over for single index and neural network STVCDP models as discussed in Section \ref{sec:IE T}. The L-STVCDP model is formulated as follows,
  To save space, we focus on linear STVCDP models (L-STVCDP) in the rest of this section. Meanwhile, the proposed estimation procedure can be extended to handle neural network STVCDP models, as in Section \ref{sec:IE T}. The proposed L-STVCDP model is given as follows,
  \begin{align}
    \label{model:STVCM DE}
    Y_{i,\tp,\iota} &= \beta_0(\tp,\iota) + S_{i,\tp,\iota}^\top\beta(\tp,\iota) + A_{i,\tp,\iota}\gamma_1(\tp,\iota)+ \bar{A}_{i,\tp,\mathcal{N}_\iota}\gamma_2(\tp,\iota) + e_{i,\tp,\iota},\\
    S_{i,\tp+1,\iota} &= \phi_0(\tp,\iota) + \Phi(\tp,\iota)S_{i,\tp,\iota} + A_{i,\tp,\iota}\Gamma_1(\tp,\iota) + \bar{A}_{i,\tp,\mathcal{N}_\iota}\Gamma_2(\tp,\iota) + \epsilon_{i,\tp,\iota}, \nonumber
  \end{align}    
  where $Z_{i,\tp,\iota} = (1, S_{i,\tp,\iota}^\top, A_{i,\tp,\iota}, \bar{A}_{i,\tp,\mathcal{N}_\iota})^\top$.
  
  Similar to \eqref{eq:DE IE T}, 
  %we can calculate 
  we can show that $\DE_{st}$ and $\IE_{st}$ are equal to the following,
  \begin{eqnarray} \label{eq:DE IE ST}
  &\DE_{st}=\sum_{\iota=1}^r\sum_{\tp=1}^m\{\gamma_1(\tp,\iota)+\gamma_2(\tp,\iota)\},\\
  &\IE_{st}=\sum_{\iota=1}^r\sum_{\tp=1}^m \beta(\tp,\iota)^\top \left[ \sum_{k=1}^{\tp-1} \left(\Phi(\tp-1,\iota)\ldots\Phi(k+1,\iota)\right) \{\Gamma_1(k,\iota)+\Gamma_2(k,\iota)\} \right],\nonumber
  \end{eqnarray}
  where the product $\Phi(\tp-1,\iota)\ldots\Phi(k+1,\iota)=1$ when $\tp-1<k+1$. 
  These two identities form the basis of our test procedure. 

\subsection{Estimation and testing procedures for DE and IE}  \label{sec:STDEIE}

  We first describe our estimation and testing procedures for DE under the spatio-temporal alternation design and present the pseudocode in Algorithms \ref{alg:DE ST} of  Section \ref{app:alg} of the supplementary document to save space.  

  Step 1 of Algorithm \ref{alg:DE ST} 
  is to independently apply Steps 1 and 2 of Algorithm \ref{alg:T DE} detailed in Section \ref{sec:DE T} to the data subset $\{(Z_{i,\tp,\iota}, Y_{i,\tp,\iota})\}_{i,\tp}$ for each region $\iota$ in order to compute a smoothed estimator $\widetilde{\bm{\theta}}_{st}^0(\iota)=\{\widetilde{\theta}_{st}^0(1,\iota)^\top,\ldots,\widetilde{\theta}^0_{st}(m,\iota)^\top\}^\top$ for $\{\theta(1,\iota)^\top,\ldots,\theta(m,\iota)^\top\}^\top$. 

  Step 2 of Algorithm \ref{alg:DE ST} 
  is to employ kernel smoothing again to spatially smooth each component of $\widetilde{\bm{\theta}}_{st}^0(\iota)$ across all $\iota\in \{1,\ldots,r\}$. %We use the same kernel function $K$ and denote the bandwidth parameter by $h_{st}$. We 
  Specifically, we compute $\widetilde{\bm{\theta}}_{st}(\iota)=\{\widetilde{\theta}_{st}(1,\iota)^\top,\ldots,\widetilde{\theta}_{st}(m,\iota)^\top\}^\top$ as the resulting refined estimator, given by
  $
  \widetilde\theta_{st}(\tp,\iota)=\sum_{\ell=1}^r\kappa_{\ell,h_{st}}(\iota)\widetilde\theta_{st}^0(\tp,\ell),
  $ 
  where $\kappa_{\ell,h_{st}}(\cdot)$ defined in (\ref{eq:omega_st})  is a normalized kernel function with bandwidth parameter $h_{st}$. %See for the detailed definition. 

  We remark that we employ kernel smoothing twice in order to estimate the varying coefficients. In the first step, we temporally smooth the least square estimator to compute $\widetilde{\bm{\theta}}_{st}^0(\iota)$. In the second step, we further spatially smooth $\widetilde{\bm{\theta}}_{st}^0(\iota)$ to compute $\widetilde{\bm{\theta}}_{st}(\iota)$. Therefore, the estimator $\widetilde{\bm{\theta}}_{st}(\iota)$ has smaller variance than $\widetilde{\bm{\theta}}_{st}^0(\iota)$, since we borrow information across neighboring  regions to improve the estimation efficiency. To elaborate this point,   the random effect in \eqref{model:STVCM DE} can be decomposed into three parts: $\eta_{i,\tp,\iota}^{\RNum{1}}+\eta_{i,\tp,\iota}^{\RNum{2}}+\eta_{i,\tp,\iota}^{\RNum{3}}$.  Temporally smoothing the varying coefficient estimator removes the random fluctuations caused by $\eta_{i,\tp,\iota}^{\RNum{3}}$ and the measurement error. Spatially smoothing the estimator further removes the random fluctuations caused by $\eta_{i,\tp,\iota}^{\RNum{2}}$. This in turn implies that the proposed test under the spatio-temporal design is more powerful than the one developed in Section \ref{sec:po_mdp} under the temporal design. Such an observation is consistent with our numerical findings in Section \ref{sec:simuts}.

  Steps 3 and 4 of  Algorithm \ref{alg:DE ST} 
  are to estimate the covariance matrix of $(\widetilde{\bm{\theta}}_{st}(1),\ldots,\widetilde{\bm{\theta}}_{st}(r))^\top$, denoted by $\widetilde{\bm{V}}_{\theta, st}$. These two steps are very similar to Steps 3 and 4 of Algorithm \ref{alg:T DE}. Specifically, we first estimate the measurement errors and random effects based on the estimated varying coefficients. We next use the sandwich formula to compute the estimated covariance matrix for the initial least-square estimator. Then the estimated covariance matrix for $\widetilde{\bm{\theta}}_{st}^0(\iota)$ can be derived accordingly. We use $\widetilde{\bm{V}}_{\theta, st}$ to denote the corresponding covariance matrix estimator.

  Step 5 of Algorithm \ref{alg:DE ST} 
  is to compute the Wald-type test statistic and its standard error estimator. Specifically, let $\widetilde\gamma_1(\tp,\iota)$ and $\widetilde\gamma_2(\tp,\iota)$ be the last two elements of $\widetilde\theta_{st}(\tp,\iota)$, we have 
  $
  \widehat{\DE}_{st}=\sum_{\iota=1}^r\sum_{\tp=1}^m\{\widetilde\gamma_1(\tp,\iota)+\widetilde\gamma_2(\tp,\iota)\}.
  $ 
  We will show in Theorem \ref{thm:st theta asymp}   that $\widehat{\DE}_{st}$ is asymptotically normal. In addition, its standard error $\widehat{se}(\widehat{\DE}_{st})$ can be derived based on $\widetilde{\bm{V}}_{\theta, st}$. 
  This yields our Wald-type test statistic $T_{st}=\widehat{\DE}_{st}/\widehat{se}(\widehat{\DE}_{st})$. We reject the null hypothesis if $T_{st}$ exceeds the upper $\alpha$th quantile of a standard normal distribution. 
  
  We next describe our estimation and testing procedures for IE. The method is very similar to the one discussed in Section \ref{sec:IE T}. We sketch an outline of the algorithm to save space. Details are presented in \ref{alg:ST IE} of  Section \ref{app:alg} of the supplementary document. 
  Specifically, we first plug in the set of smoothed estimators $\{\widetilde{\Theta}_{st}(\tp,\iota)\}_{\tp,\iota}$ and $\{\widetilde{\theta}_{st}(\tp,\iota)\}_{\tp,\iota}$ for $\{\Theta(\tp,\iota)\}_{\tp,\iota}$ and $\{\theta(\tp,\iota)\}_{\tp,\iota}$ to compute $\widehat{\IE}_{st}$, the plug-in estimator of $\IE_{st}$. We next estimate the measurement errors and random effects  and then apply the parametric bootstrap method to compute the bootstrap statistics $\{\widehat{\IE}_{st}^b\}_{b}$. Finally, we reject $H_0^{IE}$ if $\widehat{\IE}_{st}$ exceeds the upper $\alpha$th empirical quantile of $\{\widehat{\IE}_{st}^b-\widehat{\IE}\}_{b}$.

  To conclude this section, we remark that in Sections \ref{sec:po_mdp} and \ref{sec:ST method}, we focus on testing one-sided hypotheses for the direct and indirect effects. However, the proposed method can be easily extended to test two-sided hypotheses as well. %{\color{blue}Please refer to the Appendix for details.} 

\section{Theoretical Analysis}
\label{sec:theoretical analysis}

  In this section, we systematically investigate the asymptotic properties of the proposed estimators and test statistics in L-TVCDP  and derive the convergence rates of our causal estimands in NN-TVCDP. 
  We also explore the benefits of employing the swtichback design and study the theoretical properties of our estimator in  the spatio-temporal dependent experiments. %case which are similar to the temporal case.

  Firstly, we impose the following regularity assumptions for the temporal dependent experiments using L-TVCDP.
  \begin{asmp}\label{assump:kernel}
  The kernel function $K(\cdot)$ is a symmetric probability density function on $ [-1,1] $ and is Lipschitz continuous.
  \end{asmp}

  \begin{asmp}\label{assump:t}
  The covariate $\bm{Z}_i$s are i.i.d.; for $1\leq\tp\leq m$, $\Mean(Z_{i,\tp}^\top Z_{i,\tp})\in\M^{p\times p}$ is invertible; all components of $\theta(t)$ have bounded and continuous second derivatives with respect to $t$. %$Z_{i,\tp}$, $\varepsilon_{i,\tp}$, and $\eta_{i,\tp}$ are mutually independent of each other.
  \end{asmp}

  %{\color{red}
  \begin{asmp}\label{asmp:st1}
  {There exists $0<q<1$ such that the absolute values of eigenvalues of $\Phi(\tp)$ are smaller than $q$,} % and  $\Vert\Phi(\tp)\Vert_{-\infty}\geq q'>0$; 
  and there exist some constants $M_{\Gamma}$ and $M_\beta$ such that $\Vert\Gamma(\tp)\Vert_\infty\leq M_{\Gamma}$ and $\Vert\beta(\tp)\Vert_\infty\leq M_\beta$. $\{\beta(\tp)\}_{2\le \tp\le m}$, $\{\Phi(l) \}_{2\le l\le m-1}$, and $\{\Gamma(k)\}_{1\le k\le m-1}$ must not be all zero. $\Theta(\tp)$ has a continuous second-order partial derivative.
  \end{asmp}
  
  % \begin{asmp}\label{asmp:st2}
  % $\Theta(\tp)$ and the covariance function of $(\eta_{i,1},\ldots,\eta_{i,m})^\top$ have continuous second-order partial derivatives.
  % \end{asmp}

  Assumption \ref{assump:kernel} is mild as the kernel $K(\cdot)$ is user-specified. Assumption \ref{assump:t} has been commonly used in the  literature  on varying coefficient models \citep[see e.g.,][]{zhu2014spatially}.  Assumption \ref{asmp:st1} ensures that the time series is stationary, since  $\Phi(\tp)$ is the autoregressive coefficient. It is commonly imposed in the literature on time series analysis \citep{2010Time}.  

  Before presenting the theoretical properties of the proposed method for L-TVCDP, we introduce some notation. %} 
  For $1\leq\tp_1,\tp_2\leq m$, define $\bm{\Sigma}_y$ and $\bm{\Sigma}_{\eta}$ to be the $m\times m$ matrices $\{\Sigma_y(\tp_1,\tp_2)\}_{\tp_1,\tp_2}$ and $\{\Sigma_{\eta}(\tp_1,\tp_2)\}_{\tp_1,\tp_2}$, respectively. 
  We define 
  $$\bm{V}_{\widehat{\theta}}=(\Mean \bm{Z}_{i}^\top \bm{Z}_i)^{-1}\Mean (\bm{Z}_{i}^\top \bm{\Sigma}_y \bm{Z}_{i})(\Mean \bm{Z}_{i}^\top \bm{Z}_i)^{-1}\ \ \text{and}\ \ \bm{V}_{\widetilde{\theta}}=(\Mean \bm{Z}_{i}^\top \bm{Z}_i)^{-1}\Mean(\bm{Z}_i^\top \bm{\Sigma}_\eta \bm{Z}_{i})(\Mean \bm{Z}_{i}^\top \bm{Z}_i)^{-1}$$
   as the asymptotic covariance matrices of $\widehat{\bm{\theta}}$ and $\widetilde{\bm{\theta}}$,  respectively. Let  $\bm{V}_{\widehat{\theta}}(\tau,\tau)$ and $\bm{V}_{\widetilde{\theta}}(\tau,\tau)$ denote the submatrices of $\bm{V}_{\widehat{\theta}}$ and $\bm{V}_{\widetilde{\theta}}$ that correspond to the asymptotic covariance matrix of $\widehat{\bm{\theta}}$ and $\widetilde{\bm{\theta}}$, respectively. 
  {We first compare the mean squared error (MSE) of the OLS estimator $\widehat\theta(\tau)$ against that of the smoothed estimator $\widetilde\theta(\tau)$ based on L-TVCDP.

  \begin{prop}\label{prop:mse}
  Suppose $\lambda_{\min}(\bm{V}_{\widehat{\theta}}(\tau,\tau))$ and $\lambda_{\min}(\bm{V}_{\widetilde{\theta}}(\tau,\tau))$ are uniformly bounded away from zero for any $\tau$. 
  Under %TCMIA and 
  Assumptions \ref{assump:kernel} and \ref{assump:t}, we have
  $$\sum_{\tp=1}^m\text{MSE}(\widehat\theta(\tp))\asymp n^{-1}\hbox{trace}(\bm{V}_{\widehat{\theta}}),\,\ \sum_{\tp=1}^m\text{MSE}(\widehat\theta(\tp))\asymp n^{-1}\hbox{trace}(\bm{V}_{\widetilde{\theta}})+O(mh^4+m^{-1}).$$
  \end{prop}

  Proposition \ref{prop:mse} has an important implication. 
  Both trace$(\bm{V}_{\widehat{\theta}})$ and trace$(\bm{V}_{\widetilde{\theta}})$ are of the order of magnitude $O(m)$. When $m\ll \sqrt{n}$ or $h^4\gg n^{-1}$, the squared bias of $\widetilde{\bm{\theta}}$ may dominate its variance. Hence, the OLS estimator $\widehat{\bm{\theta}}$ may achieve a smaller MSE. When $m\asymp \sqrt{n}$ and $h^4=O(n^{-1} m)$, the two MSEs are of the same order of magnitude and it remains unclear which one is smaller. When $m\gg \sqrt{n}$ and $h^4=o(n^{-1})$, the variance of $\widetilde{\bm{\theta}}$ dominates its squared bias. %Since $\bm{V}_{\widehat{\theta}}-\bm{V}_{\widetilde{\theta}}$ is strictly positive, 
  Moreover,   $\bm{\Sigma}_y-\bm{\Sigma}_{\eta}$ is strictly positive definite, so is $\bm{V}_{\widehat{\theta}}-\bm{V}_{\widetilde{\theta}}$. As a result, 
  $\widetilde{\bm{\theta}}$ achieves a smaller MSE.
  In our applications, $m$ is moderately large and the condition $m\gg\sqrt{n}$ is likely to be satisfied. With properly chosen bandwidth, we expected the smoothed estimator achieves a smaller MSE. 

  Secondly,  we present the limiting distributions of $\widehat{\theta}(\tp)$ and  $\widetilde{\theta}(\tp)$ and prove the validity of our test for DE based on L-TVCDP.}

  \begin{thm}
  \label{thm:t theta asymp}
  Suppose $\lambda_{\min}(\bm{V}_{\widehat{\theta}}(\tau,\tau))$ and $\lambda_{\min}(\bm{V}_{\widetilde{\theta}}(\tau,\tau))$ 
  are uniformly bounded away from zero for any $\tau$. 
  Under %TCMIA and 
  Assumptions \ref{assump:eps t}, \ref{assump:kernel} and \ref{assump:t}, for any $(d+2)$-dimensional %nonzero %[should be $mp$-dimensional with $p=d+2$?]} 
  vectors $\bm{a}_{n,1}$, $\bm{a}_{n,2}$, with unit $\ell_2$ norm, 
  \begin{itemize}
  \item[(i)] %{\color{blue}as $n \rightarrow \infty$, we have} 
  $\sqrt{n}\bm{a}_{n,1}^\top 
  \{\widehat{\theta}(\tau) - \theta(\tau)\}\Big/\sqrt{\bm{a}_{n,1}^\top 
  \bm{V}_{\widehat{\theta}}(\tau, \tau)\bm{a}_{n,1}}
  \stackrel{d}{\longrightarrow} N(0,1)$ as $n\to \infty$ for any $\tau$; 
    \item[(ii)] %{\color{blue}as $n,m \rightarrow \infty$, $h\rightarrow0$ and $mh \rightarrow \infty$, we have} 
    Suppose $m\to \infty, h\to 0,$ and $ hm\to \infty$ as $n\to \infty$. Then 
    $\sqrt{n}\bm{a}_{n,2}^\top 
    \{\widetilde{\theta}(\tau) - \theta(\tau)\}\Big/\sqrt{\bm{a}_{n,2}^\top\bm{
    V}_{\widetilde{\theta}}(\tau, \tau)\bm{a}_{n,2}} 
    \stackrel{d}{\longrightarrow}N(b_n,1)$ as $n\to \infty$ for any $\tau$, where the bias $b_n=O(\sqrt{n}h^2+\sqrt{n}m^{-1})$. 
    \item[(iii)] Suppose $h=o(n^{-1/4})$, $m\gg \sqrt{n}$ and the sum of all elements in $m^{-2}\bm{V}_{\widetilde{\gamma}}$ is bounded away from zero where $\bm{V}_{\widetilde{\gamma}}$ denotes the submatrix of $\bm{V}_{\widetilde{\theta}}$ which corresponds to the asymptotic covariance matrix of $\widetilde{\bm{\theta}}$. Then for the hypotheses \eqref{hypo:de t}, 
    under $H_0^{DE}$ , $\prob(\widehat{\DE}/\widehat{se}(\widehat{\DE})>z_{\alpha})=\alpha+o(1)$; under $H_1^{DE}$, $\prob(\widehat{\DE}/\widehat{se}(\widehat{\DE})>z_{\alpha})\to 1$, where $z_{\alpha}$ denotes the upper $\alpha$th quantile of a standard normal distribution.  %\change{[Please change the proof accordingly.]}
  \end{itemize}
  \end{thm}

  Theorem \ref{thm:t theta asymp} has several important implications. 
  First, the bias of the smoothed estimator $\widetilde{\bm{\theta}}$ decays with $m$. In cases where $m$ is fixed, the kernel smoothing step is not preferred as it will result in an asymptotically biased estimator. Second, each $\widetilde\theta(\tp)$ converges at a rate of $O_p(n^{-{1/2}})$ under the assumption that $\lambda_{\min}(\bm{V}_{\widetilde{\theta}}(\tau,\tau))$ is bounded away from zero. The rate $O_p(n^{-1/2}m^{-1/2})$ cannot be achieved despite that we have a total of $nm$ observations, since the random errors $\{e_\tp\}_\tp$ are not independent. We also remark that in the extreme case where $\{e_\tp\}_\tp$ are independent, we can set $h\propto (nm)^{-1/5}$ and $\widetilde\theta(\tp)$ attains the classical nonparametric convergence rate $O_p((nm)^{-2/5})$. Third, %the matrix $\bm{\Sigma}_y-\bm{\Sigma}_{\eta}$ is strictly positive definite, so is 
  since $\bm{V}_{\widehat{\theta}}-\bm{V}_{\widetilde{\theta}}$ is strictly positive, this similarly implies that the smoothed estimator is more efficient when $b_n=o(1)$, or equivalently, $h=o(n^{-1/4})$ and $m\gg \sqrt{n}$. Finally, in the proof of Theorem \ref{thm:t theta asymp}, we show that the covariance estimator $\widetilde{\bm{V}}_{\theta}$ is consistent. This together with asymptotic distribution of $\widetilde{\bm{\theta}}$ yields the the consistency of our test in (iii). 
  
  Thirdly, we present  the validity of the proposed parametric bootstrap procedure for IE 
  under the temporal alternation design based on L-TVCDP.  

  \begin{thm}\label{thm:bootstrap}
  Suppose that there is some constant $0<c_1\le 1$ such that $c_1\leq \Mean \|\varepsilon_{\tp,S}\|^2$ and $\Mean e_\tp^2\leq c_1^{-1}$ for all $1\leq \tp\leq m$. Suppose that $h=o(n^{-1/4})$, $m\asymp n^{c_2}$ for some $1/2\le c_2<3/2$ and $mh\to\infty$. Then under the assumptions in Theorem \ref{thm:t theta asymp} and Assumption \ref{asmp:st1}, with probability approaching 1, we have 
  \begin{equation*}%\label{result2}
  \sup_{z}|\prob(\widehat{\IE}-\IE\leq z)-\prob(\widehat{\IE}^b-\widehat{\IE} \leq z |\textrm{Data}) \vert\leq C(\sqrt{n}h^2 + \sqrt{n} m^{-1} +n^{-1/8}),
  \end{equation*}
  where $C$ is some positive constant.
  \end{thm}

  We have several remarks. The derivation of Theorem \ref{thm:bootstrap} is non-trivial when $m$ diverges with $n$. Specifically, since $\widehat{\IE}$ is a very complicated function of the estimated varying coefficients (see Equation \eqref{eqn:IE}),    its limiting distribution is not well-defined. To prove Theorem \ref{thm:bootstrap}, we derive a nonasymptotic error bound on the difference between the distribution of $\widehat{\IE}$ and that of the bootstrap statistics conditional on the data. As a result, it ensures that the type-I error can be well-controlled and the power approaches one.  
  Please refer to the proof of Theorem \ref{thm:bootstrap} in the supplementary document for details. {Finally, we require $m$ to diverge with $n$ at certain rate. In settings with a small or fixed $m$, one can apply the proposed bootstrap procedure to the unsmoothed estimator $\widehat{\bm{\theta}}$. %
  The resulting test procedure remains valid regardless of whether $m$ is fixed or not.

  Fourthly, we illustrate the advantage  of employing the switchback design in the presence of temporal random effects. \change{As commented in the introduction, the switchback design assigns different treatments at adjacent time points $A_{i,1}=1-A_{i,2}=A_{i,3}=\ldots=A_{i,2t-1}=1-A_{i,2t}$,  whereas the alternating-day design assigns fixed treatment $A_{i,1}=A_{i,2}=A_{i,3}=\ldots=A_{i,2t-1}=A_{i,2t}$ within each day for any $i$ and $t$. In the switchback design, the random effects at adjacent time points can cancel with each other when estimating the causal effect, yielding a more efficient estimator.} To elaborate this point, we compare the mean square errors of the proposed estimators under the switchback design against those under an alternating-day design where 
 the new and old policies are daily switched back and forth. To simplify the analysis, we focus on the case where the state is one-dimensional and assume the treatment effect estimators are constructed based on the unsmoothed OLS estimators (see Section \ref{subsec:unsmooth} for details).  Let MSE($\widehat{\DE}_{sb}$) %, MSE($\widehat{\IE}_{sb}$), 
 and MSE($\widehat{\DE}_{ad}$) %and MSE($\widehat{\IE}_{ad}$) 
 denote the mean squared errors of DE %and IE 
 estimators under the switchback design and the alternating-day design, respectively. 

\begin{thm}\label{thm:switchback}
% Suppose $\Sigma_{\eta}(\tp_1,\tp_2)$ is nonnegative for any $\tp_1$ and $\tp_2$. Then under L-TVCDP,  as $n\to \infty$,  we have 
% $$n{\textrm{MSE}(\widehat{\DE}_{sb})}\le n{\textrm{MSE}(\widehat{\DE}_{ad})}+o(1),$$
% where the equality holds only when $\Sigma_{\eta}(j,k)=0$ for any $j, k$ such that $|j-k|=1,3,5,\ldots$. 
Suppose that the state is one-dimensional, $\Sigma_{\eta}(\tp_1,\tp_2)$ is nonnegative for any $\tp_1$ and $\tp_2$ and Assumptions \ref{assump:eps t} and \ref{assump:t} hold. When $\{\Phi(\tp)\}_\tp$ and $\{\Gamma(\tp)\}_\tp$ are of the same signs, respectively, i.e. for any $\tp_1,\tp_2$, $\Phi(\tp_1)\Phi(\tp_2)\ge 0$ and $\Gamma(\tp_1)\Gamma(\tp_2)\ge 0$, then as $n\to \infty$,  we have 
$$n{\textrm{MSE}(\widehat{\DE}_{sb})}\le n{\textrm{MSE}(\widehat{\DE}_{ad})}+o(1),$$
where the equality holds only when $\Sigma_{\eta}(j,k)=0$ for any $j, k$ such that $|j-k|=1,3,5,\ldots$. 
\end{thm}

To ensure that DE achieves a much smaller MSE under the switchback design, we only require that the random effects are non-negatively correlated and that  the correlation $\Sigma(j,k)$ is nonzero for some $j-k=1,3,5,\ldots$. These conditions are automatically satisfied when the random effects are positively correlated. 
We next provide a close-formed expression for the ratio of the two MSEs under an AR(1) noise structure and the constraint that $\Gamma(1)=\Gamma(2)=\cdots=\Gamma(m-1)=0$.

\begin{cor}\label{cor:switchback}
Suppose that %$\Gamma_\tp>0$ for all $\tp$ and there exists $0<\rho<0$ such that 
for any $1\le \tau_1,\tau_2 \le m$, $\Sigma_{e}(\tau_1,\tau_2)=c\rho^{|\tau_1-\tau_2|}$ for some constant $c>0$. Then under assumptions of Theorem \ref{thm:switchback}, when $\Gamma(1)=\Gamma(2)=\cdots=\Gamma(m-1)=0$,
we have as $n,m\rightarrow\infty$, 
$$\frac{\textrm{MSE}(\widehat{\DE}_{sb})}{\textrm{MSE}(\widehat{\DE}_{ad})}=\frac{(1-\rho)^2}{(1+\rho)^2}+o(1).$$
\end{cor}
It can be seen from Corollary \ref{cor:switchback} that the larger the $\rho$, the smaller the variance ratio. In particular, when $\rho=0.5$, MSE of DE under the switchback design is approximately 9 times smaller than that under the alternating-day design. 
We next consider IE. 
\begin{thm}\label{thm:switchIE}
	Suppose $m=2$. Under Assumptions \ref{assump:eps t} and \ref{assump:t},  we have
	\begin{eqnarray*}
		n\{MSE(\widehat{\IE}_{ad})-MSE(\widehat{\IE}_{sb})\}=o(1).
	\end{eqnarray*}
\end{thm}
Theorem \ref{thm:switchIE} suggests that the IE estimators under the two designs have comparable MSEs. This together with Theorem \ref{thm:switchback} underscores the superiority of the switchback design, particularly when $m=2$. However, as $m$ exceeds 2, determining the closed-form expression for $\textrm{MSE}(\widehat{\IE})$ becomes exceedingly complex, making it challenging to directly compare the two designs. Addressing this complexity and extending the comparison for cases where $m>2$  is a task we reserve for future research.

Fifth, we establish the convergence rates of the estimated DE and IE for NN-VCDP. 

\change{ \begin{thm}\label{thm:dnn} 
Suppose that $f_{\varepsilon_{\tp S}}$ is Lipschitz, meaning that for any $\tau$, there exists a constant $L_f>0$ such that $|f_{\varepsilon_{\tp S}}(x)-f_{\varepsilon_{\tp S}}(y)|\le L_f\|x-y\|_2$, where $\|\cdot\|_2$ represents the Frobenius norm. Additionally, assume that the NN-based learners satisfy $\mathbb{E}\{\widehat{G}_{a}(\tau,S_\tau)-{G}_{a}(\tau,S_\tau)\}^2\le \Delta_1^2(n,m)$ and $\mathbb{E}\{\widehat{g}_{a}(\tau,S_\tau^{a_1})-{g}_{a}(\tau,S_\tau^{a_1})\}^2\le \Delta_2^2(n,m)$, where $a\in \{0,1\}$ and $\Delta_1(n, m)$ and $\Delta_2(n, m)$ are specific functions. The density estimator should fulfill $ \int_x |f_{\varepsilon_{\tau S}}(x)-\widehat{f}_{\varepsilon_{\tau S}}(x)|dx=O_p(\Delta_3(n,m))$ for some function $\Delta_3$. Both $g_a$ and $\widehat{g}_a$ must be uniformly bounded. Moreover, the ratio of the density function of the potential state $S_{\tau}^a$ to the density of the observed state $S_{\tau}$ must be bounded by $\sqrt{\omega}$ for any $\tau$ and $a$. 
Then, as $\min(n, m)\rightarrow \infty$, we obtain the following convergence results:
\begin{eqnarray*}
\widehat{\DE}-\DE &=& O_p\left(m\sqrt{\omega}\Delta_2(n,m)+m^2\Delta_1(n,m)+m^2L_f \sqrt{\omega}\Delta_3(n,m)+\frac{m}{\sqrt{n}}\sqrt{\log (nm)}\right),\\
\widehat{\IE}-\IE &=& O_p\left(m\sqrt{\omega}\Delta_2(n,m)+m^2\Delta_1(n,m)+m^2L_f \sqrt{\omega}\Delta_3(n,m)+\frac{m}{\sqrt{n}}\sqrt{\log (nm)}\right).
\end{eqnarray*}
\end{thm}
}

  Since the convergence rates of NN-based learners have been widely studied in the literature \citep[see e.g.,][]{shen2019deep,schmidt2020dnn,shen2022optimal,yan2023dnn}, these results can be used to establish the convergence rates of $\widehat{G}_a$ and $\widehat{g}_a$.  }

  Finally,  we impose the following regularity assumptions for the proposed tests in spatio-temporal dependent experiments based on L-STVCDP.   

  \begin{asmp}\label{assump:st}
  %The covariate $\bm{Z}_i$s are i.i.d; 
  For any $\tp,\iota$, $\Mean(Z_{i,\tp,\iota}^\top Z_{i,\tp,\iota})$ is invertible; $\theta(\tp,\iota)$,  $\Sigma_{\eta^{\RNum{1}}}(\tp_1, \tp_2,\iota_1,\iota_2)$, $\Sigma_{\eta^{\RNum{2}}}(\tp_1, \iota_1, \tp_2)$,  and $\Sigma_{\eta^{\RNum{3}}}(\tp_1, \iota_1, \iota_2)$ have bounded and continuous second-order derivatives. %
  \end{asmp}

  %{\color{red}
  \begin{asmp}\label{asmp:sigma st}
  There exists $q<1$ 
  such that the absolute values of eigenvalues of $\Phi(\tp,\iota)$ are smaller than $q$. %and $\Vert\Phi(\tp,\iota)\Vert_{-\infty}\geq q'>0$; 
  In addition, there exist  $M_{\Gamma}$ and $M_\beta<\infty$ such that $\Vert\Gamma_1(\tp,\iota)+\Gamma_2(\tp,\iota)\Vert_\infty\leq M_{\Gamma}$ and $\Vert\beta(\tp,\iota)\Vert_\infty\leq M_\beta$. $\Theta(\tp,\iota)$ has a bounded and continuous second-order derivative.
  \end{asmp}%}

  % \begin{asmp}\label{asmp:para ie st}
  % $\Theta(\tp,\iota)$, $\Sigma_{\eta^{\RNum{1}}}(\tp_1, \tp_2,\iota_1,\iota_2),\Sigma_{\eta^{\RNum{2}}}(\tp_1, \iota_1, \tp_2),$ and $\Sigma_{\eta^{\RNum{3}}}(\tp_1, \iota_1, \iota_2)$ have bounded and continuous second-order derivatives.
  % \end{asmp}

  With these assumptions, we present the asymptotic properties of our DE and IE estimators and their associated test statistics for the spatio-temporal dependent experiments based on L-STVCDP.  
  Define 
  \begin{eqnarray*}
  \bm{V}_{\widetilde\theta_{st}}(\tp_1,\iota_1,\tp_2,\iota_2)=\{\Mean Z_{i,\tp_1,\iota_1}Z_{i,\tp_1,\iota_1^\top }\}^{-1}\Mean\{Z_{i,\tp_2,\iota_2}Z_{i,\tp_1,\iota_1}^\top\Sigma_{\eta^{\RNum{1}}}(\tp_1,\iota_1,\tp_2,\iota_2)\}\{\Mean Z_{i,\tp_2,\iota_2} Z_{i,\tp_2,\iota_2}^\top\}^{-1}
  \end{eqnarray*}
  as the asymptotic covariance between $\sqrt{n}\widetilde{\theta}_{st}(\tp_1,\iota_1)$ and $\sqrt{n}\widetilde{\theta}_{st}(\tp_2,\iota_2)$.

  \begin{thm}\label{thm:st theta asymp}
  %, suppose 
  Suppose $\lambda_{\min}(\bm{V}_{\widetilde\theta_{st}})$ is bounded away from zero. Under %STCMIA, 
  Assumptions \ref{assump:eps st}, \ref{assump:kernel} and \ref{assump:st}, for any set of $(d+2)$-dimensional vectors $\{B_{\tp,\iota}\}_{\tp,\iota}$, we have as $n,m,r\rightarrow\infty$, $h,h_{st}\rightarrow0$ and $mh,rh_{st}\rightarrow\infty$ that 
  \begin{itemize}
  \item[(i)] For any set of $(d+2)$-dimensional vectors $\{B_{\tp,\iota}\}_{\tp,\iota}$ with $\sum_{\tau_1,\tau_2,\iota_1,\iota_2} B_{\tp_1,\iota_1}^\top \bm{V}_{\widetilde\theta_{st}}(\tp_1,\iota_1,\tp_2,\iota_2) B_{\tp_2,\iota_2}\ge c \sum_{\tau,\iota} \|B_{\tau,\iota}\|_2^2$ for some constant $c>0$, we have 
  $$ \sqrt{n} \sum_{\tp,\iota} [B_{\tp,\iota}^\top \{\widetilde{\theta}_{st}(\tp,\iota) - {\theta}_{st}(\tp,\iota) \}]\Big/\sqrt{\sum_{\tau_1,\tau_2,\iota_1,\iota_2} B_{\tp_1,\iota_1}^\top \bm{V}_{\widetilde\theta_{st}}(\tp_1,\iota_1,\tp_2,\iota_2) B_{\tp_2,\iota_2}}\stackrel{d}{\to} N(b_{n,st}, 1), $$ where the bias $b_{n,st}=O(\sqrt{n}h^2+\sqrt{n}h_{st}^2+\sqrt{n}m^{-1}+\sqrt{n}r^{-1})$. 
  \item[(ii)] Suppose $h,h_{st}=o(n^{-1/4})$ and $m,r\gg\sqrt{n}$. Then for the hypotheses \eqref{hypo de st}, $\prob(\widehat{DE}_{st}/\widehat{se}(\widehat{DE}_{st})>z_{\alpha})=\alpha+o(1)$ under $H_0^{DE}$ and $\prob(\widehat{DE}_{st}/\widehat{se}(\widehat{DE}_{st})>z_{\alpha})\to 1$ under $H_1^{DE}$.
  \end{itemize}
  \end{thm}

  \begin{thm}\label{thm:bootstrap st}
  Suppose that there are some constants $0<c_1\le 1$ such that $c_1\leq \Mean \varepsilon_{\tp,\iota,S}^2,\Mean e_{\tp,\iota}^2\leq c_1^{-1}$ for all $1\leq \tp\leq m$, $1\le\iota\le r$, %where $\overline{\Mean}(x_{\tp,\iota}^2)=n^{-1}\sum_{i=1}^n\Mean(x_{i,\tp,\iota}^2)$,
  and that %{\color{red}
  $h,h_{st}=o(n^{-1/4})$, $m,r\gg\sqrt{n}$ and $mr\asymp n^{c_2}$ for some constant $c_2<3/2$. Then under %STCM, 
  Assumptions of Theorem \ref{thm:st theta asymp} and Assumption \ref{asmp:sigma st}, with probability approaching 1,
  \begin{equation}\label{result2}
  \sup_{z}|\prob(\widehat{\IE}_{st}-\IE_{st}\leq z)-\prob(\widehat{\IE}_{st}^b-\widehat{\IE}_{st} \le z |\textrm{Data}) \vert\leq C(\sqrt{n}h^2+\sqrt{n}h_{st}^2+\sqrt{n}m^{-1}+\sqrt{n}r^{-1}+n^{-1/8}),
  \end{equation}
  where $C$ is some positive constant.%, and $c_{st}(\alpha)$ is the $\alpha$th upper quantile of the bootstrap test statistic. 
  \end{thm}

Theorem \ref{thm:st theta asymp} establishes the limiting distribution of the proposed DE estimator for the spatio-temporal dependent experiments. Similar to Proposition \ref{prop:mse}, we can show that the smoothed estimator is more efficient when $m,r\gg \sqrt{n}$ and $h^4,h_{st}^4=o(n^{-1})$. In addition, Theorem \ref{thm:bootstrap st} allows both $m$ and $r$ to be either fixed, or diverge with $n$, and is thus applicable to a wide range of applications. 

\section{Real data based simulations}\label{sec:simulation studies}
\subsection{Temporal alternation design}
\label{sec:temporaldesign}
\begin{figure}[H]
  \centering
  \includegraphics[width=1\textwidth]{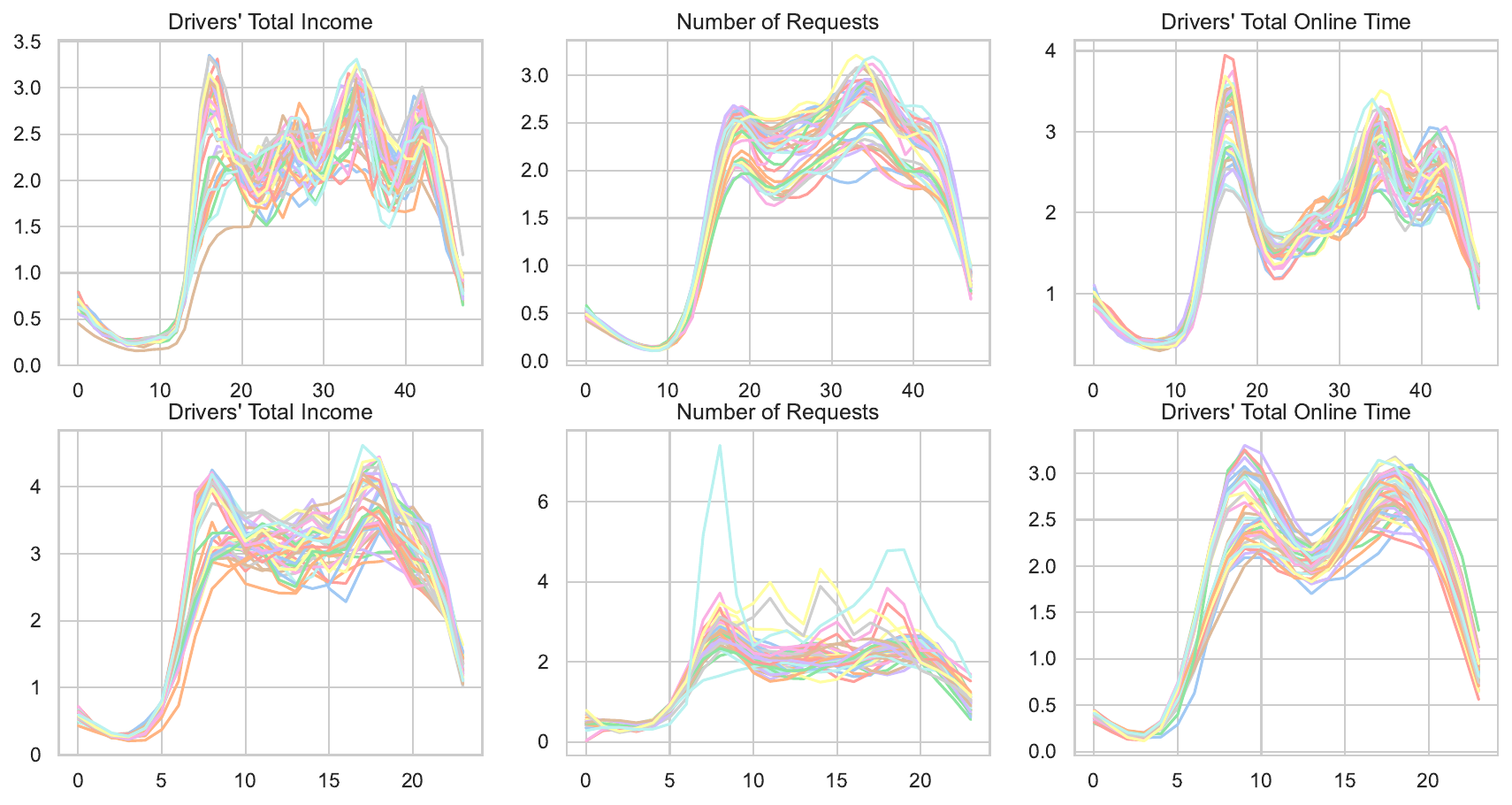}
  \caption{Scaled business metrics from City A (the first row) and City B (the second row) across 40 days, including drivers' total income, the numbers of requests and drivers' total online time.}
  \label{fig:cityAB_t_pattern}
  \end{figure}

  In this section, we conduct Monte Carlo simulations to examine the finite sample properties of the proposed test statistics based on L-TVCDP and L-STVCDP models. To generate data under the temporal alternation design, we design two simulation environments based on two real datasets obtained from Didi Chuxing. 
  The first dataset is collected from a given city A from Dec. 5th, 2018 to Jan. 13th, 2019. Thirty-minutes is defined as one time unit. 
  The second dataset is from another city B, from May 17th, 2019 to June 25th, 2019. One-hour is defined as one time unit. Both contain data for 40 days. 
  Due to privacy, we only present scaled metrics in this paper.
  Figure \ref{fig:cityAB_t_pattern} depicts the trend of some business metrics over time across 40 different days. These metrics include drivers' total income, the number of requests and drivers' total online time. Among them, the first quantity is our outcome of interest and the last two are considered as the state variables to characterize the demand and supply networks. As expected, these quantities show a similar pattern, achieving the largest values at peak time. 
  % \begin{figure}[H]
  % \begin{minipage}[t]{1\textwidth}
  % \centering
  % \includegraphics[width=1\textwidth,height=1.7in]{pattern1_t_cityA.pdf}
  % \end{minipage}%

  % \begin{minipage}[t]{1\textwidth}
  % \centering
  % \includegraphics[width=1\textwidth,height=1.7in]{pattern1_t_cityB.pdf}
  % \end{minipage}%
  % \caption{Scaled business metrics from City A (the first row) and City B (the second row) across 40 days, including drivers' total income, the numbers of requests and drivers' total online time.}
  % \label{fig:cityAB_t_pattern}
  % \end{figure}

  We next discuss how to generate synthetic data based on the real datasets. 
  The main idea is to fit the proposed L-TVCDP models  to the real dataset and apply the parametric bootstrap to simulate the data.  
  Let $\widetilde{\beta}_0(\tp)$, $\widetilde{\beta}(\tp)$, $\widetilde{\phi}_0(\tp)$,  and $\widetilde{\Phi}(\tp)$ denote the smoothed estimators for $\beta_0(\tp)$, $\beta(\tp)$, $\phi_0(\tp)$ and $\Phi(\tp)$, respectively. We set $\widetilde{\gamma}(\tp)$ and $\widetilde{\Gamma}(\tp)$ to $(\delta/100)\times (\sum_{i,\tp} Y_{i,\tp}/nm)$ and $(\delta/100)\times (\sum_{i,\tp} S_{i,\tp}/nm)$, respectively. As such, the parameter $\delta$ controls the degree of the treatment effects. Specifically, the null holds if $\delta=0$ and the alternative holds if $\delta>0$. It corresponds to the increase relative to the outcome (state). We next generate the policies according to the temporal alternation design %one-hour, threetemporal alternation design 
  and simulate the responses and states based on the fitted model. Let TI denote the time span we implement each policy under the alternation design. For instance, if TI $=3$, then we first implements one policy for three hours, then switch to the other for another three hours and then switch back and forth between the two policies. We consider three choices of $n\in \{8,14,20\}$, fives choices of $\delta\in \{0,0.25,0.5,0.75,1\}$ and three choices of TI $\in \{1,3,6\}$. This corresponds to a total of 45 cases. The bandwidth is set $h=Cn^{-1/3}$, where $C$ is selected by the 5-fold cross validation method.

\begin{figure}[H]
  \begin{minipage}[t]{1\textwidth}
  \centering
  \includegraphics[width=1\textwidth]{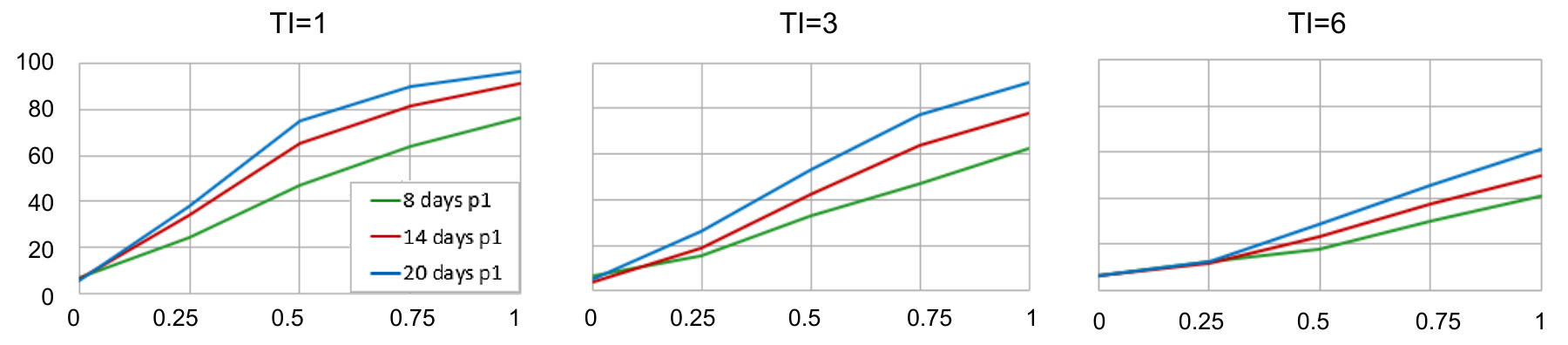}
  \end{minipage}%

  \begin{minipage}[t]{1\textwidth}
  \centering
  \includegraphics[width=1\textwidth]{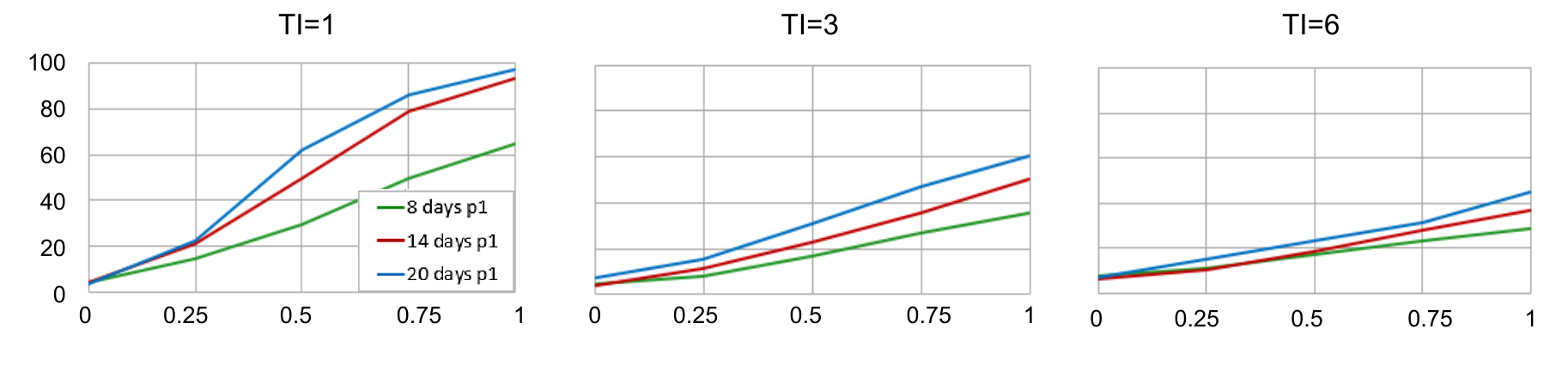}
  \end{minipage}%
  \caption{Simulation results for L-TVCDP: empirical rejection rates \change{(expressed as percentages)} of the proposed test for DE under different combinations of $(n,\delta,\textrm{TI})$ and types of outcomes. Synthetic data are simulated based on the real dataset from city A (the first row) and city B (the second row). \label{fig:cityAB_t_DE}}
  \end{figure}

  In Figure \ref{fig:cityAB_t_DE}, we depict the empirical rejection probabilities of the proposed test for DE, aggregated over 400 simulations, for all combinations. 
  It can be seen that our test controls the type-I error 
  and its power increases as $\delta$ increases. In addition, the empirical rejection rates decreases as TI increases. This phenomenon suggests that the more frequently we switch back and forth between the two policies, the more powerful the resulting test. It is due to the positive correlation between adjacent observations. To elaborate, consider the extreme case where we switch policies at each time. The policies assigned at any two adjacent time points are different. As such, the random effect cancels with each other, yielding an efficient estimator. We conduct some additional simulations using the numbers of answered requests and finished requests of cities A and B as responses (see Figure \ref{fig:cityAB_t_pattern2} in the supplement). Results are very similar and are reported in Figures \ref{fig:DE2_t_cityA}--\ref{fig:DE2_t_cityB} in the supplementary document. See also Tables \ref{tab:A DE t}--\ref{tab:B DE t} in the supplementary document.
  
\begin{figure}[H]
  \begin{minipage}[t]{1\textwidth}
  \centering
  \includegraphics[width=1\textwidth]{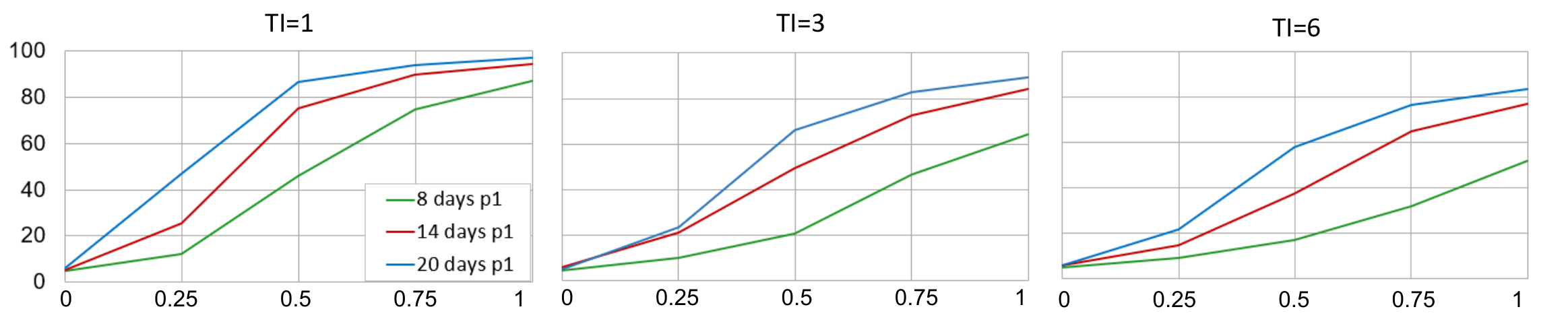}
  \end{minipage}%

  \begin{minipage}[t]{1\textwidth}
  \centering
  \includegraphics[width=1\textwidth]{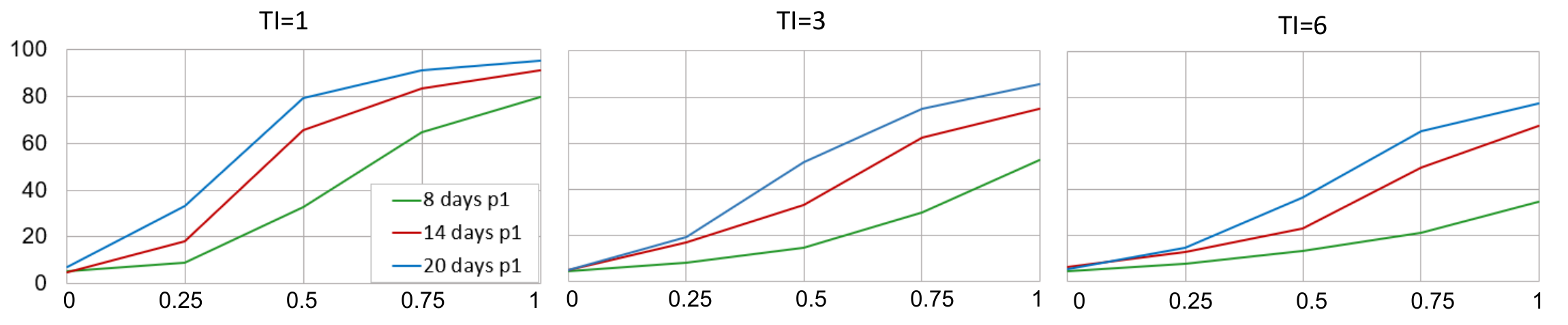}
  \end{minipage}%
  \caption{Simulation results for L-TVCDP: empirical rejection rates \change{(expressed as percentages)} of the proposed test for IE under different combinations of $(n,\delta,\textrm{TI})$. Synthetic data are simulated based on the real dataset from city A (the first row) and city B (the second row). }
  \label{fig:cityAB_t_IE}
  \end{figure}

  To infer IE, we set the outcome to drivers' total online income. The empirical rejection probabilities of the proposed test for IE are reported in Figure \ref{fig:cityAB_t_IE}.
  % (see also, Tables \ref{tab:A IE t} and \ref{tab:B IE t} in Appendix \ref{app:tab}). 
  Results are aggregated over 400 simulations. Similarly, the proposed test is consistent. Its power increases with the sample size and $\delta$. In addition, its power under  TI $=1$ is much larger than those under TI $=3$ or $6$. This suggests that we shall switch back and forth between the two policies as frequently as possible to maximize the power property of the test (see also Tables \ref{tab:A IE t}--\ref{tab:B IE t} in Supplementary document).

\subsection{Spatio-temporal  alternation design}\label{sec:simuts}
\begin{figure}[H]
     \centering
     \includegraphics[width=\textwidth,height=0.35\textwidth]{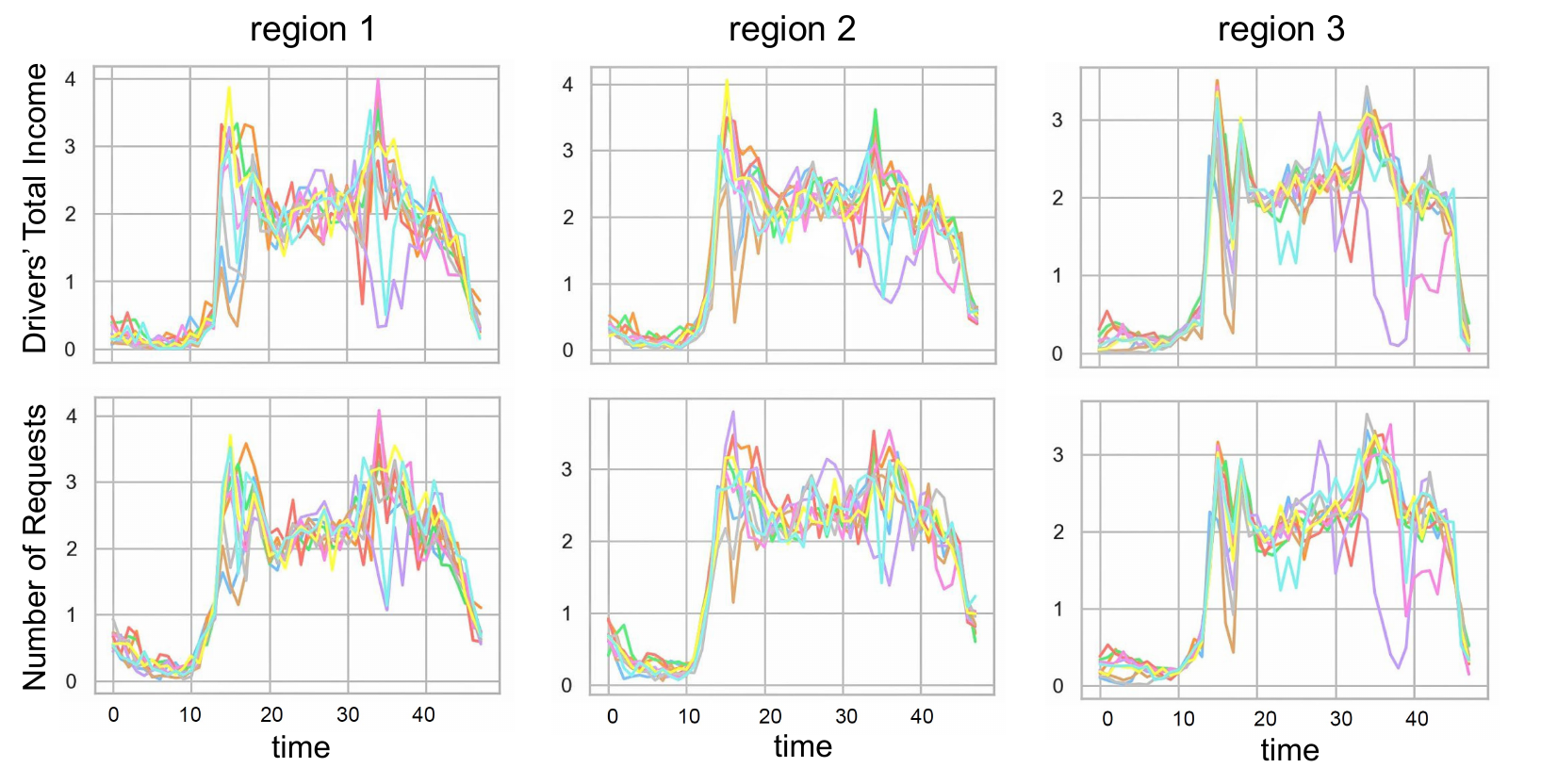}
     \caption{Number of call requests and drivers' total income across different regions and days. The values are scaled for privacy concerns.}
     \label{fig:st}
  \end{figure}

  %We also conduct Monte Carlo simulations for our spatio-temporal model. 
  To generate data under the spatio-temporal alternation design, we create a simulation environment based on the real dataset from city A. We divide the city into 10 non-overlapping regions. %{\color{red}Within each region, more than 90\% of the order requests are responded by 
  %Region division satisfies A that more than 90\% requests are responded by 
  %local drivers in the same region. [This sentence was to avoid interference, and now may be deleted?]}%We set drivers' total income to the response and set the number of order requests to the state variable. 
  %The patterns of these metrics are more unstable than those of a whole city. To clearly show the daily trends, we plot the regional scaled drivers' total income of 10 days at half-hour granularity and the regional scaled number of requests of 10 days at half-hour granularity 
  We plot these variables associated with 3 particular regions, over the first 10 days in Figure \ref{fig:st}. It can be seen that although the daily trends differ across regions, the state and the response are highly correlated. 
  %Though the patterns are rougher than the city-level ones, response and state variable in the same region are of similar shape, which enables the model to reduce the variation of the response. 

  We fit the proposed models in \eqref{model:STVCM DE} to the real dataset to estimate the varying coefficients and the variances of the random errors. Then we manually set the treatment effects $\widehat{\gamma}(\tp,\iota)$ and $\widehat{\Gamma}(\tp,\iota)$ to  $(\delta_1/100) \times (\sum_{i=1}^n\sum_{\tp=1}^m Y_{i,\tp,\iota}/nm)$ and $(\delta_2/100) \times (\sum_{i=1}^n\sum_{\tp=1}^m S_{i,\tp,\iota}/nm)$ for some constants $\delta_1$ and $\delta_2>0$. We consider both the temporal and spatio-temporal alternation designs, and
  %Finally, we generate the policies according to the spatio-temporal alternation design and 
  simulate the data via parametric bootstrap.

\begin{figure}[H]
     \centering
     \includegraphics[width=\textwidth,height=0.27\textheight]{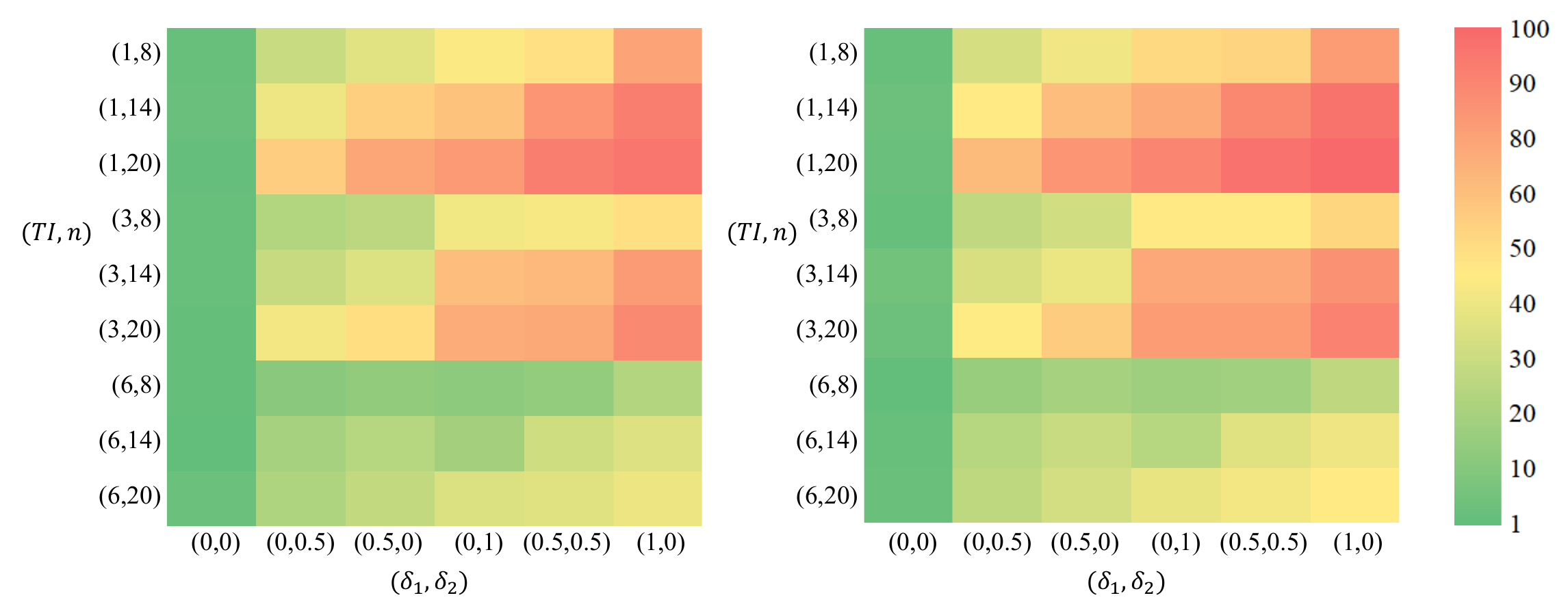}
      \caption{\label{fig:cityA_DE_st} Simulation results for L-STVCDP: the empirical rejection probabilities of the proposed test test for DE under the temporal alternation design (left panel) and the spatio-temporal alternation design (right panel).} %{\color{blue}[need to add colour bar. Added.]}}
  \end{figure}

  We also consider three choices of $n\in \{8,14,20\}$, three choices of TI $\in \{1,3,6\}$ and three choices of $\delta_1,\delta_2 \in \{0,0.5,1\}$. This yields a total of 81 combinations under each design.  
  The rejection probabilities of the proposed tests for DE and IE tests are reported in Figures \ref{fig:cityA_DE_st} and \ref{fig:cityA_IE_st} (see also Tables \ref{tab:A DE st} and \ref{tab:A IE st} in the supplementary document).
  It can be seen that the type I error rates of the proposed test are close to the nominal level under both designs. More importantly, the power under spatio-temporal alternation design is higher than that of temporal alternation design in all cases. The reason is twofold. First, under the spatio-temporal design, we independently randomize the initial policy for each region, and adjacent regions may receive different policies. Observations across adjacent areas are likely to be positively correlated. As such, the variance of the estimated treatment effects will be smaller than that under the temporal design where all regions receive the same policy at each time. Second, we employ kernel smoothing twice when computing $\widehat{\DE}_{st}$ and $\widehat{\IE}_{st}$, as discussed in Section \ref{sec:ST method}. This results in a more efficient estimator.  
  %This is consistent with our theoretical findings in Section \ref{sec:theoretical analysis}. 
  %This is because spatio-temporal-alternating design is more randomized than temporal-alternating design.
  In addition, 
  compared with the results in Tables \ref{tab:A DE t} and \ref{tab:A IE t}, 
  %one can find that both designs on multiple regions (Tables \ref{tab:A DE st}-\ref{tab:A IE st}) have higher powers than one region under all settings, which verifies the advantage of introducing spatial information into the model.
  it can be seen that the test that focuses on the entire city has better power property than the one that considers a particular region in general. 
  %We also plot the rejection probabilities for DE and IE tests as a function of $\delta$ in Figures \ref{fig:cityA_DE_st}--\ref{fig:cityA_IE_st}. Again,
  Finally, the power decreases with TI and increases with $n$, $\delta_1$ and $\delta_2$. 

   \begin{figure}[H]
     \centering
     \includegraphics[width=\textwidth,height=0.26\textheight]{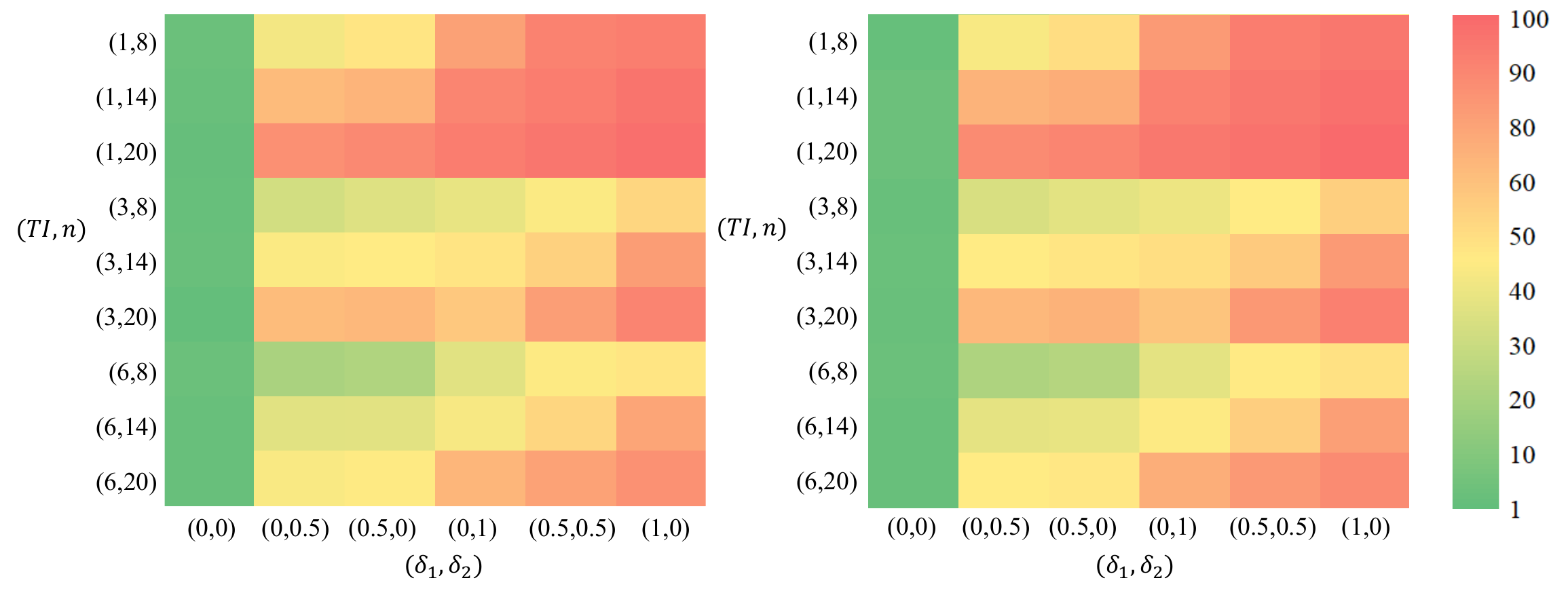}
      \caption{\label{fig:cityA_IE_st}  Simulation results for L-STVCDP: the empirical rejection probabilities of the proposed test test for IE under the temporal alternation design (left panel) and the spatio-temporal alternation design (right panel).}
  \end{figure}

\section{Real data analysis}
\label{sec:real data analysis}

  In this section, we apply the proposed tests based on L-TVCDP and L-STVCDP %tests 
  to a number of real datasets from Didi Chuxing to examine the treatment effects of some newly developed order dispatch and vehicle reposition policies. Due to privacy, we do not publicize the names of these policies.  

  We first consider four data sets collected from four online experiments under the temporal alternation design. All the experiments last for 14 days. Policies are executed based on alternating half-hourly time intervals. 
  We denote the cities, in which these experiments take place, as $C_1, C_2,$ $C_3,$ and $C_4$ and  their corresponding policies as $S_1, S_2, S_3,$ and $S_4$, respectively. For each policy, we are interested in its effect on three key business metrics, including drivers' total income, the answer rate,  and the completion rate. Similar to Section \ref{sec:temporaldesign}, we use the number of call orders and drivers' total online time to construct the time-varying state variables. 

  All the new policies are compared with some baseline policies in order to evaluate whether they improve some business outcomes.
  Specifically, in city $C_1$, policy $S_1$ is proposed to reduce the answer time (the time period between the time when an order is requested and the time when the order is responded by the driver). 
  This in turn meets more call orders requests. 
  Both policy $S_2$ in city $C_2$  and policy $S_3$ in city $C_3$ are  designed to guide drivers to regions with more orders  in order to reduce drivers' idle time ratio.
  Policies $S_2$ and $S_3$  are designed to assign more drivers to areas with more orders. This in turn reduces drivers' downtime and increase their income. 
  Policy $S_4$ aims to balance drivers' downtime and their average pick-up distance.

  We also apply our test to another four datasets collected from four A/A experiments which compare the standard policy against itself. These A/A experiments are conducted two weeks before the A/B experiments. Each lasts for 14 days and thirty-minutes is defined as one time unit. We remark that the A/A experiment is employed as a sanity check for the validity of the proposed test. We expect our test will not reject the null when applied to these datasets, since the sole standard policy  is used. 

\begin{figure}
     \centering
      \includegraphics[width=4in,height=2.7in]{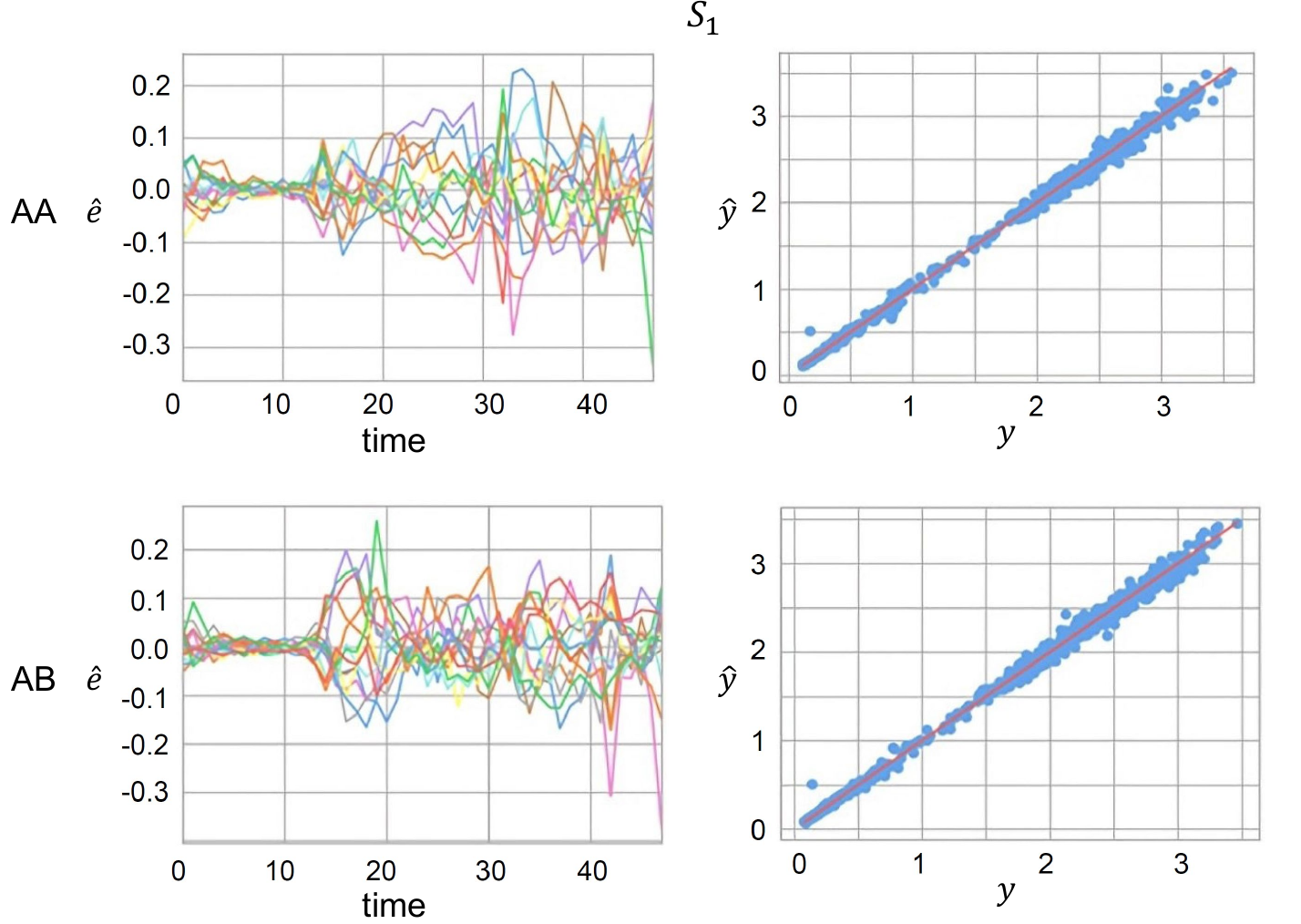}
  \caption{Plots of the fitted drivers' total income against the observed values as well as the corresponding residuals. Data are 
  %Residuals plots of the fitted drivers' total income over true drivers' total income plots for both 
  collected from an A/A or A/B experiment under the temporal alternation design.}
  \label{fig:real_DE_t}
  \end{figure}

  We fit the proposed  L-TVCDP models to each of the eight datasets. In Figures \ref{fig:real_DE_t} and  \ref{fig:real_IE_t}, we plot the predicted outcomes against the observed values and plot the corresponding residuals over time for policy $S_1$. Results for policies $S_2$--$S_4$ are represented in Figure \ref{fig:real_DE_t2} in the supplementary article. %which are similar to $S_1$. 
  It can be seen that the predicted outcomes are very close to the observed values, suggesting that the proposed model fits the data well.
  P-values of the proposed tests are reported in Tables \ref{tab:real_DE_st} and \ref{tab:real_ISE_st}. As expected, the proposed test does not reject the null hypothesis when applied to all datasets from A/A experiments. When applied to the data from A/B experiments, it can be seen that the new policy $S_1$ directly improves the answer rate and the completion rate, while increasing drivers' total income in city $C_1$. It also significantly increases drivers' income in the long run. 
  Policy $S_2$ has significant direct and indirect effects on drivers' income  as expected. Policy $S_4$ significantly increases the immediate answer rate, while  improving the overall passenger satisfaction. However, policy $S_3$ is not significantly better than the standard policy. 

  \begin{table}
  \caption{\label{tab:real_DE_st} One sided p-values of the proposed test for DE, when applied to eight datasets collected from the A/A or A/B experiment based on the temporal alternation design, \change{with  DTI, ART and CRT corresponding to drivers’ total income, the answer rate and the completion rate, respectively}.}
%  \begin{flushleft}
\centering{
  \begin{tabular}{|c|r|r|r|r|r|r|}
    \hline
    & \multicolumn{3}{c|}{AA} & \multicolumn{3}{c|}{AB} \\
    \hline
    & \multicolumn{1}{c|}{DTI(\%)} & \multicolumn{1}{c|}{ART(\%)} & \multicolumn{1}{c|}{CRT(\%)} & \multicolumn{1}{c|}{DTI(\%)} & \multicolumn{1}{c|}{ART(\%)} & \multicolumn{1}{c|}{CRT(\%)} \\
    \hline
    $S_1$    & 0.527  & 0.435  & 0.442  & 0.000  & 0.000  & 0.003  \\
    \hline
    $S_2$    & 0.232  & 0.126  & 0.209  & 0.000  & 0.763  & 0.661  \\
    \hline
    $S_3$    & 0.378  & 0.379  & 0.567  & 0.700  & 0.637  & 0.839  \\
    \hline
    $S_4$    & 0.348  & 0.507  & 0.292  & 0.198  & 0.000  & 0.133  \\
    \hline
  \end{tabular}}%
%  \end{flushleft}
  \end{table}

  \begin{table}
  \caption{\label{tab:real_ISE_st} One sided p-values of the proposed test for IE, when applied to eight datasets collected from the A/A or A/B experiment based on the temporal alternation design. Drivers' total income is set to be the outcome of interest. }
  \centering
  {
  \begin{tabular}{|c|l|r|r|r|r|r|r|r|}
    \hline
    \multicolumn{1}{|r|}{} & \multicolumn{2}{c|}{S1} & \multicolumn{2}{c|}{S2} & \multicolumn{2}{c|}{S3} & \multicolumn{2}{c|}{S4} \\
    \hline
    \multicolumn{1}{|l|}{} & \multicolumn{1}{c|}{AA} & \multicolumn{1}{c|}{AB} & \multicolumn{1}{c|}{AA} & \multicolumn{1}{c|}{AB} & \multicolumn{1}{c|}{AA} & \multicolumn{1}{c|}{AB} & \multicolumn{1}{c|}{AA} & \multicolumn{1}{c|}{AB} \\
    \hline
    p-value    & 0.334 & 0.001 & 0.341 & 0.003 & 0.254 & 0.589 & 0.427 & 0.168 \\
    \hline
  %     $p_2$    & 0.668 & 0.002 & 0.682 & 0.006 & 0.508 & 0.897 & 0.748 & 0.336 \\     \hline
  \end{tabular}%
  }
  \end{table}%

   \begin{figure}
     \centering
      \includegraphics[width=6in,height=2.7in]{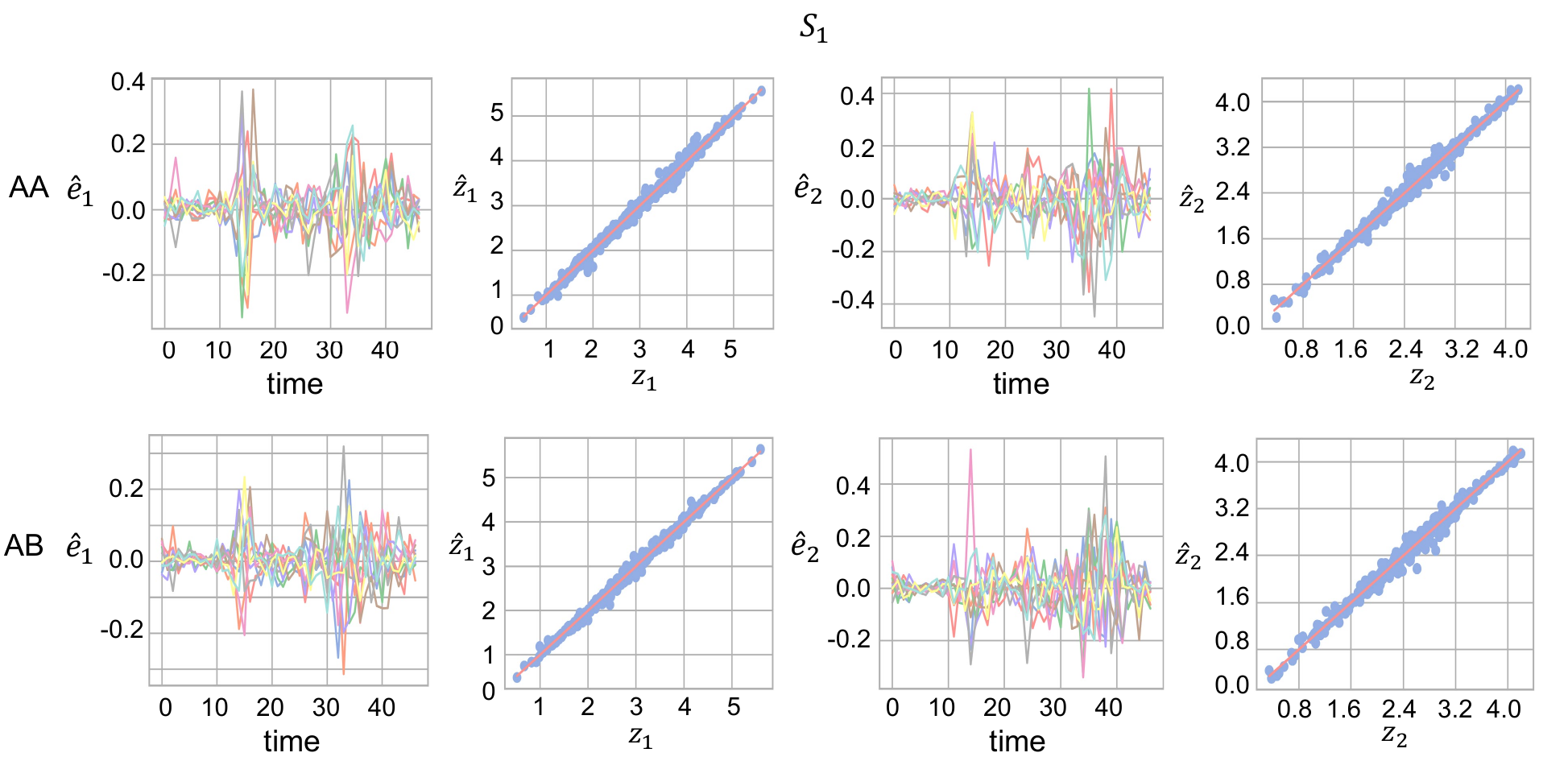}
  \caption{\label{fig:real_IE_t}Plots of the fitted number of orders ($\widehat{e}_1$) and drivers' online time ($\widehat{e}_2$) against their observed values, as well as the corresponding residuals. Data are collected from an A/A or A/B experiment under the temporal alternation design.}
  \end{figure}

\begin{figure}[H]
     \centering
      \includegraphics[width=6in,height=2.4in]{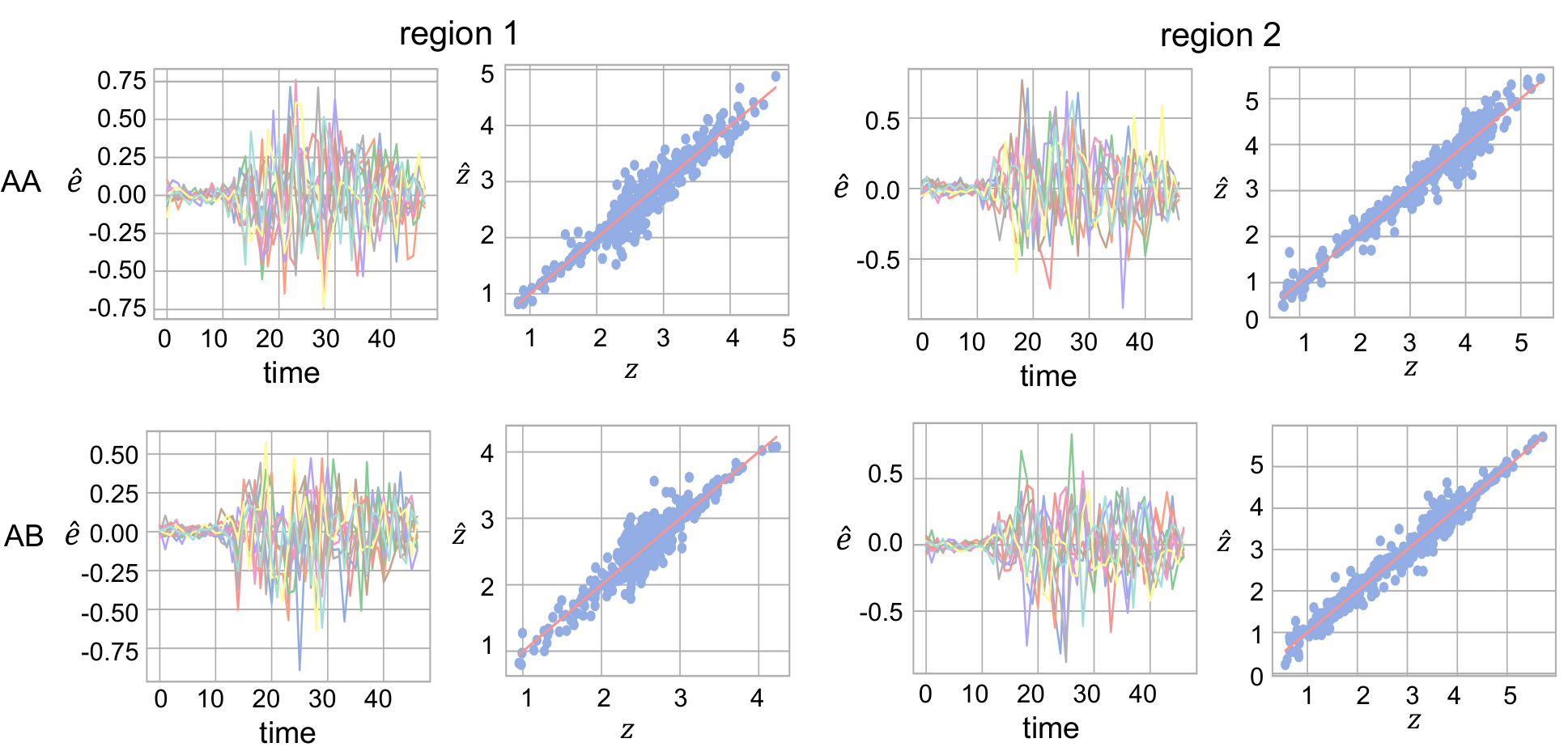}
    \caption{\label{fig:real_DE_st} Plots of the fitted drivers' income against the observed values, as well as the corresponding residuals. Data are collected from an A/A or A/B experiment under the spatio-temporal alternation design.}
  \end{figure}

\begin{figure}[H]
     \centering
      \includegraphics[width=6in,height=2.4in]{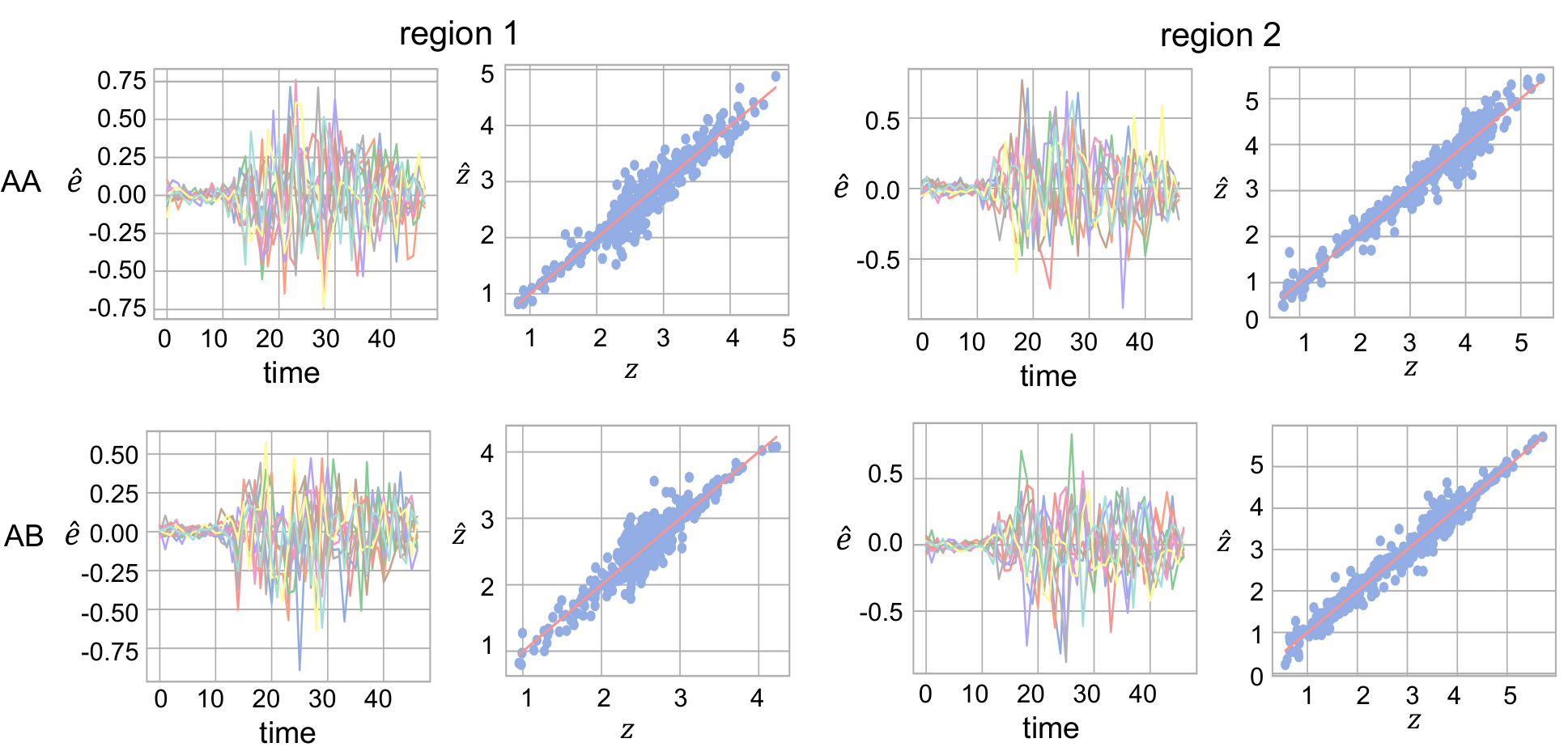}
  \caption{\label{fig:real_IE_st} Plots of the fitted number of orders against the observed values, as well as the corresponding residuals. Data are collected from an A/A or A/B experiment under the spatio-temporal alternation design.}
  \end{figure}

  We further apply the proposed test to two real datasets collected from an A/A and A/B experiment under the spatio-temporal alternation design, conducted in city $C_5$. This city is partitioned into 17 regions. Within each region, more than 90\% orders are answered by drivers in the same region. Similar to the temporal alternation design, both experiments last for 14 days and 30-minutes is set as one time unit. We take the number of requests as the state variables and drivers' total income as the outcome, as in Section \ref{sec:simuts}. In Figures \ref{fig:real_DE_st} and \ref{fig:real_IE_st}, we plot the fitted drivers' total income and the fitted number of requests against their observed values, and plot the corresponding residuals over time. We only present results associated with 2 regions in the city for space economy. The fitted values and residuals associated with other regions are similar and we do not present them to save space. It can be seen that the proposed models fit these datasets well. In addition, we report the p-values of the proposed test in Table \ref{tab:real_st}. It can be seen that the new policy significantly increases drivers' income. When applied to the dataset from the A/A experiment, it fails to reject either null hypothesis. 
  %are useful for fitting these datasets. 

 \begin{table}
  \caption{\label{tab:real_st} One sided p-values of the proposed test, when applied to two datasets collected from the A/A or A/B experiment based on the spatio-temporal alternation design. Drivers' total income is set to be the outcome of interest. }
  \centering
  {
  \begin{tabular}{|c|l|r|r|r|r|}
    \hline
    \multicolumn{1}{|r|}{} & \multicolumn{2}{c|}{DE} & \multicolumn{2}{c|}{IE} \\
    \hline
    \multicolumn{1}{|l|}{} & \multicolumn{1}{c|}{AA} & \multicolumn{1}{c|}{AB} & \multicolumn{1}{c|}{AA} & \multicolumn{1}{c|}{AB}\\
    \hline
    p-value    & 0.176
  & 0.001
  & 0.334
  & 0.000\\
    \hline
  \end{tabular}%
  }
  \end{table}%
  
\section{Discussion}\label{sec:discuss}

% In this work, motivated by policy evaluation in ride-sharing platforms,  we systematically study AB testing  in the non-stationary MDPs with weak signals. There are two important findings for power enhancement in practice. First, we  utilize the switchback design for power enhancement. 
% As mentioned earlier, by assigning different treatments to adjacent time points, 
% the random effects at these time points are likely to cancel with each other, yielding more efficient treatment effects estimators. Second, to increase the power of detecting the treatment effect in ride-sharing platforms, 
% we decompose ATE into DE and IE and test these effects separately. 
% In settings with very weak treatment effects, DE is easier to detect than ATE and IE, since  IE is a very complicated function of the estimated varying coefficients (see e.g., Equation (\ref{eqn:IE})) and is expected to have a larger variance than that of DE. Specifically, in some settings, when DE can be significant and IE maybe insignificant, 
% the signal may not be detected if we only focus on ATE.

\change{In this study, driven by the need for policy evaluation in technological companies, we thoroughly examine AB testing for temporal and/or spatial dependent experiments, particularly in scenarios characterized by weak signals, (spatio)-temporal random effects, and intricate interference structures.
Our approach offers two key benefits. Firstly, it accommodates the switchback design, which can significantly enhance testing power. As explained earlier, by applying diverse treatments to neighboring time points, we can potentially offset the impact of random effects at these times, resulting in more efficient estimations of treatment effects. 
Secondly, we break down the ATE into its DE and IE components. We then advocate for testing these effects separately. This separation aids decision-makers in gaining a clearer understanding of how different policies function and in devising more effective strategies and designs. Further details can be found in Section \ref{sec:decomp} of the supplementary document. }

   There are several intriguing avenues for future research. \change{ Firstly, considering Assumptions 1 and 2, it's worth exploring scenarios where errors in the state regression model are not necessarily independent over time. This can be achieved by incorporating random effects into the state regression model, allowing for correlated errors over time. However, this introduces dependencies between these random effects, which in turn affects the conditional independence of past and future features. Consequently, the Markov assumption is violated, and applying existing OPE methods and our proposal from Section \ref{sec:po_mdp} directly would result in biased policy value estimations. In Section \ref{sec:endogeneity} of the supplementary document, we present two approaches to mitigate this endogeneity bias. 
Secondly, we can delve into situations involving a large number of state variables. However, in ride-sharing platforms, it's reasonable to assume that the dimension of state variables is fixed. This typically involves a two-dimensional market feature, encompassing the number of call orders and the number of available drivers. We outline potential extensions to high-dimensional settings in Section \ref{subsec:highd} of the supplementary document. 
Thirdly, while the interference structure examined in this work is general, it remains relatively simple. It would be intriguing to explore more complex structural interferences across both space and time. 
Lastly, addressing statistical inference for deep neural networks remains an open challenge. This could represent a significant step toward incorporating deep learning into causal inference, offering promising directions for future research.
}

\bibliography{references}
\appendix
\section{Algorithms, Assumptions and Lemmas}
\label{app:alg}
  %{\color{blue}[I suggest to move the appendix to the supplementary article due to the page limit. Reply: agree]}
  Let $\widetilde{\bm{V}}_\theta(\tp_1,\tp_2)$ and $\bm{V}_{\widetilde\theta}(\tp_1,\tp_2)$ be the submatrices of $\widetilde{\bm{V}}_\theta$ and $\bm{V}_{\widetilde\theta}$, respectively, formed by rows in $\{(\tp_1-1)(d+2)+1,(\tp_1-1)(d+2)+2,\ldots,\tp_1(d+2)\}$ and columns in $\{(\tp_2-1)(d+2)+1,(\tp_2-1)(d+2)+2,\ldots,\tp_2(d+2)\}$. We first introduce some auxiliary lemmas. 

  \begin{lem}\label{lem:cov terms t}
  Under TCMIA and Assumptions \ref{assump:kernel} - \ref{assump:t}, as $n,m \rightarrow \infty$, $h\rightarrow0$, and $mh \rightarrow \infty$, we have $\sup_{\tp_1,\tp_2}|\widetilde{\bm{V}}_\theta(\tp_1,\tp_2) - \bm{V}_{\widetilde\theta}(\tp_1,\tp_2)| = o_p(1)$. %where $\widetilde{\bm{V}}_\theta$ and $\bm{V}_{\widetilde\theta}$ are defined in \eqref{eq:tilde V de} and \eqref{eq:V theta}, respectively.
  \end{lem}

  \begin{lem}\label{lem:cov terms st}
  Under STCMIA, Assumptions \ref{assump:kernel} and \ref{assump:st}, as $n,m,r\rightarrow\infty$, $h,h_{st}\rightarrow0$ and $mh,rh_{st}\rightarrow\infty$, then $\sup_{\tp_1,\iota_1,\tp_2,\iota_2}|\widetilde{\bm{V}}_{\theta,st}(\tp_1,\iota_1,\tp_2,\iota_2)-\bm{V}_{\widetilde\theta_{st}}(\tp_1,\iota_1,\tp_2,\iota_2)|=o_p(1)$.
  \end{lem}
  We describe our inference procedure for DE under the spatio-temporal case here. A pseudocode summarizing our algorithm is given in Algorithm \ref{alg:DE ST}.
  We denote for $\iota=1,\ldots,r$,
  \begin{eqnarray}
  &\bm{Y}_i=\diag\{Y_{i,1,1},\ldots,Y_{i,m,1},\ldots,Y_{i,1,r},\ldots,Y_{i,m,r}\},\nonumber\\
  &\bm{Z}_i=\diag\{Z_{i,1,1}^\top,\ldots,Z_{i,m,1}^\top,\ldots,Z_{i,1,r}^\top,\ldots,Z_{i,m,r}^\top\}.
  \end{eqnarray}
  Denote the longitude and latitude (scaled to be $[0,1]$) of region $\iota$ by $(u_\iota,v_\iota)$,
  % and let $K_{h_{st}}(k-\iota) = K_{h_{st}}(u_k-u_\iota) \times K_{h_{st}}(v_k-v_\iota)$,
  \begin{eqnarray}
  \label{eq:omega_st}
  &\displaystyle \kappa_{\ell,h_{st}}(\iota)=\frac{K\{(u_\iota-u_\ell)/h_{st}\}K\{(v_\iota-v_\ell)/h_{st}\}}{\sum_{j=1}^rK\{(u_\iota-u_j)/h_{st}\}K\{(v_\iota-v_j)/h_{st}\}}.
  \end{eqnarray}
  Let $\mathcal{K}=\mathcal{K}_1\mathcal{K}_2$,  where $\mathcal{K}_1$ is a block matrix whose $(\iota,\ell)$th block is $\kappa_{\ell,h_{st}}(\iota)\bm{J}_{pm}$ for $1\leq\iota,\ell\leq r$ and $\mathcal{K}_2=\diag\{\Omega,\ldots,\Omega\}$. The estimation and inference procedure of DE in the spatio-temporal case is given as follows.

  \begin{breakablealgorithm}
  \renewcommand{\algorithmicrequire}{\textbf{Input:}}
  \renewcommand{\algorithmicensure}{\textbf{Output:}}
  \caption{Inference of DE under the spatio-temporal design\label{alg:DE ST}}
  \begin{algorithmic}[1]
  \STATE Compute $\widehat\theta_{st}^0(\tp,\iota)=\left(\sum_{i=1}^nZ_{i,\tp,\iota}^\top Z_{i,\tp,\iota}\right)^{-1}\left(\sum_{i=1}^nZ_{i,\tp,\iota}^\top Y_{i,\tp,\iota}\right)$ and $\widetilde\theta^0_{st}(\tp,\iota)=\sum_{j=1}^m\omega_{j,h}(\tp)\widehat\theta(j,\iota)$ for each $\tp,\iota$.
  \STATE Compute $\widetilde\theta_{st}(\tp,\iota)=\sum_{\ell=1}^r\kappa_{\ell,h_{st}}(\iota)\widetilde\theta(\tp,\ell)$.
  \STATE Estimate the covariance $\Sigma_y$ by the following steps:

  \noindent \quad (i). estimate the combined noise by $\widehat{e}_{i,\tp,\iota}= Y_{i,\tp,\iota} - Z_{i,\tp,\iota}^\top\widetilde{\theta}_{st}(\tp,\iota)$; 

  \noindent \quad (ii). estimate the subject effects and measurement errors by
  \begin{eqnarray}\label{eq:hat eta eps st}
  &\widehat{\eta}_{i,\tp,\iota}^{\RNum{1}} =\sum_{\ell=1}^r \kappa_{\ell,h_{st}}(\iota)\sum_{j=1}^m\omega_{j,h}(\tp)\widehat{e}_{i,j,\ell},\quad\widehat{\eta}_{i,\tp,\iota}^{\RNum{2}} =\sum_{\ell=1}^r \kappa_{\ell,h_{st}}(\iota)\widehat{e}_{i,j,\ell}-\widehat{\eta}_{i,\tp,\iota}^{\RNum{1}},\nonumber\\
  &\widehat{\eta}_{i,\tp,\iota}^{\RNum{3}} =\sum_{j=1}^m\omega_{j,h}(\tp)\widehat{e}_{i,j,\ell}-\widehat{\eta}_{i,\tp,\iota}^{\RNum{1}},\quad\widehat\varepsilon_{i,\tp,\iota}=\widehat{e}_{i,\tp,\iota}-\widehat{\eta}_{i,\tp,\iota}^{\RNum{1}}-\widehat{\eta}_{i,\tp,\iota}^{\RNum{2}}-\widehat{\eta}_{i,\tp,\iota}^{\RNum{3}}.
  \end{eqnarray}
  \quad (iii). the covariances of $\eta$ and $\varepsilon$ are estimated by
  \begin{eqnarray}\label{eq:hat var eta eps st}
  &\widehat{\Sigma}_{\eta^{\RNum{1}}}(\tp_1,\iota_1, \tp_2,\iota_2) = \frac{1}{n-1}\sum_{i=1}^n\widehat{\eta}_{i,\tp_1,\iota_1}^{\RNum{1}}\widehat{\eta}_{i,\tp_2,\iota_2}^{\RNum{1}},\quad\widehat{\Sigma}_{\eta^{\RNum{2}}}(\tp_1,\iota_1, \tp_2) = \frac{1}{n-1}\sum_{i=1}^n\widehat{\eta}_{i,\tp_1,\iota_1}^{\RNum{2}}\widehat{\eta}_{i,\tp_2,\iota_1}^{\RNum{2}},\nonumber\\
  &\widehat{\Sigma}_{\eta^{\RNum{3}}}(\tp_1,\iota_1,\iota_2) = \frac{1}{n-1}\sum_{i=1}^n \widehat{\eta}_{i,\tp_1,\iota_1}^{\RNum{3}}\widehat{\eta}_{i,\tp_1,\iota_2}^{\RNum{3}},\quad\widehat{\sigma}_{\varepsilon}^2(\tp_1,\iota_1) = \frac{1}{n-1}\sum_{i=1}^n \widehat{\varepsilon}_{i,\tp_1,\iota_1}^2;
  \end{eqnarray}
  \quad (iv). the covariance of outcome is estimated by
  \begin{eqnarray*}
  &\widehat\Sigma_y(\tp_1,\iota_1,\tp_2,\iota_2)=&\widehat\Sigma_{\eta^{\RNum{1}}}(\tp_1,\iota_1,\tp_2,\iota_2)+\widehat\Sigma_{\eta^{\RNum{2}}}(\tp_1,\iota_1,\tp_2)I(\iota_1=\iota_2)\\
  & &+\widehat\sigma_{\varepsilon^{\RNum{1}}}^2(\tp_1,\iota_1,\iota_2)I(\tp_1=\tp_2)+\widehat\sigma_{\varepsilon^{\RNum{2}}}^2(\tp_1,\iota_1)I(\tp_1=\tp_2,\iota=\iota_2).
  \end{eqnarray*}
  \STATE Compute 
  \begin{eqnarray*}
  \widehat{\bm{V}}_{\theta_{st}}=\left\{\sum_{i=1}^n \bm{Z}_i^\top \bm{Z}_i\right\}^{-1} 
  \left\{\sum_{i=1}^n \bm{Z}_i^\top\widehat{\bm{\Sigma}}^{-1}\bm{Z}_i\right\} 
  \left\{\sum_{i=1}^n \bm{Z}_i^\top \bm{Z}_i\right\}^{-1}
  \end{eqnarray*}
  where $\widehat{\bm{\Sigma}}=\{\widehat\Sigma_y(\tp_1,\iota_1,\tp_2,\iota_2)\}_{\tp_1,\iota_1,\tp_2,\iota_2}$ and $\widetilde{\bm{V}}_{\theta_{st}}=\mathcal{K}\widehat{\bm{V}}_{\theta_{st}}\mathcal{K}^\top$.
  \STATE Calculate $\widehat{\DE}_{st}$ and the standard error $\widehat{se}(\widehat{\DE}_{st})$ based on $\widetilde{\bm{V}}_{\theta_{st}}$. %The one sided 
  \STATE Reject $H_{0}^{DE}$ if $\widehat{\DE}_{st}/\widehat{se}(\widehat{\DE}_{st})$ exceeds the upper $\alpha$th quantile of a standard normal distribution.
  \end{algorithmic}  
  \end{breakablealgorithm}

  \begin{breakablealgorithm}
  \renewcommand{\algorithmicrequire}{\textbf{Input:}}
  \renewcommand{\algorithmicensure}{\textbf{Output:}}
  \caption{Inference of IE under the spatio-temporal design\label{alg:ST IE}}
  \begin{algorithmic}[1]
  \STATE Compute the OLS estimator 
  \begin{eqnarray*}
  \widehat{\bm{\Theta}} = \left\{\sum_{i=1}^n \bm{Z}_{i,(-m)} \bm{Z}_{i,(-m)}^\top\right\}^{-1}
     \left\{\sum_{i=1}^n \bm{Z}_{i,(-m)} \bm{S}_{i,(-1)}^\top\right\}.
  \end{eqnarray*}
  \STATE Compute ${\widetilde\Theta_{st}=\mathcal{K}\widetilde\Theta}$.
  \STATE Plug-in the parameter estimates $\widetilde{\Theta}_{st}$ and $\widetilde{\theta}_{st}$ to obtain $\widehat\IE_{st}$.
  \STATE Compute the residuals $\widehat{E}_{i,\tp,\iota}=S_{i,\tp,\iota}-Z_{i,\tp,\iota}^\top \widetilde{\Theta}(\tp,\iota)$.
  \FOR{$b=1,\ldots,B$}
  \STATEx generate i.i.d. standard normal random variables $\{\xi_i^b\}_{i=1}^n$;
  \STATEx generate pseudo outcomes $S_{i,\tp,\iota}^b$ and $Y_{i,\tp,\iota}^b$ by $S_{i,\tp+1,\iota}^b=Z_{i,\tp,\iota} \widetilde{\Theta}(\tp,\iota)+\xi_i\widehat{E}_{i,\tp,\iota}$ and $Y_{i,\tp,\iota}^b=Z_{i,\tp,\iota} \widetilde{\theta}_{st}(\tp,\iota)+\xi\widehat{e}_{i,\tp,\iota}$, where $Z_{i,\tp,\iota}^b=\{1,(S_{i,\tp,\iota}^b)^\top,A_{i,\tp,\iota},\bar{A}_{i,\tp,\mathcal{N}_\iota}\}^\top$;
  \STATEx substitute $Y_{i,\tp,\iota}$ and $S_{i,\tp,\iota}$ with $Y_{i,\tp,\iota}^b$ and $S_{i,\tp,\iota}^b$, and repeat the procedures in Steps 1-3 to obtain the plug-in estimator $\widehat{\IE}_{st}^{b}$.
  \ENDFOR
  \STATE Reject $H_0^{IE}$ if $\widehat{\IE}_{st}$ exceeds the upper $\alpha$th empirical quantile of $\{\widehat{\IE}_{st}^b-\widehat{\IE}_{st}\}_b$.
  %Compute the $p$-values
  % \begin{equation*}
  % p_1=\frac{1}{B}\sum_{b=1}^B\mathbbm{1}\{\widehat{\IE}_{st}^{b}>\widehat{\IE}_{st}\},\quad p_2=\frac{2}{B}\sum_{b=1}^B\mathbbm{1}\{\IE_{st}^{b}>|\widehat{\IE}_{st}|\}.
  % \end{equation*}
  \end{algorithmic}  
  \end{breakablealgorithm}

\section{Proof of Lemma \ref{lemma:identification}}
{We first prove \eqref{eqn:reward}. %Without loss of generality, assume both the outcome and the state are discrete. 
%Notice that
It follows from the law of total expectation that
\begin{eqnarray*}
    \Mean(Y_{\tp}|\bar{A}_{\tp},\bar{S}_{\tp})=\Mean^{\bar{Y}_{\tp-1}|\bar{A}_{\tp},\bar{S}_{\tp}} \{  \Mean(Y_{\tp}|\bar{A}_{\tp}, \bar{S}_{\tp}, \bar{Y}_{\tp-1})\},
\end{eqnarray*}
where $\bar{A}_{\tp}$, $\bar{S}_{\tp}$ and $\bar{Y}_{\tp-1}$ denote the history of actions, states and outcomes, respectively. %$\bar{H}_{\tp-1}$ denotes the union of historical actions, states, outcomes, and 
The first expectation on the right-hand-side (RHS) is taken with respect to the conditional distribution of $\bar{Y}_{\tp-1}$ given that $(\bar{A}_{\tp},\bar{S}_{\tp})$. 

Without loss of generality, assume both the outcome and the state are discrete. Let $p^{\bar{Y}_{\tp-1}|\bar{A}_{\tp},\bar{S}_{\tp}}$ denotes the conditional probability mass function of $\bar{Y}_{\tp-1}$ given $\bar{A}_{\tp},\bar{S}_{\tp}$, we have %It follows from CA that $\Mean(Y_{\tp}|\bar{A}_{\tp}, \bar{S}_{\tp}, \bar{Y}_{\tp-1})=$
\begin{eqnarray*}
    \Mean(Y_{\tp}|\bar{A}_{\tp}=\bar{a}_{\tp},\bar{S}_{\tp}=\bar{s}_{\tp})=\sum_{ \bar{y}_{\tp-1}} p^{\bar{Y}_{\tp-1}|\bar{A}_{\tp}=\bar{a}_{\tp},\bar{S}_{\tp}=\bar{s}_{\tp}}(\bar{y}_{\tau-1})\{\Mean(Y_{\tp}|\bar{A}_{\tp}=\bar{a}_{\tp}, \bar{S}_{\tp}=\bar{s}_{\tp}, \bar{Y}_{\tp-1}=\bar{y}_{\tp-1}) \}.
\end{eqnarray*}
According to CA, the second term on the RHS is equal to
\begin{eqnarray*}
	\Mean[Y_{\tp}^*(\bar{a}_{\tp})|\bar{A}_{\tp}=\bar{a}_{\tp},\bar{S}_{\tp}^*(\bar{a}_{\tp-1})=\bar{s}_{\tp},\bar{Y}_{\tp-1}^*(\bar{a}_{\tp-1})=\bar{y}_{\tp-1}],
\end{eqnarray*}
where $\bar{S}_{\tp}^*(\bar{a}_{t-1})$ and $\bar{Y}_{\tp-1}^*(\bar{a}_{\tp-1})$ denote the sets of potential states and outcomes up to time $\tp$ and $\tp-1$, respectively. It follows that
\begin{eqnarray*}
	 \Mean(Y_{\tp}|\bar{A}_{\tp}=\bar{a}_{\tp},\bar{S}_{\tp}=\bar{s}_{\tp})=\sum_{ \bar{y}_{\tp-1}} p^{\bar{Y}_{\tp-1}|\bar{A}_{\tp}=\bar{a}_{\tp},\bar{S}_{\tp}=\bar{s}_{\tp}}(\bar{y}_{\tau-1})\{\Mean[Y_{\tp}^*(\bar{a}_{\tp})|\bar{A}_{\tp}=\bar{a}_{\tp},\bar{S}_{\tp}^*(\bar{a}_{\tp-1})=\bar{s}_{\tp},\bar{Y}_{\tp-1}^*(\bar{a}_{\tp-1})=\bar{y}_{\tp-1}]\}.
\end{eqnarray*}
Under SRA and PA, the conditional expectation on the right-hand-side is independent of the actions. In addition, it is equal to $R_{\tp}(\bar{a}_{\tp},\bar{s}_{\tp})$, independent of $\bar{y}_{\tp-1}$. This yields \eqref{eqn:reward}.

We next show \eqref{eqn:rtau}. Using similar arguments, we can show that
\begin{eqnarray*}
    \Mean \{R_{\tp}(a_{\tp},S_{\tp}^*(\bar{a}_{\tp-1}),\cdots,S_1)\}=\Mean [\Mean \{R_{\tp}(a_{\tp},S_{\tp}^*(\bar{a}_{\tp-1}), \cdots,S_1)|A_1=a_1, \bar{S}_{\tp-1}^*(\bar{a}_{\tp-2}), \bar{Y}_{\tp-1}^*(\bar{a}_{\tp-1})\}].
\end{eqnarray*}
Under CA, we can replace $Y_1^*(a_1)$ and $S_2^*(a_1)$ with $Y_1$ and $S_2$, respectively. Under SRA and PA, the event $A_2=a_2$ can be included in the conditioning set. This yields that 
\begin{eqnarray*}
    &&\Mean \{R_{\tp}(a_{\tp},S_{\tp}^*(\bar{a}_{\tp-1}),\cdots,S_1)\}\\&=&\Mean [\Mean \{R_{\tp}(a_{\tp},S_{\tp}^*(\bar{a}_{\tp-1}), \cdots,A_1,S_1)|A_2=a_2, A_1=a_1, \bar{S}_{\tp-1}^*(\bar{a}_{\tp-2}), \bar{Y}_{\tp-1}^*(\bar{a}_{\tp-1}),S_1,Y_1\}].
\end{eqnarray*}
Iteratively applying this argument allows us to repeatedly replace the counterfactual variables with the observed ones. At the end, all the potential outcomes/states will be replaced with the observed versions conditional on the actions. The proof is hence completed.}
%yields the desired assertion.}
%yields the desired assertion. The proof is hence completed.

\section{Proof of Proposition \ref{concl:linear}}\label{pf:prop1}
{Recall that 
\begin{eqnarray*}
\DE=\sum_{\tp=1}^m \Mean \{R_{\tp}(1,S_\tp^*(\bm{0}_{\tp-1}),0, S_{\tp-1}^*(\bm{0}_{\tp-2}),\ldots,S_1)-R_\tp(0,S_\tp^*(\bm{0}_{\tp-1}),0, S_{\tp-1}^*(\bm{0}_{\tp-2}),\ldots,S_1)\},\\ 
\IE=\sum_{\tp=1}^m \Mean  \{R_{\tp}(1,S_\tp^*(\bm{1}_{\tp-1}),1, S_{\tp-1}^*(\bm{1}_{\tp-2}),\ldots,S_1)-R_\tp(1,S_\tp^*(\bm{0}_{\tp-1}),0, S_{\tp-1}^*(\bm{0}_{\tp-2}),\ldots,S_1)\}.
\end{eqnarray*}
%By the fact that 
Under Model \ref{exmp:linear}, each summand in DE equals $\gamma(\tp)$. It follows that
$$\DE=\sum_{\tp=1}^m\gamma(\tp).$$
Similarly, for IE, we have
\begin{align*}
&\Mean \{R_{\tp}(1,S_\tp^*(\bm{1}_{\tp-1}),1, S_{\tp-1}^*(\bm{1}_{\tp-2}),\ldots,S_1)-R_\tp(1,S_\tp^*(\bm{0}_{\tp-1}),0, S_{\tp-1}^*(\bm{0}_{\tp-2}),\ldots,S_1)\}\\
=&\Mean\{\beta_0(\tp)+S_\tp^*(\bm{1}_{\tp-1})^\top\beta(\tp)+\gamma(\tp)\}-\Mean\{\beta_0(\tp)+S_\tp^*(\bm{0}_{\tp-1}))^\top\beta(\tp)+\gamma(\tp)\}\\ 
=&\Mean\{S_\tp^*(\bm{1}_{\tp-1})-S_\tp^*(\bm{0}_{\tp-1})\}^\top\beta(\tp)\\ 
=&\Mean[\Phi(\tp-1)\{S_{\tp-1}^*(\bm{1}_{\tp-2})-S_{\tp-1}^*(\bm{0}_{\tp-2})\}+\Gamma(\tp-1)]^\top\beta(\tp)\\
=&\Mean[\Phi(\tp-1)\Phi(\tp-2)\{S_{\tp-2}^*(\bm{1}_{\tp-3})-S_{\tp-2}^*(\bm{0}_{\tp-3})\}+\Phi(\tp-1)\Gamma(\tp-2)+\Gamma(\tp-1)]^\top\beta(\tp)\\
\cdots\\
=&\beta(\tp)^\top \left\{ \sum_{k=1}^{\tp-1} \left(\prod_{l=k+1}^{\tp-1} \Phi(l)\right) \Gamma(k) \right\},
\end{align*}
which completes the proof.
 }
%$\bar{a}_{t-1}$

\section{Proofs of Lemmas \ref{lem:cov terms t} and \ref{lem:cov terms st}}

The proof of Lemma \ref{lem:cov terms st} is similar to that of Lemma \ref{lem:cov terms t}.  Hence, we focus on proving Lemma \ref{lem:cov terms t} for space economy.

\begin{pf}:
We first prove that $\sup_{\tp_1,\tp_2}\vert\widehat\Sigma_y(\tp_1,\tp_2)-\Sigma_y(\tp_1,\tp_2)\vert=o_p(1)$. It suffices to show that $n^{-1}\sum_{i=1}^n \widehat{\eta}_{i,\tp_1} \widehat{\eta}_{i,\tp_2}$ and $n^{-1}\sum_{i=1}^n \widehat{\varepsilon}_{i,\tp}^2$ are consistent estimators of $\Sigma_\eta(\tp_1,\tp_2)$ and $\sigma_{\varepsilon,\tp}^2$. According to Section \ref{sec:T method}, we have $\widehat{e}_{i,\tp} = Y_{i,\tp} - Z_{i,\tp}^\top \widetilde{\theta}(\tp)$. Notice that
\begin{equation*}
\widehat\eta_{i,\tp}=\sum_{j=1}^m \omega_{j,h}(\tp)\widehat{e}_i(j).
\end{equation*}
We follow notations in \cite{zhu2014spatially} and write
\begin{eqnarray*}
&\bar{\varepsilon}_{i,\tp} = \sum_{j=1}^m \omega_{j,h}(\tp) \varepsilon_{i,j}, \quad
\Delta_K \eta_{i,\tp} = \sum_{j=1}^m \omega_{j,h}(\tp) \left\{ \eta_{i,j} - \eta_{i,\tp} \right\},\\
&\Delta_K \theta(\tp) = \sum_{j=1}^m \omega_{j,h}(\tp) \left\{ \theta(j) - \widehat\theta(j) \right\}, \quad
\Delta_{\eta_i}(\tp) = \bar{\varepsilon}_{i,\tp} + \Delta_K \eta_{i,\tp} + Z_{i,\tp}^\top \Delta_K \theta (\tp)
.
\end{eqnarray*}
Then we have 
\begin{equation*}
\widehat{\eta}_{i,\tp} - {\eta}_{i,\tp} = \Delta_{\eta_i}(\tp),
\end{equation*}
which gives
\begin{align}
n^{-1}\sum_{i=1}^n \widehat{\eta}_{i,\tp_1} \widehat{\eta}_{i,\tp_2} \, = \,
& n^{-1}\sum_{i=1}^n \eta_{i,\tp_1}\eta_{i,\tp_2} 
+ n^{-1}\sum_{i=1}^n \Delta_{\eta_i}(\tp_1)\Delta_{\eta_i}(\tp_2) \nonumber\\ 
& + n^{-1}\sum_{i=1}^n \eta_{i,\tp_1}\Delta_{\eta_i}(\tp_2)  
+ n^{-1}\sum_{i=1}^n \Delta_{\eta_i}(\tp_1)\eta_{i,\tp_2}.\nonumber
\end{align} 
The first term $n^{-1}\sum_{i=1}^n \eta_{i,\tp_1}\eta_{i,\tp_2}$ converges to $\Phi_{\eta}(\tp_1, \tp_2)$ according to the Law of Large Number.
We next show
\begin{itemize}
\item[(a)] $I_1= n^{-1} \sum_{i=1}^n \Delta_{\eta_i}(\tp_1)\Delta_{\eta_i}(\tp_2) $ converges to zero for any $(\tp_1,\tp_2) \in \mathcal{T}^2$.
\item[(b)] $I_2= n^{-1} \sum_{i=1}^n \eta_{i,\tp_1}\Delta_{\eta_i}(\tp_2) + n^{-1}\sum_{i=1}^n \Delta_{\eta_i}(\tp_1)\eta_{i,\tp_2}$ converges to zero for any $(\tp_1,\tp_2) \in \mathcal{T}^2$.
\end{itemize}

By mutually multiplying the three terms in the summation form of $\Delta_{\eta_i}(\tp)$, we have % nine terms in total. 
\begin{align*}
I_1 &= n^{-1} \sum_{i=1}^n \bar{\varepsilon}_{i,\tp_1} \bar{\varepsilon}_{i,\tp_2} +
n^{-1}\sum_{i=1}^n \Delta_K \eta_{i,\tp} \Delta_K \eta_{i,\tp_2}  + 
n^{-1}\sum_{i=1}^n  Z_{i,\tp_1}^\top \Delta_K \theta(\tp_1) \Delta_K \theta(\tp_2)^\top Z_{i,\tp_2}   \\
& + n^{-1}\sum_{i=1}^n  \bar{\varepsilon}_{i,\tp_1} \Delta_K \eta_i(\tp_2)  +
n^{-1}\sum_{i=1}^n \Delta_K \eta_{i,\tp_1} \bar{\varepsilon}_{i,\tp_2} +
n^{-1}\sum_{i=1}^n  \bar{\varepsilon}_{i,\tp_1}  \Delta_K \theta(\tp_2)^\top Z_{i,\tp_2} \\
& + n^{-1}\sum_{i=1}^n  Z_{i,\tp_1}^\top \Delta_K \theta(\tp_1)\bar{\varepsilon}_{i,\tp_2} +
n^{-1}\sum_{i=1}^n \Delta_K \eta_{i,\tp_1} \Delta_K \theta(\tp_2)^\top Z_{i,\tp_2} +
n^{-1}\sum_{i=1}^n  Z_{i,\tp_1}^\top \Delta_K \theta(1)\Delta_K \eta_{i,\tp_2}
\end{align*}
By the independence between $\varepsilon_{i,\tp_1} $ and $\varepsilon_{i,\tp_2} $, the first term  $n^{-1} \sum_{i=1}^n \bar{\varepsilon}_{i,\tp_1} \bar{\varepsilon}_{i,\tp_2}$ converges to zero. As for the second term, using standard arguments in establishing theoretical properties of kernel estimators\footnote{See e.g., \url{http://www.stat.cmu.edu/~larry/=sml/NonparRegression.pdf}.}, the bias term satisfies $\Mean\sum_{j=1}^m\omega_{j,h}(\tp) \left\{ \eta_{i,j}-\eta_{i,\tp} \right\} = O(h^2+m^{-1})$, whereas the variance term satisfies $\Var [\sum_{j=1}^m\omega_{j,h}(\tp) \left\{ \eta_{i,j}-\eta_{i,\tp} \right\}] = O(m^{-\textbf{}1}h^{-1})$. It follows that
\begin{align*}
&n^{-1}\sum_{i=1}^n \Delta_K \eta_{i,\tp_1} \Delta_K \eta_{i,\tp_2} \\
=& n^{-1}\sum_{i=1}^n \left[ \sum_{j=1}^m\omega_{j,h}(\tp_1) \left\{ \eta_{i,j}-\eta_{i,\tp_1}\right\} \right] 
\left[ \sum_{j=1}^m \omega_{j,h}(\tp_2) \{ \eta_{i,j}-\eta_{i,\tp_2}\} \right] \\
=&O_p(h^4+m^{-1}h^{-1}).
\end{align*}
%following the facts that $\Mean\sum_{j=1}^m\omega_{j,h}(\tp) \left\{ \eta_{i,j}-\eta_{i,\tp} \right\} = O_p(h^2+m^{-1})$ and $\Var\sum_{j=1}^m\omega_{j,h}(\tp) \left\{ \eta_{i,j}-\eta_{i,\tp} \right\} = O_p(m^{-1}h^{-1})$. } 
As for the third term, notice that $ \left\{ \widehat\theta(\tp) - \theta(\tp):\tp \right\}$ converges uniformly to zero, $\left\{\Delta_K \theta(\tp):\tp\right\}$ converges uniformly to zero as well. Under the given conditions, $n^{-1}\sum_{i=1}^n Z_{i,\tp_1}^\top Z_{i,\tp_2}$ is $O_p(1)$. It follows that the third term is $o_p(1)$. 
%$$ %n^{-1}\sum_{i=1}^n  Z_{i,\tp}^\top \Delta_K \theta(\tp) \Delta_K \theta(t^{\prime})^\top Z_i(t^\prime) =  n^{-1} \left\{ \sum_{i=1}^n Z_{i,\tp}^\top Z_{i,\tp} \right\} \Delta_K \theta(\tp) \Delta_K \theta(t^{\prime})^\top = o(1),$$
%since $ \left\{ \widehat\theta(\tp) - \theta(\tp) \right\}$ converges to zero and then $\Delta_K \theta(\tp)$ converges to zero. 
The remaining six cross products converges to zero according to the Law of Large Number and the mutual independence of $Z_i$, $\varepsilon_i $, and $\eta_i$ imposed in Assumption \ref{assump:t} .
This completes the proof of (a).

To prove (b), we only need to prove $ n^{-1} \sum_{i=1}^n \eta_{i,\tp_1}\Delta_K \eta_{i,\tp_2} = o_p(1)$ since that $\eta_i$ is independent of $Z_i$ and $\varepsilon_i $. 
This follows from the fact that 
\begin{align}\label{222}
 & n^{-1}\sum_{i=1}^n \eta_{i,\tp_1} \left[ \sum_{j=1}^m \omega_{j,h}(\tp_2) \{  \eta_{i,j}-\eta_{i,\tp_2} \} \right]\nonumber\\
 =& \sum_{j=1}^m \omega_{j,h}(\tp_2) n^{-1} \left\{ \sum_{i=1}^n \eta_{i,j}\eta_{i,\tp} - \sum_{i=1}^n \eta_{i,\tp_1}\eta_{i,\tp_2} \right\}  \nonumber \\
 =& \sum_{j=1}^m \omega_{j,h}(\tp_2)\{\Sigma_\eta(j,\tp_1) - \Sigma_\eta(t,\tp_2)\} + o_p(1), 
\end{align}   
where the first two term on the right hand of (\ref{222}) is $ O(h^2) $ according to the assumption on the distribution of $\eta_{i,\tp}$; see the equation (26) in the supplementary materials of \cite{zhu2014spatially}.

We next prove the consistency of $n^{-1}\sum_{i=1}^n \widehat{\varepsilon}_{i,\tp}^2$. Notice that 
$$\widehat{\varepsilon}_{i,\tp} = \widehat{e}_{i,\tp} - \widehat{\eta}_{i,\tp} = y_{i,\tp} - Z_{i,\tp}^\top \widehat{\theta}(\tp) -  \widehat{\eta}_{i,\tp}. $$ 
Similarly to the proof of (a),
% Lemma \ref{lem:cov terms st} (1), 
we denote 
$\Delta_{\theta}(\tp) = \widehat\theta(\tp) - \theta(\tp)$, and $ \Delta_{\varepsilon_i}(\tp) = - Z_{i,\tp}^\top \Delta_{\theta}(\tp) - \Delta_{\eta_i}(\tp)$. It follows that  
$$  n^{-1}\sum_{i=1}^n \widehat{\varepsilon}_{i,\tp}^2 = 
n^{-1}\sum_{i=1}^n \varepsilon_{i,\tp}^2  
+ n^{-1}\sum_{i=1}^n \Delta_{\varepsilon_i}^2(\tp) 
+ 2n^{-1}\sum_{i=1}^n \varepsilon_{i,\tp}\Delta_{\varepsilon_i}^2(\tp).
$$
The first term $ n^{-1}\sum_{i=1}^n \varepsilon_{i,\tp}^2 $ converges to $\sigma_\varepsilon^2(\tp)$ according to the Law of Large Number, and the other two terms both converge to zero based on the same arguments used before. We omit the details to save space.

Finally, recall that $\widehat{\bm{V}}_\theta$ is the sandwich estimator of $\bm{V}_{\widehat\theta}$ defined in \eqref{eqn:sandwich}. It is straightforward to show that $\sup_{\tp_1,\tp_2}\vert\widehat{\bm{V}}_{\theta}(\tp_1,\tp_2)-\bm{V}_{\widehat\theta}(\tp_1,\tp_2)\vert=o_p(1)$ based on $\sup_{\tp_1,\tp_2}\vert\widehat\Sigma_y(\tp_1,\tp_2)-\Sigma_y(\tp_1,\tp_2)\vert=o_p(1)$. %Note that $\bm{V}_{\widehat\theta}$ has a ridge of $\sigma_{\varepsilon,\tp}^2$ on the diagonal compared to $\bm{V}_{\widetilde\theta}$, and multiplying $\Omega$ is the operation of applying kernel smoothing to $\widehat{\bm{V}}$ (more specific, smoothing out the ridge). By the consistency of $\widehat{\bm{V}}$, 
Similarly, we can derive that $\sup_{\tp_1,\tp_2}\vert\widetilde{\bm{V}}(\tp_1,\tp_2)-\bm{V}_{\widetilde\theta}(\tp_1,\tp_2)\vert=o_p(1)$. We omit the details to save space.  
\end{pf}

\section{Proof of Theorem \ref{thm:t theta asymp}}\label{pf:de t}
\begin{pf}:
Argument (i) in Theorem \ref{thm:t theta asymp} can be directly proven based on the properties of the ordinary least square estimator. We focus on proving Argument (ii).
Notice that $\widetilde\theta(\tp)$ can essentially rewritten as a linear combination of $\{\widehat{\theta}(k)\}_k$, i.e.,
\begin{align*}
\widetilde\theta(\tp)&=\sum_{k=1}^m\omega_{k,h}(\tp)\widehat\theta(k)=\sum_{k=1}^m\omega_{k,h}(\tp)\{\widehat\theta(k)-\theta(k)+\theta(k)-\theta(\tp)+\theta(\tp)\}\\
&=\theta(\tp)+\sum_{k=1}^m\omega_{k,h}(\tp)\{\widehat\theta(k)-\theta(k)\}+\sum_{k=1}^m\omega_{k,h}(\tp)\{\theta(k)-\theta(\tp)\}.
\end{align*}
It follows that
$$
\begin{aligned}
& E\{\widetilde{\theta}(\tau)-\theta(\tau)\} \\
=& \sum_{k=1}^{m} \omega_{k, h}(\tau)\{\widehat{\theta}(k)-\theta(\tau)\} \\
=& \sum_{k=1}^{m} \omega_{k, h}(\tau)\{\theta(k)-\theta(\tau)\} \\
=&\left\{\sum_{k=1}^{m} \frac{1}{m h} K\left(\frac{\tau-k}{m h}\right)\right\}^{-1} \cdot\left[\sum_{k=1}^{m} \frac{1}{m h} K\left(\frac{\tau-k}{m h}\right)\{\theta(k)-\theta(\tau)\}\right]
\end{aligned}
$$
Denote
$$
\begin{aligned}
&{f}(\tau)=\sum_{k=1}^{m} \frac{1}{m h} K\left(\frac{\tau-k}{m h}\right) \\
&{g}_{1}(\tau)=\sum_{k=1}^{m} \frac{1}{m h} K\left(\frac{\tau-k}{m h}\right)\{\theta(k)-\theta(\tau)\}
\end{aligned}
$$
Note that $f(\tau) {\longrightarrow} 1$, %i.e. $f(\tau)=O(1)$, 
it suffices to bound $|{g}_{1}(\tau)|$. Define
$$
{g}_{2}(\tau)=\int_{0}^{1} \frac{1}{h} K\left(\frac{u m-\tau}{m h}\right)\{\theta(u m)-\theta(\tau)\} d u.
$$
By decomposing ${g}_{1}(\tau)={g}_{2}(\tau)+\left\{{g}_{1}(\tau)-{g}_{2}(\tau)\right\}$, we first show ${g}_{2}(\tau)=O\left(h^{2}\right)$, and then prove ${g}_{1}(\tau)-{g}_{2}(\tau)=O\left(m^{-1}\right)$.
The time domain of interest is fixed, and the increment of $m$ equals the encryption of grids. Define a function $\theta_0$ such that $\theta_{0}(\cdot)$ such that $\theta(\tau)=\theta_{0}\left(\frac{\tau}{m}\right)$ for any $\tau$. It follows that
$$
\theta(s)-\theta(t)=\theta_{0}\left(\frac{s}{m}\right)-\theta_{0}\left(\frac{t}{m}\right)=\theta_{0}^{\prime}\left(\frac{t}{m}\right)\left(\frac{s-t}{m}\right)+\frac{1}{2} \theta_{0}^{\prime \prime}\left(\frac{t}{m}\right)\left(\frac{s-t}{m}\right)^{2}+O\left(m^{-3}\right).
$$
Then we have
$$
\begin{aligned}
{g}_{2}(\tau) &=\int_{0}^{1}  \frac{1}{h} K\left(\frac{u-\tau / m}{h}\right)\left\{\theta_{0}(u)-\theta_{0}\left(\frac{\tau}{m}\right)\right\} d u \\
&=\int_{0}^{1} \frac{1}{h} K\left(\frac{u-\tau / m}{h}\right)\left\{\theta_{0}^{\prime}\left(\frac{\tau}{m}\right)\left(u-\frac{\tau}{m}\right)+\theta_{0}^{\prime \prime}\left(\frac{\tau}{m}\right)\left(u-\frac{\tau}{m}\right)^{2}\right\} d u \\
&=\int_{0}^{1} K\left(\frac{u-\tau / m}{h}\right) \cdot\left(\frac{u-\tau / m}{h}\right)^{2} \cdot \theta_{0}^{\prime \prime}\left(\frac{\tau}{m}\right) h^{2} d\left(\frac{u-\tau / m}{h}\right) \\
&=O\left(h^{2}\right).
\end{aligned}
$$
Note that for any second-order continuous function $f_0$,
$$
\int_{a}^{b} f_0(x) d x=\frac{1}{2}(b-a)\{f_0(a)+f_0(b)\}-\frac{1}{12}(b-a)^{3} f_0^{\prime \prime}(\xi)
$$
for some $\xi \in(a, b)$. Let
$$
s(u)=\frac{1}{h} K\left(\frac{u-\tau / m}{h}\right)\left\{\theta_{0}(u)-\theta_{0}\left(\frac{\tau}{m}\right)\right\}.
$$
Then where exists some $\xi_{k} \in(k-1, k)$ such that
$$
\begin{aligned}
{g}_{2}(\tau) &=\sum_{k=1}^{m} \int_{{(k-1)}/{m}}^{k/m} s(u) d u \\
&=\sum_{k=1}^{m} \frac{1}{2 m}\{s(k)+s(k-1)\}-\frac{1}{12 m} \sum_{k=1}^{m} s^{\prime \prime}\left(\xi_{k}\right) \\
&={g}_{1}(\tau)+\frac{1}{2 m}\{s(0)-s(m)\}-\frac{1}{12 m} \sum_{k=1}^{m} s^{\prime \prime}\left(\xi_{k}\right)
\end{aligned}
$$
Hence
$$
{g}_{2}(\tau)-{g}_{1}(\tau)=\frac{1}{2 m}\{s(0)-s(m)\}-\frac{1}{12 m} \sum_{k=1}^{m} s^{\prime \prime}\left(\xi_{k}\right).
$$
We can represent $(12 m)^{-1} \sum_{k=1}^{m} s^{\prime \prime} (\xi_{k})$ as the summation of the follow three quantities:
{\small\begin{align*}
   & \frac{1}{12m^3h^3}\sum_{k=1}^mK''\left(\frac{\xi_k-\tp}{mh}\right)\left\{\theta_0\left(\frac{\xi_k}{m}\right)-\theta_0\left(\frac{\tp}{m}\right)\right\}
   \approx\frac{1}{12m^2h^2}\int_0^1\frac{1}{h}K''\left(\frac{u-\tp/m}{h}\right)\left\{\theta_0(u)-\theta_0\left(\frac{\tp}{m}\right)\right\}
   =O(m^{-2}),\\
   &\frac{1}{12m^3h^2}\sum_{k=1}^mK'\left(\frac{\xi_k-\tp}{mh}\right)\theta_0'(\xi_k/m)
   \approx\frac{1}{12m^2h}\int_0^1\frac{1}{h}K'\left(\frac{u-\tp/m}{h}\right)\theta_0'(u)du
   =O(m^{-2}h^{-1}),\\
   &\frac{1}{12m^3h}\sum_{k=1}^mK\left(\frac{\xi_k-\tp}{mh}\right)\theta_0''(\xi_k/m)
   \approx\frac{1}{12m^2}\int_0^1K\left(\frac{u-\tp/m}{h}\right)\theta_0''(u)du
   =O(m^{-2}).
\end{align*}}
It follows that
${g}_{2}(\tau)-{g}_{1}(\tau)=O(m^{-1})$
and the bias term satisfies
\begin{equation}{\label{eq:gap}}
g_1(\tp)=O(m^{-1}+h^2).
\end{equation}
 As for the covariance matrix, we have that
\begin{align*} 
& \operatorname{Cov}\{\tilde{\theta}(\tau), \tilde{\theta}(s)\} \\
=& \operatorname{Cov}\left\{\sum_{k=1}^{m} w_{h}(\tau-k) \widehat{\theta}(k), \Sigma_{l=1}^{m} w_{h}(s-l) \widehat{\theta}(l)\right\} \\
=& E\left[\sum_{k=1}^{m}\sum_{l=1}^{m} w_{h}(\tau-k) w_{h}(s-l)\{\widehat{\theta}(k)-\theta(k)\}\{\widehat{\theta}(l)-\theta(l)\}\right]\\
=& \frac{1}{n} \sum_{k=1}^{m} \sum_{l=1}^{m} w_{h}(\tau-k) w_{h}(s-l) V_{\widehat{\theta}}(k, l) \\=& \frac{1}{n} \cdot \frac{\widehat{g}(\tau, s)}{f(\tau) \cdot {f}(s)}, 
\end{align*}
where $V_{\widehat{\theta}}(k, l)=\operatorname{Cov}\{\widehat{\theta}(k), \widehat{\theta}(l)\} \in \mathbb{R}^{p \times p}$ and that 
 \begin{align*} 
& \widehat{g}(\tau, s)=\frac{1}{n m^{2} h^{2}}\left[\sum_{k=1}^{m} \sum_{l=1}^{m} K\left(\frac{\tau-k}{m h}\right) K\left(\frac{s-l}{m h}\right) V_{\widehat{\theta}}(k, l)\right].\\ 
\end{align*}
 Let
 \begin{align*} 
& V_{\varepsilon}=V_{\widehat{\theta}}-V_{\widetilde{\theta}} \\ &=\left(E Z_{i}^{\top} Z_{i}\right)^{-1} \cdot E\left(Z_{i}^{\top} \Sigma_{\varepsilon} Z_{i}^{\top}\right) \cdot\left(E Z_{i}^{\top} Z_{i}\right)^{-1} \\
=&\operatorname{diag}\left\{\sigma_{j}^{2}\left(E Z_{i j} Z_{i j}^{\top}\right)^{-1}\right\}_{j=1, \ldots, m},
\end{align*}
and $V_{\varepsilon}(k)=\sigma_{k}^{2}\left(E Z_{i k} Z_{i k}^{\top}\right)^{-1}$.
Then we can represent
$$
\widehat{g}(\tau, s)=\widehat{g}_{1}(\tau, s)+\widehat{g}_{2}(\tau, s),
$$
where
\begin{align*}
&\widehat{g}_{1}(\tau, s)=\frac{1}{n m^{2} h^{2}}\left[\sum_{k=1}^{m} \sum_{l=1}^{m} K\left(\frac{\tau-k}{m h}\right) K\left(\frac{s-l}{m h}\right) V_{\tilde{\theta}}(k, l)\right], \\
&\widehat{g}_{2}(\tau, s)=\frac{1}{n m^{2} h^{2}}\left[\sum_{k=1}^{m} K\left(\frac{\tau-k}{m h}\right) K\left(\frac{s-k}{m h}\right) V_{\varepsilon}(k)\right].
\end{align*}
Using the same arguments in \eqref{eq:gap}, we have
\begin{align*}
&\widehat{g}_{1}(\tau, s)=\frac{1}{n}V_{\tilde{\theta}}(\tau,s)+O(n^{-1}m^{-1}+n^{-1}h^2), \\
&\widehat{g}_{2}(\tau, s)=O(n^{-1}m^{-1}).
\end{align*}
The above arguments implies that for any vector $\bm{a}_{n,2}$ with unit $\ell_2$ norm, the asymptotic bias of $\sqrt{n}\bm{a}_{n,2}^\top(\widetilde{\bm{\theta}}-\bm{\theta})$ is upper bounded by $n^{-1/2}\|\bm{a}_{n,2}\|_2 \| \Mean \widetilde{\bm{\theta}}-\bm{\theta} \|_2=O(\sqrt{n}h^2+\sqrt{n}m^{-1})$, using Cauchy-Schwarz inequality, and that its asymptotic variance is given by $\bm{a}_{n,2}^\top \bm{V}_{\widetilde\theta} \bm{a}_{n,2}$. Under the assumption that  $\lambda_{\min}(\bm{a}_{n,2}^\top \bm{V}_{\widetilde\theta} \bm{a}_{n,2})$ is bounded away from zero, the bias of $\sqrt{n}\bm{a}_{n,2}^\top(\widetilde{\bm{\theta}}-\bm{\theta})/\sqrt{\bm{a}_{n,2}^\top \bm{V}_{\widetilde\theta} \bm{a}_{n,2}}$ is bounded by $O(\sqrt{n}h^2+\sqrt{n}m^{-1})$ as well.

%{
%\color{blue}
It remains to prove the asymptotic normality of $\sqrt{n}\bm{a}_{n,2}^\top(\widetilde{\bm{\theta}}-\bm{\theta})$. Let $\bm{a}_{n,2}=(a_{n,2,1}^\top, a_{n,2,2}^\top,\cdots,a_{n,2,m}^\top)^\top$ where each $a_{n,2,\tp}$ corresponds to a $(d+2)$-dimensional vector. 
The key observation is that, $\widetilde{\bm{\theta}}-\bm{\theta}$ is a linear transformation of $\widehat{\bm{\theta}}-\bm{\theta}$, which is equivalent to a sum of independent random vectors, given by 
\begin{eqnarray*}
  n^{-1/2}\sum_{i=1}^n \sum_{\tp=1}^m \sum_{k=1}^m \omega_{k,h}(\tp) a_{n,2,\tp}^\top (\Mean Z_{i,k} Z_{i,k}^\top)^{-1} Z_{i,k} \eta_{i,k} +o_p(1).%\\=n^{-1/2}\sum_{i=1}^n \sum_{\tp=1}^m a_{n,2,\tp}^\top (\Mean Z_{i,\tp} Z_{i,\tp}^\top)^{-1} Z_{i,\tp} \eta_{i,\tp} +o_p(1).
\end{eqnarray*}
%[Could you provide the detailed formula?] 
We aim to apply Lindeberg central limit theorem to show the asymptotic normality. It remains to verify the Lindeberg condition:
\begin{eqnarray*}
  (\bm{a}_{n,2}^\top \bm{V}_{\widetilde\theta} \bm{a}_{n,2})^{-1}\Mean |\sum_{\tp=1}^m \sum_{k=1}^m \omega_{k,h}(\tp) a_{n,2,\tp}^\top (\Mean Z_{i,k} Z_{i,k}^\top)^{-1} Z_{i,k} \eta_{i,k} |^2\\ \times \mathbb{I}(|\sum_{\tp=1}^m \sum_{k=1}^m \omega_{k,h}(\tp) a_{n,2,\tp}^\top (\Mean Z_{i,k} Z_{i,k}^\top)^{-1} Z_{i,k} \eta_{i,k} |>\epsilon \sqrt{n\bm{a}_{n,2}^\top \bm{V}_{\widetilde\theta} \bm{a}_{n,2}} )\to 0,
\end{eqnarray*}
for any $\epsilon>0$. The left-hand-side is uniformly bounded by 1. As such, it suffices to show
\begin{eqnarray*}
  \prob\left( |\sum_{\tp=1}^m \sum_{k=1}^m \omega_{k,h}(\tp) a_{n,2,\tp}^\top (\Mean Z_{i,k} Z_{i,k}^\top)^{-1} Z_{i,k} \eta_{i,k} |>\epsilon \sqrt{n\bm{a}_{n,2}^\top \bm{V}_{\widetilde\theta} \bm{a}_{n,2}}\right)\to 0.
\end{eqnarray*}
However, this follows directly by the Chebyshev's inequality. 
%[Could you provide the detailed formula?]
%}

%{\color{red}The asymptotic normality of $\widetilde{\bm{\theta}}$ is a direct result of the relationship $\widetilde{\bm{\theta}}=\Omega\widehat{\bm{\theta}}$ as $\Omega$ is a linear functional.}

%Notice that $f(\tp)\to 1$ as $m\rightarrow1$, we have $\sqrt{n}(\widetilde{\bm{\theta}}-\bm{\theta})\stackrel{d}{\to}N(b_n,\bm{V}_{\widetilde\theta})$, and hence
%\begin{eqnarray*}
%\Mean m^{-1/2}\sqrt{n}\bm{a}_{n,1}^\top(\widetilde{\bm{\theta}}-\bm{\theta})\leq\left(\sum_{i=1}^ma_{n,1,i}^2\right)^{1/2}\left(m^{-1}\sum_{j=1}^m(\widetilde{\theta}(j)-{\theta}(j))^2\right)^{1/2}=O(h^2+m^{-1/2}h^{-1/2}).
%\end{eqnarray*}
Finally, it is proven in Lemma 1 that $\widetilde{\bm{V}}_\theta$ is a consistent estimate of $\bm{V}_{\widetilde\theta}$. As such, $\widehat{se}(\widehat{\DE})$ is a consistent estimate of ${se}(\widehat{\DE})$. Argument (iii) thus follows.
\end{pf}

\section{Proof of Theorem \ref{thm:bootstrap}}
We focus on provide an upper error bound for 
\begin{equation*}
\rho^*(z)=\Big|\prob\left(\frac{1}{m}\widehat{\textrm{IE}}-\frac{1}{m}\textrm{IE}\leq z\right)-\prob\left(\frac{1}{m}\widehat{\textrm{IE}}^b-\frac{1}{m}\widehat{\textrm{IE}} \leq z \Big|\textrm{Data}\right) \Big|.
\end{equation*}
We begin with some notations. Note that $\widetilde\theta(\tp)$ can be expressed as 
\begin{equation*}
\widetilde\theta(\tp)=\theta_s(\tp)+\frac{1}{n}\sum_{i=1}^n \left(\sum_{k=1}^mB_{i,k}(\tp)e_{i,k}\right),
\end{equation*}
where% $\theta_s(\tp)=\sum_{k=1}^m\omega_{k,h}(\tp)\theta(k)$  and
\begin{equation*}
B_{i,k}(\tp)=\omega_{k,h}(\tp)\left(\frac{1}{n}\sum_{i'=1}^nZ_{i',k}^\top Z_{i',k}\right)^{-1}Z_{i,k}
\end{equation*}
are independent of the random part $e_i$, and $\theta_s(\tp)=\sum_k \omega_{k,h}(\tp) \theta(k)$. Let $e_{i,\tp}^\theta=\sum_{k=1}^mB_{i,k}(\tp)e_{i,k}=\{e_{i,\tp}^{\beta_0},(e_{i,\tp}^\beta)^\top,e_{i,\tp}^\gamma\}^\top$ and $e_\tp^\theta=n^{-1/2}\sum_{i=1}^n e_{i,\tp}^\theta$. 
Similarly, we can represent $\widetilde{\Theta}(\tp)$ as %let $\Theta_s(\tp)=\sum_{k=1}^m\omega_{k,h}(\tp)\Theta(k)$, for $\tp=2,\ldots,m$,
\begin{equation*}
\widetilde\Theta(\tp)=\Theta_s(\tp)+\frac{1}{n}\sum_{i=1}^n \left(\sum_{k=1}^{m-1}B_{i,k}(\tp)E_{i,k}\right),
\end{equation*}
where $\Theta_s(\tp)=\sum_k \omega_{k,h}(\tp) \Theta(k)$. 
Let $E_{i,\tp}^\Theta=\sum_{k=1}^mB_{i,k}(\tp)E_{i,k}=\{E_{i,\tp}^{\phi_0},(E_{i,\tp}^\Phi)^\top,E_{i,\tp}^\Gamma\}^\top$ and $E_\tp^\Theta=n^{-1/2}\sum_{i=1}^nE_{i,\tp}^\Theta$. It follows that
\begin{eqnarray*}
\widetilde\beta(\tp)=\beta_s(\tp)+\frac{1}{\sqrt{n}}e_\tp^\beta,\ 
\widetilde\Phi(\tp)=\Phi_s(\tp)+\frac{1}{\sqrt{n}}E_\tp^\Phi,\
\widetilde\Gamma(\tp)=\Gamma_s(\tp)+\frac{1}{\sqrt{n}}E_\tp^\Gamma.
\end{eqnarray*}
The OLS estimation corresponds to the special case $h=0$. We remark that $E_\tp^\Theta$ is asymptotically normal when $h=0$ and degenerates to a point distribution when $mh\rightarrow\infty$. To make the following analysis hold for the OLS-based test statistic, we view $E_\tp^\Theta$ as a random variable in the discussion below.

For simplicity, let $\vc(\cdot)$ be the operator that reshapes a matrix into a vector by stacking its columns on top of one another. Denote 
\begin{align}\label{eq:GARx}
&x_{i,\tp} = \Big[
(e_{i,\tp}^\beta)^\top,
\{\vc(E_{i,\tp}^\Phi)\}^\top,
(E_{i,\tp}^\Gamma)^\top
\Big]^\top\in\R^{2d(d+2)},\nonumber\\
&x_i=\big(x_{i,2}^\top,x_{i,3}^\top,\ldots,x_{i,m}^\top\big)^\top\in\R^{p_x},\quad p_x=2(m-1)dp,\ d=p-2.
\end{align}
%Recall $x_{ij}$ defined in \eqref{eq:GARx}, we first assume $\Mean x_ix_i^\top$ is known. 
Let $\{y_i\}_i$ be independent mean zero Gaussian vectors with $\Mean y_iy_i^\top=\Mean x_ix_i^\top$. We similarly represent $y_i$ as
\begin{align}\label{eq:GARy}
&y_{i,\tp} = \Big[
(\overline e_{i,\tp}^\beta)^\top,
\{\vc(\overline E_{i,\tp}^\Phi)\}^\top,
(\overline E_{i,\tp}^\Gamma)^\top
\Big]^\top\in\R^{2d(d+2)},\nonumber\\
&y_i=\big(y_{i,2}^\top,y_{i,3}^\top,\ldots,y_{i,m}^\top\big)^\top\in\R^{p_x}.
\end{align}
Let $\left\{e_{i,j}^b,E_{i,j}^b\right\}$ be the empirical Gaussian analogs of $\left\{e_{i,j},E_{i,j}\right\}$. In other words, for $i=1,\ldots,n$, $j=1,\ldots,m$, let
\begin{equation*}
e_{i,j}^b=\widehat e_{i,j}\xi_i,\ E_{i,j}^b=\widehat E_{i,j}\xi_i,
\end{equation*}
where $\xi_1,\ldots,\xi_n$ are i.i.d standard normal random variables. We next define
\begin{align}\label{eq:GARw}
&w_{i,\tp} = \Big[
(e_{i,\tp}^{\beta,b})^\top,
\{\vc(E_{i,\tp}^{\Phi,b})\}^\top,
(E_{i,\tp}^{\Gamma,b})^\top
\Big]^\top\in\R^{2d(d+2)},\nonumber\\
&w_i=\big(w_{i,2}^\top,w_{i,3}^\top,\ldots,w_{i,m}^\top\big)^\top\in\R^{p_x}.
\end{align}
Let
\begin{align*}
&X=(X_2^\top,X_3^\top,\ldots,X_m^\top)=\frac{1}{\sqrt{n}}\sum_{i=1}^nx_i,\\
&Y=(Y_2^\top,Y_3^\top,\ldots,Y_m^\top)=\frac{1}{\sqrt{n}}\sum_{i=1}^ny_i,\\
&W=(W_2^\top,W_3^\top,\ldots,W_m^\top)=\frac{1}{\sqrt{n}}\sum_{i=1}^nw_i.
\end{align*}
Define the following function
{\small\begin{equation*}
F_\IE(X;\theta,\Theta)\equiv \frac{1}{m}\sum_{l=2}^m\left[\left(\beta(l)+\frac{e_{l}^\beta}{\sqrt{n}}\right)^\top\sum_{j=1}^{l-1}\left\{\prod_{k=j+1}^{l-1}\left(\Phi(k)+\frac{E_{k}^\Phi}{\sqrt{n}}\right)\left(\Gamma(j)+\frac{E_{j}^\Gamma}{\sqrt{n}}\right)\right\}\right].
\end{equation*}}

%Note that empirically, the true values of $\theta,\Theta$ are unavailable, and the test statistics are generated based on the estimates. 
We next represent the proposed test statistic and the bootstrap samples based on $F_{\IE}$. 
Recall that $\Theta_s(\tp)=\sum_k \omega_{k,h}(\tp) \Theta(k)$ and $\theta_s(\tp)=\sum_k \omega_{k,h}(\tp) \theta(k)$ are the smoothed parameters, and $\widetilde\theta,\widetilde\Theta$ correspond to the estimates. The difference between the proposed test statistic and the oracle indirect effect $m^{-1}(\widehat{\textrm{IE}}-\textrm{IE})$ can be represented as $T_{0}^*=F_\IE(X;\theta_s,\Theta_s)-F_\IE(0;\theta,\Theta)$. Similarly, we can represent $m^{-1}(\widehat{\textrm{IE}}^b-\widehat{\textrm{IE}})$ by $W_0^*=F_\IE(W;\widetilde\theta,\widetilde\Theta)-F_\IE(0;\widetilde\theta,\widetilde\Theta)$. 
%Define $T_{01}^*=F_\IE(X;\theta_s,\Theta_s)-F_\IE(0;\theta_s,\Theta_s)$ and $\Delta_{T0}=F_\IE(0;\theta_s,\Theta_s)-F_\IE(0;\theta,\Theta)$. It follows that $T_{0}^*=T_{01}^*+\Delta_{T0}$.
By definition, we have
\begin{align}\label{rho*}
\rho^*(z)&=\Big|P\left\{ T_0^*\le z\right\}-P\left\{ W_0^*\le z\right\}\Big|.
\end{align}
We also define the oracle statistics:  $T_0=F_\IE(X;\theta,\Theta)-F_\IE(0;\theta,\Theta)=F_\IE(X)-F_\IE(0)$, $Z_0=F_\IE(Y;\theta,\Theta)-F_\IE(0;\theta,\Theta)=F_\IE(Y)-F_\IE(0)$, $W_0=F_\IE(W;\theta,\Theta)-F_\IE(0;\theta,\Theta)=F_\IE(Z)-F_\IE(0)$ by replacing $\theta_s$, $\widetilde{\theta}$, $\Theta_s$ and $\widetilde{\Theta}$ with the oracle values. This yields an upper bound for
\begin{align}\label{rho}
\rho(z)=\Big|P\left\{ T_0\le z\right\}-P\left\{ W_0\le z\right\}\Big|.
\end{align}

The proof is divided into two parts. We first provide an upper error bound for $\sup_z\rho(z)$, showing that $T_0$ can be well-approximated by $W_0$. See Lemma \ref{lem:T0W0} below. Then, we provide upper error bounds for the difference between $W_0$ and $W_0^*$, and the difference between $T_0$ and $T_0^*$. This yields the error bound for $\sup_z \rho^*(z)$.
%show that $W_0$ and $W_0^*$ are close, and $T_0^*$ and $T_0$ are close as well. Hence we can approximate the quantiles of $T_0^*$ by those of $W_0^*$.

% \setcounter{lem}{3}
\begin{lem}
\label{lem:T0W0}
Under the conditions of Theorem \ref{thm:bootstrap}, $\sup_z\rho(z)\le Cn^{-1/8}$ for some constant $C>0$.
\end{lem}
%\noindent\textit{Sketch of proof of Lemma \ref{lem:T0W0}.} 
We first outline the main idea of the proof. We then present the details. 
The proof is based on the high-dimensional Gaussian approximation theory developed by \cite{chernozhukov_gaussian_2013}. In their paper, they developed a coupling inequality for maxima of sums of high-dimensional random vectors. They began by approximating the maximum function using a smooth surrogate and then developed a coupling inequality for the smooth function of the high-dimensional random vector. 
%which studied the maximum of a sum of high dimensional random vectors by using a smooth approximation of the maximum function. 

In our setup, the statistic $T_0$ can be represented as a smooth function of sums of random vectors whose dimension is allowed to diverge with the sample size. Such an observation allows us to employ the coupling inequality to establish the size and power property of the proposed test. 
% say $x_i,i=1,\ldots,n$ as given in \eqref{eq:GARx}, the main idea of proof is similar to \cite{chernozhukov_gaussian_2013} but the calculation details are different, as the functions of interest are different. The 
The proof of Lemma \ref{lem:T0W0} contains two main parts. In the first part, we assume the covariance of the time-varying covariates is known and employ Slepian interpolation, Stein’s leave-one-out method as well as a truncation method to bound the Kolmogorov distance between the distributions of $T_0$ and its Gaussian analog $Z_0$. In the second part, we establish the validity of the multiplier bootstrap for estimating quantiles of $Z_0$ when the covariance matrix is unknown, i.e., $W_0$. The detailed proof is given as follows.

\vspace{0.1in}
\begin{pf}{\it of Lemma \ref{lem:T0W0}}:
Define function $g(s)=g_0\big(\psi(s-t)\big)$ for some constant $\psi>0$ and some thrice differentiable function $g_0$ that satisfies $g_0(s)=1$ when $s\leq0$, $g_0(s)=0$ when $s\geq1$ and $g_0(s)\geq0$ otherwise.
Let $m=g\circ F_\IE$. We also introduce the following notations: 
$\mathbbm{E}_n(\cdot)=n^{-1}\sum_{i=1}^n(\cdot)$;
$\overline{\Mean}(\cdot)=\mathbbm{E}_n\Mean(\cdot)$;
$C^k$ denotes the class of $k$ times continuously differentiable functions;
$C_b^k$ denotes the class of functions $f\in C^k$ and $\sum_z|\partial^jf(z)/\partial z^j|$ for $j=0,\ldots,k$;
$a\lesssim b$ if $a$ is smaller than or equal to $b$ up to a universal positive constant;
$a\simeq b$ if $a\lesssim b$ and $b\lesssim a$.
We define the Slepian interpolation $Z(t)$ between $Y$ and $X$, Stein's leave-one-out version $Z^{(i)}(t)$ of $Z(t)$, and other useful terms as follows:
\begin{eqnarray*}
Z(t)=\sqrt{t}X+\sqrt{1-t}Y=\sum_{i=1}^nZ_i(t),\quad z_i(t)=n^{-1/2}(\sqrt{t}x_i+\sqrt{1-t}y_i),\\
Z^{(i)}(t)=Z(t)-Z_i(t),\quad \dot{z}_{ij}(t)=\frac{1}{2\sqrt{n}}\left(\frac{1}{\sqrt{t}}x_{ij}-\frac{1}{\sqrt{1-t}}y_{ij}\right).
\end{eqnarray*}

We first prove
\begin{equation}\label{result1}
\sup_{t\in\R}\vert P(T_0\leq t)-P(Z_0\leq t)\vert\leq C^\prime n^{-1/8},
\end{equation}
where $C^\prime >0$ is a constant.
From the construction of $g(\cdot)$, we have $G_k {\lesssim}\psi^k,k=0,1,2,3$ where $G_k=\sup_{z\in\R}\vert\partial^k g(z)\vert,k\geq0$, and
\begin{align*}
&P(T_0\leq t)=P\big(F_{\IE}(X)\leq t\big)\leq \Mean g\big(F_{\IE}(X)\big),\\
&\Mean g\big(F_{\IE}(Y)\big)\leq P\big(F_{\IE}(Y)\leq t+\psi^{-1}\big),\\
&P(Z_0\leq t+\psi^{-1})=P\big(F_{\IE}(Y)\leq t+\psi^{-1}\big)\geq\Mean g\big(F_{\IE}(Y)\big),
\end{align*}
which give the decompose
$$P(T_0\leq t)-P(Z_0\leq t)\leq\underbrace{\big\{\Mean g\big(F_{\IE}(X)\big)-\Mean g\big(F_{\IE}(Y)\big)\big\}}_{(a)}+\underbrace{\big\{P(Z_0\leq t+\psi^{-1})-P(Z_0\leq t)\big\}}_{(b)}.$$
In the following, we calculate (a) in Steps 1-2 and derive the bound for (b) in Step 3.

\textbf{Step 1.} We first calculate the upper bounds of (a).
We have by Taylor's expansion,
\begin{equation*}
\Mean\{m(X)-m(Y)\}=\sum_{j=1}^{p_x}\sum_{i=1}^n\int_0^1\Mean\{\partial_jm(Z(t))\dot{Z}_{ij}(t)\}dt=\RNum{1}+\RNum{2}+\RNum{3},
\end{equation*}
where
\begin{align*}
&\RNum{1}=\sum_{j=1}^{p_x}\sum_{i=1}^n\int_0^1\Mean\{\partial_jm(Z^{(i)}(t))\dot{Z}_{ij}(t)\}dt,\\
&\RNum{2}=\sum_{j,k=1}^{p_x}\sum_{i=1}^n\int_0^1\Mean\{\partial_j\partial_km(Z^{(i)}(t))\dot{Z}_{ij}(t)Z_{ik}(t)\}dt,\\
&\RNum{3}=\sum_{j,k,l=1}^{p_x}\sum_{i=1}^n\int_0^1\int_0^1(1-s)\Mean\{\partial_j\partial_k\partial_lm(Z^{(i)}(t)+s Z_{i,t})\dot{Z}_{ij}(t)Z_{ik}(t)Z_{il}(t)\}ds dt.
\end{align*}
By independence of $Z^{(i)}(t)$ and $\dot{Z}_{ij}(t)$ together with the fact that $\Mean\{\dot{Z}_{ij}(t)\}=0$, we have $\RNum{1}=0$.
Note that $Z^{(i)}(t)$ is independent of $\dot{Z}(t)Z_{ik}(t)$, and $\Mean\{\dot{Z}_{ij}(t)Z_{ik}(t)\}=n^{-1}\Mean\{x_{ij}x_{ik}-y_{ij}y_{ik}\}$,
\begin{equation*}
\RNum{2}=\sum_{j,k=1}^{p_x}\sum_{i=1}^n\int_0^1\Mean\{\partial_j\partial_km(Z^{(i)}(t))\}\Mean\{\dot{Z}_{ij}(t)Z_{ik}(t)\}dt=0.
\end{equation*}
We now prove $(a)\leq\vert\RNum{3}\vert\lesssim\psi^3n^{-2}+\psi^2n^{-2}+\psi n^{-2}$ in Step 2. 

\textbf{Step 2.} Note that
\begin{align*}
\RNum{3}=&\sum_{j,k,l=1}^{p_x}\sum_{i=1}^n\int_0^1\left[\Mean\left\{\int_0^1 \partial_j\partial_k\partial_lm(Z^{(i)}(t)+s Z_{i}(t))ds\right\}\dot{Z}_{ij}(t)Z_{ik}(t)Z_{il}(t)\right] dt\\
\simeq&\sum_{j,k,l=1}^{p_x}\sum_{i=1}^n\int_0^1\Mean \partial_j\partial_k\partial_lm(Z(t))\dot{Z}_{ij}(t)Z_{ik}(t)Z_{il}(t) dt,
\end{align*}
where $$\partial_j\partial_k\partial_lm(Z)\simeq\psi^3\partial_jF_{\IE}(Z)\partial_kF_{\IE}(Z)\partial_lF_{\IE}(Z)+\psi^2\partial_jF_{\IE}(Z)\partial_k\partial_lF_{\IE}(Z)+\psi\partial_j\partial_k\partial_lF_{\IE}(Z).$$
Note that
\begin{align}
\vert\RNum{3}\vert&\leq\sum_{j,k,l=1}^{p_x}\sum_{i=1}^n \int_0^1\sqrt{\Mean\vert \partial_j\partial_k\partial_lm(Z(t))\vert^2}\sqrt{\Mean\vert \dot{Z}_{ij}(t)Z_{ik}(t)Z_{il}(t)\vert^2} dt
\nonumber\\
&\leq\int_0^1\left(\sum_{j,k,l=1}^{p_x}\sqrt{\Mean\vert \partial_j\partial_k\partial_lm(Z(t))\vert^2}\right)\left(\max_{1\leq j,k,l\leq p_x}n\overline\Mean\vert \dot{Z}_{ij}(t)Z_{ik}(t)Z_{il}(t)\vert\right) dt.
\label{III decomp}
\end{align}
% $\partial_j\partial_k\partial_lm\big(Z^{(i)}(t)+s Z_i(t)\big)\lesssim\partial_j\partial_k\partial_lm\big(Z(t)\big)\simeq\partial_j\partial_k\partial_lm\big(Z(1)\big)$.

% We first compute $\sum_{j,k,l=1}^{p_x} \Mean \partial_j\partial_k\partial_lm(X)$.
We first compute $\sum_{j,k,l=1}^{p_x}\sqrt{\Mean\vert \partial_j\partial_k\partial_lm(Z(t))\vert^2}$.
Define function
$$\mathcal{G}=\mathbbm{1}\left\{\max_{1\leq j\leq p_x/2}\left\vert u_j/\sqrt{n}\right\vert<(1-q)/2\right\},$$
where   
\begin{align*}
u&=\left(
( e_2^\beta)^\top,
\{\vc(E_2^\Phi)\}^\top,
(E_2^\Gamma)^\top,\cdots,
(e_m^\beta)^\top,
\{\vc(E_m^\Phi)\}^\top,
(E_m^\Gamma)^\top
\right)^\top\\
&=(u_1,u_2,\cdots,u_{p_x/2})^\top.
\end{align*}
Then we have
\begin{align*}
\sqrt{\Mean \{\partial_j\partial_k\partial_lm(Z)\}^2}
=&\sqrt{\Mean \{\partial_j\partial_k\partial_lm(Z)\}^2\mathcal{G}+\Mean\{\partial_j\partial_k\partial_lm(Z)\}^2\{1-\mathcal{G}\}}\\
\simeq&\psi^3\Mean(\partial_jF_\IE\partial_kF_\IE\partial_lF_\IE\mathcal{G})+\psi^3\Mean\{\partial_jF_\IE\partial_kF_\IE\partial_lF_\IE(1-\mathcal{G})\}\\
&+\psi^2\Mean(\partial_j\partial_kF_\IE\partial_lF_\IE\mathcal{G})+\psi^2\Mean\{\partial_j\partial_kF_\IE\partial_lF_\IE(1-\mathcal{G})\}\\
&+\psi\Mean(\partial_j\partial_k\partial_lF_\IE\mathcal{G})+\psi\Mean\{\partial_j\partial_k\partial_lF_\IE(1-\mathcal{G})\}.
\end{align*}
In the following, we focus on establishing the upper error bounds for $\sum_{j,k,l}\Mean(\partial_jF_\IE\partial_kF_\IE\partial_lF_\IE\mathcal{G})$ and $\sum_{j,k,l}\Mean\{\partial_jF_\IE\partial_kF_\IE\partial_lF_\IE(1-\mathcal{G})\}$. The other bounds can be derived similarly. 
\begin{enumerate}
\item[2.1]\textit{The bound of $\sum_{j,k,l}\Mean(\partial_jF_\IE\partial_kF_\IE\partial_lF_\IE\mathcal{G})$.}\\
Let $\overline{q}=(1+q)/2$.
Notice that 
$$\sum_{j,k,l}\Mean(\partial_jF_\IE\partial_kF_\IE\partial_lF_\IE\mathcal{G})\lesssim m^3\Mean\vert\partial_jF_\IE\mathcal{G}\vert^3.$$
We next compute $\Mean\vert\partial_jF_\IE\mathcal{G}\vert$, which belongs to either one of the following three categories:
\begin{align*}
\left\vert\frac{\partial F_\IE}{\partial e_\tp^\beta}\mathcal{G}\right\vert
&=m^{-1}n^{-1/2}\left\vert\sum_{j=1}^{t-1}\left\{\prod_{k=j+1}^{t-1}\left(\Phi(k)+\frac{E_{k}^\Phi}{\sqrt{n}}\right)\left(\Gamma(j)+\frac{E_j^\Gamma}{\sqrt{n}}\right)\right\}\mathcal{G}\right\vert\\
&\lesssim m^{-1}n^{-1/2}\sum_{j=1}^{t-1}\overline{q}^{t-j-1}\{M_{\Gamma}+(1-q)/2\}\\
&\simeq m^{-1}n^{-1/2};
\end{align*}
\begin{align*}
\left\vert\frac{\partial F_\IE}{\partial E_j^\Gamma}\mathcal{G}\right\vert
&=m^{-1}n^{-1/2}\left\vert\sum_{t=j+1}^m\left(\beta(\tp)+\frac{e_\tp^\beta}{\sqrt{n}}\right)^\top\prod_{k=j+1}^{t-1}\left(\Phi(k)+\frac{E_{k}^\Phi}{\sqrt{n}}\right)\mathcal{G}\right\vert\\
&\leq m^{-1}n^{-1/2}\{M_\beta+(1-q)/2\}\sum_{t=j+1}^m\overline{q}^{t-1-j}\\
&\simeq m^{-1}n^{-1/2};
\end{align*}
\begin{align*}
\left\vert\frac{\partial F_\IE}{\partial E_l^\Phi}\mathcal{G}\right\vert
&=m^{-1}n^{-1/2}\left\vert\sum_{t=2}^m\left(\beta(\tp)+\frac{e_\tp^\beta}{\sqrt{n}}\right)^\top\right.\\
&\hspace{1in}\left.\cdot\sum_{j=1}^{t-1}\left\{\prod_{k\neq l\atop k=j+1}^{t-1}\left(\Phi(k)+\frac{E_{k}^\Phi}{\sqrt{n}}\right)\left(\Gamma(j)+\frac{E_j^\Gamma}{\sqrt{n}}\right)\right\}\mathcal{G}\right\vert\\
&\lesssim  m^{-1}n^{-1/2}\sum_{t=l+1}^m\sum_{j=1}^{l-1}\overline{q}^{t-2-j}\{M_\beta+(1-q)/2\}\\
&\simeq m^{-1}n^{-1/2}.
\end{align*}
It follows that $\sum_{j,k,l}\Mean(\partial_jF_\IE\partial_kF_\IE\partial_lF_\IE\mathcal{G})\lesssim n^{-3/2}$.

\item[2.2]\textit{The bound of $\sum_{j,k,l}\Mean\{\partial_jF_\IE\partial_kF_\IE\partial_lF_\IE(1-\mathcal{G})\}$.}\\
Similarly, we have
\begin{equation*}
\sum_{j,k,l}\Mean\{\partial_jF_\IE\partial_kF_\IE\partial_lF_\IE(1-\mathcal{G})\}\lesssim m^3\Mean\vert\partial_jF_\IE(1-\mathcal{G})\vert^3.
\end{equation*}
We consider the derivative with respect to $\eta_\tp^\beta$ as an example. Notice that
\begin{align*}
\Mean\left\{\frac{\partial F_\IE}{\partial \eta_\tp^\beta}(1-\mathcal{G})\right\}
&=\Mean\left\vert\sum_{j=1}^{t-1}\left\{\prod_{k=j+1}^{t-1}\left(\Phi(k)+\frac{E_{k}^\Phi}{\sqrt{n}}\right)\left(\Gamma(j)+\frac{E_j^\Gamma}{\sqrt{n}}\right)\right\}(1-\mathcal{G})\right\vert\\
&\lesssim m^{-1}n^{-1/2}\left[\Mean \left\vert\sum_{j=1}^{t-1}\left\{\prod_{k=j+1}^{t-1}\left(\Phi(k)+\frac{E_{k}^\Phi}{\sqrt{n}}\right)\left(\Gamma(j)+\frac{E_j^\Gamma}{\sqrt{n}}\right)\right\}\right\vert^2\right]^{1/2}\\
&\hspace{1in} \cdot P\left\{\max_{1\leq j\leq p_x/2}\left\vert u_j/\sqrt{n}\right\vert\geq(1-q)/2\right\}.
\end{align*} 
By Lemma 2.2.10 in \cite{1996Weak}, we have $\Mean\vert\max_j u_j\vert\lesssim \log m$. It follows that
\begin{align*}
&\left[\Mean \left\vert\sum_{j=1}^{t-1}\left\{\prod_{k=j+1}^{t-1}\left(\Phi(k)+\frac{E_{k}^\Phi}{\sqrt{n}}\right)\left(\Gamma(j)+\frac{E_j^\Gamma}{\sqrt{n}}\right)\right\}\right\vert^2\right]^{1/2}\\
\leq&\left[\sum_{j=1}^{t-1}\prod_{k=j+1}^{t-1}\Mean\left\vert\Phi(k)+\frac{\max_j u_j}{\sqrt{n}}\right\vert^2\cdot\Mean\left\vert\Gamma(j)+\frac{\max_j u_j}{\sqrt{n}}\right\vert^2\right]^{1/2}\\
\lesssim&\left[\sum_{j=1}^{t-1}\left(1+\frac{\log m}{\sqrt{n}}\right)^{2j}\right]^{1/2}\\
\simeq&\left(1+\frac{\sqrt{n}}{\log m}\right)\left(1+\frac{\log m}{\sqrt{n}}\right)^m\\
\simeq&n^{1/2}(\log m)^{-1}\exp(n^{-1/2}m\log m).
\end{align*} 
Let $t_0=n^{1/2}(1-q)/2$ and $t_1=t_0-\Mean\max_j u_j$. Notice that
\begin{align*}
P\{\max_j\vert u_j\vert>t_0\}=&P\big(\{\max_j u_j>t_0\}\cap\{\max_j\vert u_j\vert=\max_j u_j\}\big)\\
&+P\big(\{\min_ju_j<-t_0\}\cap\{\max_j\vert u_j\vert=-\min_j u_j\}\big)\\
\leq&2P\{\max_j u_j>t_0\}\\
\lesssim&P\{\vert\max_ju_j-\Mean\max_j u_j\vert>t_1\}.
\end{align*}
By Borell TIS inequality and Lemma 2.2.10 in \cite{1996Weak}, we have
\begin{eqnarray}
P\{\max_j\vert u_j\vert>t_0\}\lesssim\exp(-t_1^2)\simeq\exp\{-n+2n^{1/2}\log m-(\log m)^2\}.\label{eq:1-G}
\end{eqnarray}
Hence 
\begin{equation*}
\sum_{j,k,l}\Mean\{\partial_jF_\IE\partial_kF_\IE\partial_lF_\IE(1-\mathcal{G})\}\lesssim n^{-3/2}\delta^3,
\end{equation*}
where
\begin{equation}\label{eq:rho}
\delta=n^{1/2}(\log m)^{-1}\exp\{-n+2n^{1/2}\log m-(\log m)^2+n^{-1/2}m\log m\}.
\end{equation}
\end{enumerate}

Combine the above arguments, we obtain
\begin{equation*}
\sum_{j,k,l}\Mean(\partial_jF_\IE\partial_kF_\IE\partial_lF_\IE\mathcal{G})+\sum_{j,k,l}\Mean\{\partial_jF_\IE\partial_kF_\IE\partial_lF_\IE(1-\mathcal{G})\}\lesssim n^{-3/2}(1+\delta^3),
\end{equation*}
Using similar arguments, we can show that
\begin{align*}
&\sum_{j,k,l}\Mean(\partial_j\partial_kF_\IE\partial_lF_\IE\mathcal{G})+\sum_{j,k,l}\Mean\{\partial_j\partial_kF_\IE\partial_lF_\IE(1-\mathcal{G})\}\lesssim n^{-3/2}(1+\delta^2),\\
&\sum_{j,k,l}\Mean(\partial_j\partial_k\partial_lF_\IE\mathcal{G})+\sum_{j,k,l}\Mean\{\partial_j\partial_k\partial_lF_\IE(1-\mathcal{G})\}\lesssim n^{-3/2}(1+\delta).
\end{align*}
It follows that
\begin{equation}\label{III1}
\sum_{j,k,l=1}^{p_x}\sqrt{\Mean\vert \partial_j\partial_k\partial_lm(Z(t))\vert^2}\lesssim\psi^3n^{-3/2}(1+\delta^3)+\psi^2n^{-3/2}(1+\delta^2)+\psi n^{-3/2}(1+\delta),
\end{equation}
%$$$$
where $\delta$ depends on $m,n$ through \eqref{eq:rho}.

Let $\omega(t)=1/\min\{\sqrt{t},\sqrt{1-t}\}$. We observe that
\begin{align}
&\quad\int_0^1\max_{j,k,l}n\overline\Mean\vert\dot{Z}_{ij}(t)Z_{ik}(t)Z_{il}(t)\vert dt\nonumber\\
&=\int_0^1\omega(t)\max_{j,k,l}n\overline\Mean\vert\{\dot{Z}_{ij}/\omega(t)\}(t)Z_{ik}(t)Z_{il}(t)\vert dt\nonumber\\
&\leq_{\tiny\textcircled{1}} n\int_0^1\omega(t)\max_{j,k,l}\left(\overline\Mean\vert\dot{Z}_{ij}/\omega(t)\vert^3(t)\overline\Mean\vert Z_{ik}(t)\vert^3 \overline\Mean\vert Z_{il}(t)\vert^3\right)^{1/3} dt\nonumber\\
&\leq_{\tiny\textcircled{2}} n^{-1/2}\max_j\overline\Mean(|x_{ij}|+|y_{ij}|)^3\int_0^t\omega(t)dt\nonumber\\
&\lesssim n^{-1/2}\max_j\overline\Mean|x_{ij}|^3,\label{III2}
\end{align}
where \textcircled{1} is by H{\"o}lder inequality and \textcircled{2} follows from the fact that $\vert\dot{Z}_{ij}/\omega(t)\vert\leq n^{-1/2}(|x_{ij}|+|y_{ij}|)$, $|Z_{ik}(t)|\leq n^{-1/2}(|x_{ik}|+|y_{ik}|)$. 

The condition $m=O(n^{c_2})$ for some $c_2<3/2$ implies that $\delta=o(1)$. This 
together with \eqref{III decomp}, \eqref{III1} and \eqref{III2} yields that %with the conditions that $m=o(n^{3/2}),\ \delta=o(1)$, we obtain
\begin{equation}\label{(a)}
% (a)=\vert\RNum{3}\vert\lesssim\psi^3n^{-2}(1+\delta^3)+\psi^2n^{-2}(1+\delta^2)+\psi n^{-2}(1+\delta),
(a)=\vert\RNum{3}\vert\lesssim\psi^3n^{-2}+\psi^2n^{-2}+\psi n^{-2}.
\end{equation}

\textbf{Step 3.} We now derive the upper bound of $(b)\equiv P(Z_0\leq t+\psi^{-1})-P(Z_0\leq t)$. Let $t'=t+F_\IE(0)$. Recall that $\overline e_\tp^\beta$ is defined in \eqref{eq:GARy}. Denote $\bar{1}=(1,\ldots,1)^\top\in\R^{d}$. Using %the truncation and 
similar arguments in Step 2.2, we have
\begin{align}
P(Z_0\leq t)&\leq P(Z_0\mathcal{G}\leq t)+\Mean(1-\mathcal{G})\nonumber\\
&\lesssim P\left(\frac{1}{m}\sum_{t=2}^m\left(\beta(\tp)+\frac{\overline e_\tp^\beta}{\sqrt{n}}\right)^\top\bar{1}\leq t'\right)+\exp\{-n+2n^{1/2}\log m-(\log m)^2\}\nonumber\\
&\simeq P\left(\frac{1}{m}\sum_{t=2}^m\left(\beta(\tp)+\frac{\overline e_\tp^\beta}{\sqrt{n}}\right)^\top\bar{1}\leq t'\right),\label{PZ0}
\end{align}
where the second inequality is due to the conclusion \eqref{eq:1-G} and the third inequality follows from the condition $ m=O(n^{c_2})$ for some $c_2<3/2$. Notice that $\overline e_\tp^\beta$ is a Gaussian random vector, we have 
\begin{equation*}
\sup\vert P(Z_0\leq t+\psi^{-1})-P(Z_0\leq t)\vert\simeq n^{1/2}\psi^{-1}.
\end{equation*}

To summarize, we have shown that
$$P(T_0\leq t)-P(Z_0\leq t)\lesssim \psi^3n^{-2}+\psi^2n^{-2}+\psi n^{-2}+n^{1/2}\psi^{-1}.$$
Take $\psi\simeq n^{5/8}$, we have
$$P(T_0\leq t)-P(Z_0\leq t)\lesssim n^{-1/8}.$$ 

By Lemma 3.2 of \cite{chernozhukov_gaussian_2013}, we have shown that for $\alpha\in(0,1)$ and $\vartheta>0$,
\begin{align*}
&P(c_{W_0}(\alpha)\leq c_{Z_0}(\alpha+\vartheta^{1/2}))\geq1-P(\Delta>\vartheta),\\
&P(c_{Z_0}(\alpha)\leq c_{W_0}(\alpha+\vartheta^{1/2}))\geq1-P(\Delta>\vartheta),
\end{align*}
where $c_{W_0}(\alpha)$ and $c_{z_0}(\alpha)$ denote the critical values of $W_0$ and $Z_0$ under the significance level $\alpha$, respectively. 
Define
$$\rho_\ominus=\sup_{\alpha\in(0,1)}P\Big(\{c_{Z_0}(\alpha)<T_0\leq c_{W_0}(\alpha)\}\cup\{c_{W_0}(\alpha)<T_0\leq c_{Z_0}(\alpha)\}\Big).$$
Note that 
\begin{align*}
&P\Big(c_{Z_0}(\alpha)<T_0\leq c_{W_0}(\alpha)\Big)\\
=&P\Big(c_{Z_0}(\alpha)<T_0\leq c_{Z_0}(\alpha+\vartheta^{1/2})\Big)+P\Big(\{c_{Z_0}(\alpha+\vartheta^{1/2})<T_0\leq c_{W_0}(\alpha)\}\cap\{ c_{W_0}(\alpha)>c_{Z_0}(\alpha+\vartheta^{1/2})\}\Big)\\
&-P\Big(\{c_{W_0}(\alpha)<T_0\leq c_{Z_0}(\alpha+\vartheta^{1/2})\}\cap\{ c_{W_0}(\alpha)\leq c_{Z_0}(\alpha+\vartheta^{1/2})\}\Big)\\
\leq&P\Big(c_{Z_0}(\alpha)<T_0\leq c_{Z_0}(\alpha+\vartheta^{1/2})\Big)+P\Big(c_{W_0}(\alpha)> c_{Z_0}(\alpha+\vartheta^{1/2})\Big)\\
\leq&P\Big(c_{Z_0}(\alpha)<Z_0\leq c_{Z_0}(\alpha+\vartheta^{1/2})\Big)+\rho+P(\Delta>\vartheta)\\
\leq&\vartheta^{1/2}+\rho+P(\Delta>\vartheta).
\end{align*}
Similarly, we can show
$$P\Big(c_{W_0}(\alpha)<T_0\leq c_{Z_0}(\alpha)\Big)\leq\vartheta^{1/2}+\rho+P(\Delta>\vartheta).$$
By the definition of $\rho_\ominus$, we have 
$$\rho_\ominus\leq 2\vartheta^{1/2}+2P(\Delta>\vartheta)+2\rho.$$
On the other hand, 
\begin{align*}
&\vert P\big(T_0\leq c_{W_0}(\alpha)\big)-\alpha\vert\\
\leq&\vert P\big(T_0\leq c_{W_0}(\alpha)\big)-P\big(T_0\leq c_{Z_0}(\alpha)\big)\vert+\rho\\
\leq&P\Big(\{c_{Z_0}(\alpha)<T_0\leq c_{W_0}(\alpha)\}\cup\{c_{W_0}(\alpha)<T_0\leq c_{Z_0}(\alpha)\}\Big)+\rho\\
\leq&\rho_\ominus+\rho.
\end{align*}
Notice that $\Delta=O(n^{-1/2})$ when $\vartheta=O(n^{-1/4})$. %one can obtain \eqref{rho}.
The proof of Lemma \ref{lem:T0W0} is thus completed.
\end{pf}

\vspace*{0.2in}
With Lemma \ref{lem:T0W0}, we next present the proof of Theorem \ref{thm:bootstrap}.
\vspace*{0.2in}

\begin{pf}{\it of Theorem \ref{thm:bootstrap}}:
Define $T_{01}^*=F_\IE(X;\theta_s,\Theta_s)-F_\IE(0;\theta_s,\Theta_s)$ and $\Delta_{T0}=F_\IE(0;\theta_s,\Theta_s)-F_\IE(0;\theta,\Theta)$. It follows that $T_{0}^*=T_{01}^*+\Delta_{T0}$. 
Notice that
\begin{align*}
\rho^*(z)&=\Big|P\left\{ T_0^*\le z\right\}-P\left\{ W_0^*\le z\right\}\Big|\\ 
&\le \Big|P\left\{ T_0^*\le z\right\}-P\left\{ T_0\le z\right\}\Big|+\Big|P\left\{ T_0\le z\right\}-P\left\{ W_0\le z\right\}\Big|\\ 
&\hspace{1.5in}+\Big|P\left\{ W_0^*\le z\right\}-P\left\{ W_0\le z\right\}\Big|\\ 
&\le \Big|P\left\{ T_{0}^*\le z\right\}-P\left\{ T_{01}^*\le z\right\}\Big|+\Big|P\left\{ T_{01}^*\le z\right\}-P\left\{ T_0\le z\right\}\Big|\\ 
&\hspace{0.2in}+\Big|P\left\{ T_0\le z\right\}-P\left\{ W_0\le z\right\}\Big|+\Big|P\left\{ W_0^*\le z\right\}-P\left\{ W_0\le z\right\}\Big|.
\end{align*}
Similar to the proof of Lemma \ref{lem:T0W0}, we have 
\begin{align*}
&P\left\{ T_{01}^*\le z\right\}-P\left\{ T_0\le z\right\}\\
\leq&{\big\{\Mean g\big(F_{\IE}(X;\theta_s,\Theta_s)-F_{\IE}(0;\theta_s,\Theta_s)\big)-\Mean g\big(F_{\IE}(X;\theta,\Theta)-F_{\IE}(0;\theta,\Theta)\big)\big\}}\\
&\quad+{\big\{P(T_0\leq t+\psi^{-1})-P(T_0\leq t)\big\}}\\
=&{\big\{\Mean g\big(F_{\IE}(X;\theta_s,\Theta_s)-F_{\IE}(X;\theta,\Theta)\big)-\Mean g\big(F_{\IE}(0;\theta_s,\Theta_s)-F_{\IE}(0;\theta,\Theta)\big)\big\}}\\
&\quad+{\big\{P(T_0\leq t+\psi^{-1})-P(T_0\leq t)\big\}},
\end{align*}
and
\begin{align*}
&P\left\{ W_{0}^*\le z\right\}-P\left\{ W_0\le z\right\}\\
\leq&{\big\{\Mean g\big(F_{\IE}(W;\widetilde\theta,\widetilde\Theta)-F_{\IE}(0;\widetilde\theta,\widetilde\Theta)\big)-\Mean g\big(F_{\IE}(W;\theta,\Theta)-F_{\IE}(0;\theta,\Theta)\big)\big\}}\\
&\quad+{\big\{P(W_0\leq t+\psi^{-1})-P(W_0\leq t)\big\}}\\
=&{\big\{\Mean g\big(F_{\IE}(W;\widetilde\theta,\widetilde\Theta)-F_{\IE}(W;\theta,\Theta)\big)-\Mean g\big(F_{\IE}(0;\widetilde\theta,\widetilde\Theta)-F_{\IE}(0;\theta,\Theta)\big)\big\}}\\
&\quad+{\big\{P(W_0\leq t+\psi^{-1})-P(W_0\leq t)\big\}}.
\end{align*}
Denote $\delta_{\theta s}=\theta_s-\theta$, $\delta_{\Theta s}=\Theta_s-\Theta$, $\delta_{\widetilde\theta}=\widetilde\theta-\theta$,  and $\delta_{\widetilde\theta}=\widetilde\theta-\theta$. To bound these differences, the biases $\delta_{\theta s},\delta_{\Theta s},\delta_{\widetilde\theta},\delta_{\widetilde\theta}$ can be treated in the same position as $X$ or $W$. Take $F_\IE(W;\widetilde\theta,\widetilde\Theta)$ as an instance, we have
\begin{align*}
&F_\IE(W;\widetilde\theta,\widetilde\Theta)\\
=&\frac{1}{m}\sum_{l=2}^m\left[\left(\beta(l)+\delta_{\widetilde\beta}(l)+\frac{e_{l}^\beta}{\sqrt{n}}\right)^\top\sum_{j=1}^{l-1}\left\{\prod_{k=j+1}^{l-1}\left(\Phi(k)+\delta_{\widetilde\Phi}(k)+\frac{E_{k}^\Phi}{\sqrt{n}}\right)\left(\Gamma(j)+\delta_{\widetilde\Gamma}(j)+\frac{E_{j}^\Gamma}{\sqrt{n}}\right)\right\}\right].
\end{align*}
According to Theorem \ref{thm:t theta asymp}, $\delta_{\widetilde\theta}$ and $\delta_{\widetilde\theta}$ are asymptotic normal with variance of order $n^{-1}$ (mean is negligible compared to the variance), i.e., that same order as $e_l^\theta/\sqrt{n}$. Hence by the same techniques as in proof of Lemma \ref{lem:T0W0}, one can obtain
\begin{equation*}
\Big|P\left\{ W_0^*\le z\right\}-P\left\{ W_0\le z\right\}\Big|\le Cn^{-1/8}.
\end{equation*}
The biases $\delta_{\theta s},\delta_{\Theta s}$ are of order $O(h^2+m^{-1})=o(n^{-1/2})$. They are not random given $m$ and $h$. Then $\max_k\Vert \delta_{\Theta s}(k)\Vert_\infty\asymp\max_k\Vert \delta_{\theta s}(k)\Vert_\infty=o(n^{-1/2})$. Using similar arguments in proving Lemma \ref{lem:T0W0}, we can show that
\begin{equation*}
\Big|P\left\{ T_{01}^*\le z\right\}-P\left\{ T_0\le z\right\}\Big|\le Cn^{-1/8}.
\end{equation*}
We omit the details to save space. 
%shall be similar but easier compared to the proof of Lemma \ref{lem:T0W0}, which we do not repeat.

%  $T_{0}^*=F_\IE(X;\theta_s,\Theta_s)-F_\IE(0;\theta,\Theta)$ and $W_0^*=F_\IE(W;\widetilde\theta,\widetilde\Theta)-F_\IE(0;\widetilde\theta,\widetilde\Theta)$.
% Define $T_{01}^*=F_\IE(X;\theta_s,\Theta_s)-F_\IE(0;\theta_s,\Theta_s)$, $\Delta_{T0}=F_\IE(0;\theta_s,\Theta_s)-F_\IE(0;\theta,\Theta)$ and then $T_{0}^*=T_{01}^*+\Delta_{T0}$.

Finally, it remains to bound $\Delta_{T0}$. Notice that
\begin{align*}
\Delta_{T0}&=F_\IE(0;\theta_s,\Theta_s)-F_\IE(0;\theta,\Theta)\\ 
&=\frac{1}{m}\sum_{l=2}^m\left[\left(\beta(l)+\delta_{\beta s}(l)\right)^\top\sum_{j=1}^{l-1}\left\{\prod_{k=j+1}^{l-1}\left(\Phi(k)+\delta_{\Phi s}(k)\right)\left(\Gamma(j)+\delta_{\Gamma s}(j)\right)\right\}\right]\\ 
&\qquad-\frac{1}{m}\sum_{l=2}^m\left[\beta(l)^\top\sum_{j=1}^{l-1}\left\{\prod_{k=j+1}^{l-1}\Phi(k)\Gamma(j)\right\}\right]\\
&=\frac{1}{m}\sum_{l=2}^m\left[\beta(l)^\top\sum_{j=1}^{l-1}\left\{\prod_{k=j+1}^{l-1}\left(\Phi(k)+\delta_{\Phi s}(k)\right)\left(\Gamma(j)+\delta_{\Gamma s}(j)\right)-\prod_{k=j+1}^{l-1}\Phi(k)\Gamma(j)\right\}\right]\\ 
&\qquad +\frac{1}{m}\sum_{l=2}^m\left[\delta_{\beta s}(l)^\top\sum_{j=1}^{l-1}\left\{\prod_{k=j+1}^{l-1}\left(\Phi(k)+\delta_{\Phi s}(k)\right)\left(\Gamma(j)+\delta_{\Gamma s}(j)\right)\right\}\right].
\end{align*}
%Note that $\delta_{\theta s},\delta_{\Theta s}$ are fixed quantities, 
Let $\delta=\max\{\max_k\Vert \delta_{\theta s}(k)\Vert_\infty,\max_k\Vert \delta_{\Theta s}(k)\Vert_\infty\}=O(h^2+m^{-1})$. It follows that
 % When $n$ tends large, there exists $q^*<0$ such that $|\Phi(k)+\delta_{\Phi s}(k)|\le|q+\delta|\le q^*<1$. Denote $\Phi(k)+\delta=\Phi(k)^*$, then $|\Phi(k)^*|\le q^*$ and as $n\rightarrow\infty$, we have 
\begin{align*}
&\left\vert\sum_{j=1}^{l-1}\left\{\prod_{k=j+1}^{l-1}\left(\Phi(k)+\delta_{\Phi s}(k)\right)\left(\Gamma(j)+\delta_{\Gamma s}(j)\right)-\prod_{k=j+1}^{l-1}\Phi(k)\Gamma(j)\right\}\right\vert\\
\le &M_\Gamma\sum_{j=1}^{l-1}\left\vert\prod_{k=j+1}^{l-1}\left(\Phi(k)+\delta\right)-\prod_{k=j+1}^{l-1}\Phi(k)\right\vert
+\sum_{j=1}^{l-1}\prod_{k=j+1}^{l-1}\left\vert\left(\Phi(k)+\delta\right)\delta_{\Gamma s}(j)\right\vert\\ 
% \le& M_\Gamma\sum_{j=1}^{l-1}\left\vert\prod_{k=j+1}^{l-1}\Phi(k)^*-\prod_{k=j+1}^{l-1}(\Phi(k)^*-\delta)\right\vert
% +\sum_{j=1}^{l-1}\prod_{k=j+1}^{l-1}\left\vert\Phi(k)^*\delta_{\Gamma s}(j)\right\vert\\ 
%\lesssim & \sum_{j=1}^{l-1}\left\vert\prod_{k=j+1}^{l-1}\Phi(k)\left\{\prod_{k=j+1}^{l-1}(1+\delta/q')-1\right\}\right\vert+\sum_{j=1}^{l-1}\prod_{k=j+1}^{l-1}\left\vert\left(\Phi(k)+\delta\right)\delta_{\Gamma s}(j)\right\vert\\ 
\lesssim&\sum_{j=1}^{l-1}\left\vert \sum_{k=1}^{l-1-j} \delta^k {l-1-j \choose k} q^{l-1-j-k} \right\vert+\delta\sum_{j=1}^{l-1}\prod_{k=j+1}^{l-1} \bar{q}\\
=&\sum_{j=1}^{l-1}\left\vert  (\delta+q)^{l-1-j}-q^{l-1-j}\right\vert+\delta=\sum_{j=1}^{l-1}\{ (\delta+q)^{l-1-j}-q^{l-1-j}\}+\delta\\
\lesssim &\left\vert\frac{1-q^l}{1-q}-\frac{1-(q+\delta)^l}{1-q-\delta}\right\vert+\delta\lesssim \delta
\lesssim h^2+m^{-1}.
\end{align*}
Then, we have $\Delta_{T0}=O(h^2+m^{-1})$.
Hence, $\Big|P\left\{ T_{0}^*\le z\right\}-P\left\{ T_{01}^*\le z\right\}\Big|=\Big|P\left\{ T_{01}^*+ \Delta_{T0}\le z\right\}-P\left\{ T_{01}^*\le z\right\}\Big|\le P\{z- |\Delta_{T0}|\le T_{01}^*\le z+ |\Delta_{T0}|\}=O(n^{1/2}h^2+n^{1/2}m^{-1})$ holds with probability 1 as $n\rightarrow\infty$. The proof is hence completed. 
%When $h=O(n^{-5/16})$ and $m\asymp n^{c_2}$ for some $5/8\le c_2<3/2$, we obtain $\Big|P\left\{ T_{0}^*\le z\right\}-P\left\{ T_{01}^*\le z\right\}\Big|=O(n^{-1/8})$, which completes the proof.
\end{pf}

% \section{Proof of Corollary \ref{cor:SI}}
% The proof of Corollary \ref{cor:SI} is based on the conclusion of \cite{luo2016single} following the procedure of proof of Theorem \ref{thm:bootstrap}.
% We present the proof outline for $\textrm{DE}_1$ and $\textrm{IE}_1$ can be derived similarly. From Theorem 2 in \cite{luo2016single}, we have 
% $$\Mean\widehat{\theta}(\tp)=\theta(\tp)+\frac{1}{2}A_n(\tp)^{-1}h^2\overset{\triangle}{=}\theta_s(\tp),$$
% $$\Cov\{\widehat{\theta}(s),\widehat{\theta}(\tp)\}=n^{-1}A_n(s)^{-1}\left\{n^{-1}\sum_{i=1}^nB_i(s)\Sigma_\eta(s,\tp)B_i(\tp)^\top\right\}A_n(\tp)^{-1},$$
% where the concrete expressions of $A_n(\cdot)$ and $B_n(\cdot)$ can be found in the Appendix of \cite{luo2016single}. This implies that there exists some random vector $\bm{e}_{SI}\in\mathbb{R}^{(d+1)m}$ with mean 0 and covariance $$\Sigma_{SI}=A_n(s)^{-1}\left\{n^{-1}\sum_{i=1}^nB_i(s)\Sigma_\eta(s,\tp)B_i(\tp)^\top\right\}A_n(\tp)^{-1}$$ such that 
% $$\widehat{\bm{\theta}}=\widehat{\bm{\theta}}_s+\frac{1}{\sqrt{n}}\bm{e}_{SI}.$$
% Then we can define the independent mean zero Gaussian vectors $\{\overline{e}_{SI,i,\tp}\}_i$ with $\Mean\overline{\bm{e}}_{SI,i}\overline{\bm{e}}_{SI,i}^\top=\Mean{\bm{e}}_{SI,i}{\bm{e}}_{SI,i}^\top$ and the empirical Gaussian analogs ${e}_{SI,i,\tp}^b=\widehat{e}_{SI,i,\tp}\xi_i$ where $\{\xi_i\}_i$ are i.i.d standard normal random variables. Then step 1-2 of proof of Lemma \ref{lem:T0W0} hold under the assumption that $g_{2,\tp}$ and $g_{1,\tp}^\prime$ are bounded for all $1\le\tp\le m$, and the remains can be straightforwardly carried over.

\section{Proof of Theorems \ref{thm:switchback}, \ref{thm:switchIE}, Corollary 1 and More on the Switchback Design} \label{subsec:switchback}

\begin{pf}{ of Theorem \ref{thm:switchback}:}
From Model 1, we can derive that for $i=1,\ldots,n$, $t=1,\ldots,m$,
\begin{equation}\label{eq:x expan}
S_{t}=\Lambda^*_{t-1}+\mathbbm{B}_{0t}S_{1}+\sum_{j=1}^{t-1}\mathbbm{B}_{ j t}\Gamma_{ j}A_{j}+\varepsilon_{t-1 S}^*,
\end{equation}
where $\mathbbm{B}_{jt}=\prod_{k=j+1}^{t-1}\Phi(k)$, $\Lambda^*_{t-1}=\sum_{j=1}^t\mathbbm{B}_{jt}\phi_{0}(j)$ and $\varepsilon_{t-1 S}^*=\sum_{j=1}^{t-1}\mathbbm{B}_{j t}\varepsilon_{jS}$.
Define
$$  
M_{ t}^c=
\begin{pmatrix}
1 & A_{1 t} & S_{1  t} \\
1 & A_{2 t}  & S_{2  t} \\
\vdots & \vdots & \vdots \\
1 & A_{n t}  & S_{n  t}
\end{pmatrix},
$$  
$$\varphi_{ t}=\frac{\Mean A_{t}S_{t}-\Mean A_{t}\Mean S_{t}}{1-\Mean A_{t}}=\frac{\Cov(A_t,S_t)}{\Var(A_t)/\Mean A_t}\quad\hbox{and}\quad 
\mathcal{S}_{0 t}=\left[\Var\left(S_t-\frac{1}{\Mean A_{t}}\varphi_{ t}A_t\right)\right]^{-1}.$$
Under the two designs, $A_t$ depends on the observed data only through the past actions. It follows from \eqref{eq:x expan} that
$$\varphi_t=2\sum_{j=1}^{t-1}\mathbbm{B}_{jt}\Gamma_j\Cov(A_{j},A_{t}),$$
%Plug this into the expression of $\mathcal{S}_{0 t}$, we have
and hence
$$\mathcal{S}_{0 t}=\{\Var(\mathbbm{B}_{0t}S_{i1}+\varepsilon_{i,t-1,S}^*)\}^{-1}.$$
Direct algebra gives
\begin{align}\label{eqn:directalgebra}
\left\{\frac{1}{n}(M_{ t}^c)^{\top} M_{ t}^c\right\}^{-1}=\left(\begin{array}{ccc}
   1 &  \Mean A_t & \Mean S_t \\
   \Mean A_t & \Mean A_t & \Mean A_t S_t \\
   \Mean S_t & \Mean A_t S_t & \Mean S_t^2
\end{array}\right)^{-1}+o_p(1).
\end{align}
Consider the matrix on the right-hand-side. Notice that
\begin{eqnarray*}
   \left(\begin{array}{ccc}
   1 &  &  \\
     &    1  & \\
    & -\varphi_t/\Mean A_t  & 1
\end{array}
\right)\left(\begin{array}{ccc}
   1 &  &  \\
   -\Mean A_t  &    1  & \\
   -\Mean S_t &  & 1
\end{array}
\right) \left(\begin{array}{ccc}
   1 &  \Mean A_t & \Mean S_t \\
   \Mean A_t & \Mean A_t & \Mean A_t S_t \\
   \Mean S_t & \Mean A_t S_t & \Mean S_t^2
\end{array}\right)\left(\begin{array}{ccc}
   1 & -\Mean A_t & -\Mean S_t \\
     &    1  &  \\
    & & 1
\end{array}
\right)\\\times \left(\begin{array}{ccc}
   1 &  & \\
     &    1  & -\varphi_t/\Mean A_t \\
    & & 1
\end{array}
\right) =\left(\begin{array}{ccc}
    1 & & \\
     & \Var(A_t) & \\
     &  & \Var[S_t-\Var^{-1}(A_t)\Cov(A_t,S_t)A_t]
\end{array}
\right).
\end{eqnarray*}
It follows that
\begin{align*}
    \left(\begin{array}{ccc}
   1 &  \Mean A_t & \Mean S_t \\
   \Mean A_t & \Mean A_t & \Mean A_t S_t \\
   \Mean S_t & \Mean A_t S_t & \Mean S_t^2 
\end{array}\right)^{-1}=\left(\begin{array}{ccc}
   1 & -\Mean A_t & -\Mean S_t \\
     &    1  & \\
    & & 1
\end{array}
\right)\left(\begin{array}{ccc}
   1 &  &  \\
     &    1  & -\varphi_t/\Mean A_t \\
    & & 1
\end{array}
\right)\\\times \left(\begin{array}{ccc}
   1 &  & \\
     &   \frac{1}{(\Mean A_t)(1-\Mean A_t)}  & \\
    & & \mathcal{S}_{0,t}
\end{array}
\right)
\left(\begin{array}{ccc}
   1 &  &  \\
     &    1  &  \\
    &  -\varphi_t/\Mean A_t & 1
\end{array}
\right)\left(\begin{array}{ccc}
   1 &  &  \\
   -\Mean A_t  &    1  & \\
   -\Mean S_t & & 1
\end{array}
\right).
\end{align*}
With some calculations, it follows from \eqref{eqn:directalgebra}
 that \begin{align*}
\left\{\frac{1}{n}(M_{ t}^c)^{\top} M_{ t}^c\right\}^{-1}=&
\begin{pmatrix}
*  & *  & *  \\
-\frac{1}{1-\Mean A_{t}}-\frac{\varphi_{ t} \mathcal{S}_{0 t}(\varphi_{ t}-\Mean S_{t})}{\Mean A_{t}}  &  \frac{1}{\Mean A_{  t}(1-\Mean A_{ t})}+\frac{\varphi_{ t}^2 \mathcal{S}_{0 t}}{\Mean^2 A_{t}}  &  -\frac{\varphi_{ t} \mathcal{S}_{0 t}}{\Mean A_{t}}\\ 
\mathcal{S}_{0 t}\left(\varphi_{ t}-\Mean S_{t}\right)  &  -\frac{\mathcal{S}_{0 t}\varphi_{t}}{\Mean A_{t}}  & \mathcal{S}_{0t}
\end{pmatrix}  
  +o_p(1).
\end{align*}
Consequently, the resulting OLS estimator satisfies
\begin{align*}
\widehat{\gamma}(t)=&{\gamma}(t)+\frac{1}{n}\left\{-\frac{1}{1-\Mean A_{t}}-\frac{\varphi_{ t} \mathcal{S}_{0 t}(-\Mean S_{  t}+\varphi_{ t})}{\Mean A_{t}}\right\}\sum_ie_{it} \\ 
&+\frac{1}{n}\left\{\frac{1}{\Mean A_{ t}(1-\Mean A_{  t})}+\frac{\varphi_{ t}^2 \mathcal{S}_{0 t}}{\Mean^2 A_{  t}}\right\}\sum_iA_{it}e_{it}-\frac{1}{n}\frac{\varphi_{ t} \mathcal{S}_{0 t}}{\Mean A_t}\sum_iS_{it}e_{it}+o_p(n^{-1/2})\\ 
=&{\gamma}(t)+\frac{1}{n}\left\{\frac{1}{\Mean A_t(1-\Mean A_t)}+\frac{\varphi_{ t}^2 \mathcal{S}_{0 t}}{\Mean^2 A_{  t}}\right\}\sum_i(A_{it}-\Mean A_t)e_{it}-\frac{1}{n}\frac{\varphi_{ t} \mathcal{S}_{0 t}}{\Mean A_t}\sum_i(S_{it}-\Mean S_{t})e_{it}+o_p(n^{-1/2})\\ 
=&{\gamma}(t)+\frac{4}{n}\sum_{i=1}^n(A_{it}-\Mean A_t)e_{it}-\frac{2}{n}\varphi_t \mathcal{S}_{0t}\sum_{i=1}^n\{\mathbbm{B}_{0t}(S_{i1}-\Mean S_{1})+\varepsilon_{i,t-1,S}^*\}e_{it}+o_p(n^{-1/2}) \\ 
&+\frac{4}{n}\varphi_t^2\mathcal{S}_{0t}\sum_{i=1}^n(A_{it}-\Mean A_t)e_{it}-\frac{2}{n}\varphi_t \mathcal{S}_{0t}\sum_{j=1}^{t-1}\mathbbm{B}_{jt}\Gamma_j\sum_{i=1}^n(A_{ij}-\Mean A_{j})e_{it}.
\end{align*}
It can be verified that under either the alternating-day or the switchback design, the last line equals 0. Specifically, under the alternating-day design, we have $A_{ij}=A_{it}$. Hence, $\varphi_t=2^{-1}\sum_{j=1}^{t-1}\mathbbm{B}_{jt}\Gamma_j$ and the last term becomes $4n^{-1}\varphi_t^2\mathcal{S}_{0t}\sum_{i=1}^n (A_{it}-\Mean A_t)e_{it}$, equal to the first term on the last line. Under the switchback design, noting the relationship $A_{ij}-\Mean A_j=(-1)^{t-j}(A_{it}-\Mean A_t)$, we can obtain that the last line equals 0. Hence, we have
\begin{align*}
\widehat{\gamma}(t)=&{\gamma}(t)+\frac{4}{n}\sum_{i=1}^n(A_{it}-\Mean A_t)e_{it}-\frac{2}{n}\varphi_t^\top\mathcal{S}_{0t}\sum_{i=1}^n\{\mathbbm{B}_{0t}(S_{i1}-\Mean S_{i1})+\varepsilon_{i,t-1,S}^*\}e_{it}+o_p(n^{-1/2}).
\end{align*}
This gives
$$\widehat{\DE}=\DE+\frac{4}{n}\sum_{t=1}^m\sum_{i=1}^n(A_{it}-\Mean A_t)e_{it}-\frac{2}{n}\sum_{t=2}^m\left[\varphi_t\mathcal{S}_{0t}\sum_{i=1}^n\{\mathbbm{B}_{0t}(S_{i1}-\Mean S_{i1})+\varepsilon_{i,t-1,S}^*\}e_{it}\right]+o_p(n^{-1/2}).$$
Then
\begin{align*}
MSE(\widehat{\DE})=&\frac{1}{n}\Mean \left\{4\sum_{t=1}^m(A_{it}-\Mean A_t)e_{it}\right\}^2+\frac{1}{n}\Mean\left[2\sum_{t=2}^m\varphi_t\mathcal{S}_{0t}\{\mathbbm{B}_{0t}(S_{i1}-\Mean S_{i1})+\varepsilon_{i,t-1,S}^*\}e_{it}\right]^2.
% =&\frac{1}{n}\Mean \left\{4\sum_{t=1}^m(A_{it}-\Mean A_t)e_{it}\right\}^2+\frac{1}{n}\Mean\left\{2\sum_{t=2}^m\varphi_t^\top\mathcal{S}_{0t}\varepsilon_{i,t-1,S}^*e_{it}\right\}^2\\ 
% &+\frac{1}{n}\Mean\left\{2\sum_{t=2}^m\varphi_t^\top\mathcal{S}_{0t}\mathbbm{B}_{0t}(S_{i1}-\Mean S_{i1})e_{it}\right\}^2.
\end{align*}
In the switchback design, $A_{i1}=1-A_{i2}=\cdots=A_{i,\tp-1}=1-A_{i\tp}$. %whereas in the alternating-day design, $A_{i1}=A_{i2}=\cdots=A_{i\tp}=A_i$. 
Denote $E_{it}=\mathcal{S}_{0t}\{\mathbbm{B}_{0t}(S_{i1}-\Mean S_{i1})+\varepsilon_{i,t-1,S}^*\}e_{it}$. Then we have%for $S_{it}\in\mathbb{R}$,
\begin{align*}
&nMSE(\widehat{\DE}_{ad})-nMSE(\widehat{\DE}_{sb})\\
=&4\Var\left(\sum_{k=1}^{m}e_{k}\right)-
4\Var\left\{\sum_{k=1}^{m/2}(e_{2k-1}-e_{2k})\right\}+\Mean\left[2\sum_{t=2}^m\varphi_{t,ad} E_{it}\right]^2-\Mean\left[2\sum_{t=2}^m\varphi_{t,sb} E_{it}\right]^2+o(1)
\\
=&4\sum_{j, k}\Sigma_{e}(j,k)-4\sum_{j, k}(-1)^{|j-k|}\Sigma_{e}(j,k)+4\sum_{t_1,t_2=2}^m\Cov(E_{it_1},E_{it_2})(\varphi_{t_1,ad}-\varphi_{t_1,sb})(\varphi_{t_2,ad}+\varphi_{t_2,sb})+o(1).
\end{align*}
Under the given conditions, $E_{it_1}$ and $E_{it_2}$ are positively correlated.
When $\{\Phi(\tp)\}_\tp$ and $\{\Gamma(\tp)\}_\tp$ are of the same signs, respectively, we have 
\begin{align*}
&\sum_{t_1,t_2=2}^m\Cov(E_{it_1},E_{it_2})(\varphi_{t_1,ad}-\varphi_{t_1,sb})(\varphi_{t_2,ad}+\varphi_{t_2,sb})\\
=&\sum_{t_1,t_2=2}^m\Cov(E_{it_1},E_{it_2})\left(\sum_{1\le j_1\le t_1-1 \atop t_1-j_1=1,3,\ldots}\prod_{k=j_1+1}^{t_1-1}\Phi(k)\Gamma(j_1)\right)
\left(\sum_{1\le j_2\le t_2-1 \atop t_2-j_2=2,4,\ldots}\prod_{k=j_2+1}^{t_2-1}\Phi(k)\Gamma(j_2)\right)\ge0.
\end{align*}
Hence
$$nMSE(\widehat{\DE}_{ad})-nMSE(\widehat{\DE}_{sb})\ge8\sum_{|j-k|=1,3,\ldots}\Sigma_{e}(j,k)+o(1).$$
This completes the proof.
\end{pf}

\begin{pf}{ of Corollary 1:} %For DE, it is straightforward to compute that
{Without loss of generality, assume the constant $c$ equals one. With some calculations, we have that
\begin{align*}
MSE(\widehat{\DE}_{sb})&\asymp\frac{4}{n}\sum_{j, k}\sum_{j, k}(-1)^{|j-k|}\Sigma_{e}(j,k)\\ 
& = 4n^{-1}\left\{ m+2\sum_{k=1}^{m/2}(m-2k)\rho^{2k}-2\sum_{k=1}^{m/2}(m-2k+1)\rho^{2k-1}\right\}
\\ &=4n^{-1}\left\{m-2\sum_{k=1}^{m/2}(m-2k)(\rho^{2k-1}-\rho^{2k})-2\sum_{k=1}^{m/2}\rho^{2k-1}\right\}
\\ &= 4n^{-1}\left\{m-2(1-\rho)\sum_{k=1}^{m/2}(m-2k)\rho^{2k-1}\right\}
\\ & = 4n^{-1}\left\{m-2m(1-\rho)\sum_{k=1}^{m/2}\rho^{2k-1}\right\}= \frac{1-\rho}{1+\rho}4n^{-1}m,
\\ 
MSE(\widehat{\DE}_{ad})&\asymp\frac{4}{n}\Sigma_{e}(j,k)\\ 
&=  4n^{-1}\left\{m+2\sum_{k=1}^{m/2}(m-2k)\rho^{2k}+2\sum_{k=1}^{m/2}(m-2k+1)\rho^{2k-1}\right\}
\\&= 4n^{-1}\left\{m+2m(1+\rho)\sum_{k=1}^{m/2}\rho^{2k-1}\right\}+o(n^{-1})=\frac{1+\rho}{1-\rho}4n^{-1}m,
\end{align*}
which yields that $\text{MSE}(\widehat{\DE}_{sb})/\text{MSE}(\widehat{\DE}_{ad})\asymp(1-\rho)^2/(1+\rho)^2$.
}
\end{pf}

\begin{pf}{ of Theorem \ref{thm:switchIE}:}
When $m=2$, IE essentially equals $\beta(2)\Gamma(1)$. It follows that
\begin{eqnarray*}
    \widehat{\textrm{IE}}=\textrm{IE}+[\widehat{\beta}(2)-\beta(2)]\Gamma(1)+\beta(2)[\widehat{\Gamma}(1)-\Gamma(1)]+o_p(n^{-1/2}).
\end{eqnarray*}
Similar to the proof of Theorem \ref{thm:switchback}, it can be shown that
\begin{eqnarray}\nonumber
    \widehat{\beta}(2)-\beta(2)&=&\frac{\mathcal{S}_{0 2}}{n}\sum_{i=1}^n (S_{i2}-\Mean S_2)e_{i2}-\frac{2\mathcal{S}_{0 2}\varphi_2}{n}\sum_{i=1}^n (A_{i2}-\Mean A_2)e_{i2}+o_p(n^{-1/2})\\\nonumber
    &=&\frac{\mathcal{S}_{0 2}}{n}\sum_{i=1}^n [\Phi(1)(S_{i,1}-\Mean S_1)+\Gamma(1)(A_{i,1}-\Mean A_1)+\varepsilon_{i,1S} ]e_{i2}\\\nonumber
    &-&\frac{2\mathcal{S}_{0 2}\varphi_2}{n}\sum_{i=1}^n (A_{i2}-\Mean A_2)e_{i2}+o_p(n^{-1/2}).\\\label{eqn0}
    \widehat{\Gamma}(1)-\Gamma(1)&=&\frac{4}{n}\sum_{i=1}^n (A_{i1}-\Mean A_1)\varepsilon_{i,1S}+o_p(n^{-1/2}).
\end{eqnarray}
Under both designs, we can show that
\begin{eqnarray}\label{eqn1}
    \widehat{\beta}(2)-\beta(2)=\frac{\mathcal{S}_{0 2}}{n}\sum_{i=1}^n [\Phi(1)(S_{i,1}-\Mean S_1)+\varepsilon_{i,1S} ]e_{i2}+o_p(n^{-1/2}).
\end{eqnarray}
Notice that the i.i.d. sums in both \eqref{eqn0} and \eqref{eqn1} are design independent. Consequently, the IE estimators under the two designs achieve the same asymptotic MSE. The proof is hence completed.
\end{pf}

\noindent\textbf{Comparison against the regular switchback design}. %\change{Consider the regular switchback design \citep{bojinov2020design} which administers independent Bernoulli treatments across time. Similar to the proof of Theorem \ref{thm:switchback}, we can show that the direct estimator estimator under the regular switchback design satisfies $$nMSE(\widehat{\DE}_{ad})-nMSE(\widehat{\DE}_{rs})\ge4\sum_{j\neq k}\Sigma_{e}(j,k)+o(1).$$ This suggests that when the errors in the reward regression model are positively correlated, the regular switchback design can also improve the efficiency of estimating the direct effects. 
We next compare our switchback design against the regular switchback design \citep{bojinov2020design} which administers independent Bernoulli treatments across time. Consider the case where there exists some $0<\rho<1$ such that for any $1\le j,k \le m$, $\Cov(\eta_j,\eta_k)=\Sigma_{\eta,jk}=\rho^{|j-k|}$, $\Var(\varepsilon_{j})=\{\sigma_{j}^2\}_j$. %and $\Var(\varepsilon_{j,S})=\sigma_{j,S}^2$ for some positive constants $\sigma_j$ and . Then 
It follows that $\Cov(e_j,e_k)=\rho^{|j-k|}+\sigma_{j}^2\mathbb{I}\{j=k\}$. We focus on the variance of  DE under the settings of Corollary 1. 
We first calculate the covariance of highest resolution covariance  $\Sigma_e$ with $\rho=0.8$, $\sigma_{j}^2=0.36$ and $m=144$, and then generate covariances of $m=72,48,36,24,12,6$ by computing the corresponding sub-matrices from $\Sigma_e$. As shown in Figure \ref{fig:illustration} below, the proposed switchback design is more efficient than the regular one for any $m$. It also implies that the variances decrease with $m$ in both designs. 
% We mention that the classical A/B testing design that randomly assigns treatment globally over the day corresponds to the reduced case $m=1$ whose variance of treatment effect estimate is 1307.84. 
%Hence we remark that it is a general principle that the test power increases as the number of time slices enlarges in the switchback design. Though such feature has made the design prevalent in practice, the theoretical guarantees of the test when $m$ tends large have not been established, which motivates this work.

  \begin{figure}[htbp]
     \centering
      \includegraphics[width=4in]{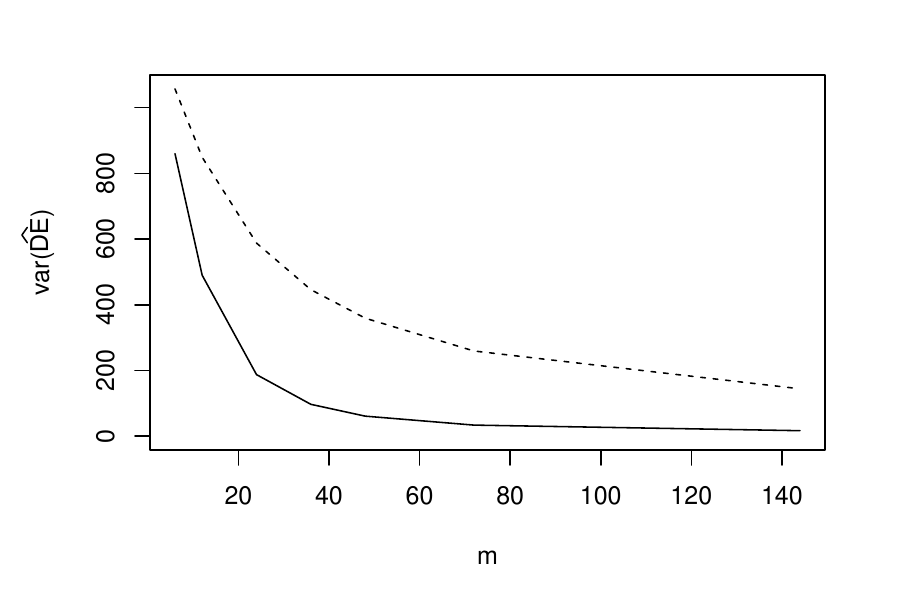}
  \caption{\label{fig:illustration} The solid line represents the variance of the DE under the proposed switchback design and whereas the dash line represents the one under the regular switchback design.}
  \end{figure}

\section{Proof of Theorem \ref{thm:dnn}}

\change{We first establish the error bound for $|\widehat{\DE}-\DE|$. 
Recall that
$$\widehat{\DE}-\DE=\frac{1}{nM}\sum_{i=1}^n\sum_{k=1}^M\sum_{\tau=1}^m\Mean^*\left[\left\{\widehat{g}_1(\tau,\widehat{S}_{i\tau k}^0)-\widehat{g}_0(\tau,\widehat{S}_{i\tau k}^0)\right\}-\mathbb{E}\left\{g_1(\tau,{S}_{\tau}^0)-g_0(\tau,{S}_{\tau}^0)\right\}\right].$$
%where $\mathbb{E}^*$ is taken with respect to the simulated random errors. 
It follows that
{\footnotesize\begin{eqnarray}\label{eq:de nn gap}
\begin{split}
    &|\widehat{\DE}-\DE|\\ 
    \le &\sum_{a=0}^1 \left|\sum_{\tau=1}^m\frac{1}{nM}\sum_{i=1}^n\sum_{k=1}^M\left\{\widehat{g}_a(\tau,\widehat{S}_{i\tau k}^0)-\mathbb{E} g_a(\tau,{S}_{\tau}^0)\right\}\right|
    \\
    =&\sum_{a=0}^1\left|\sum_{\tau=1}^m \frac{1}{nM}\sum_{i=1}^n\sum_{k=1}^M\left\{\widehat{g}_a(\tau,\widehat{S}_{i\tau k}^0)-{g}_a(\tau,\widehat{S}_{i\tau k}^0)\right\}+\frac{1}{nM}\sum_{\tp=1}^m\sum_{i=1}^n\sum_{k=1}^M\left\{{g}_a(\tau,\widehat{S}_{i\tau k}^0)-\mathbb{E} g_a(\tau,{S}_{\tau}^0)\right\}\right|\\
    \le & \sum_{a=0}^1\sum_{\tau=1}^m\left|\frac{1}{n}\sum_{i=1}^n\mathbb{E}^*\left\{\widehat{g}_a(\tau,\widehat{S}_{i\tau k}^0)-{g}_a(\tau,\widehat{S}_{i\tau k}^0)\right\}\right|+\sum_{a=0}^1\sum_{\tau=1}^m\left|\frac{1}{n}\sum_{i=1}^n\mathbb{E}^*{g}_a(\tau,\widehat{S}_{i\tau k}^0)-\mathbb{E}g_a(\tau,{S}_{\tau}^0)\right|\\
     &+O_p\big(\sqrt{m}(nM)^{-1/2}\big)
    % \le 2m\Delta_2(n,m)+\sum_{a=0}^1\sum_{\tau=1}^m\left|\frac{1}{n}\sum_{i=1}^n\mathbb{E}^* g_a(\tau,\widehat{S}_{i\tau k}^0)-\mathbb{E} g_a(\tau,{S}_{\tau}^0)\right|+O_p\big(\sqrt{m}(nM)^{-1/2}\big),
\end{split}    
\end{eqnarray}}
where the expectation $\mathbb{E}^*$ is taken with respect to the simulated random errors. 

We next calculate the bound $\left|n^{-1}\sum_{i=1}^n\left\{\mathbb{E}^*{g}_a(\tau,\widehat{S}_{i\tau k}^0)-\mathbb{E}g_a(\tau,{S}_{\tau}^0)\right\}\right|$ for $1\le\tau\le m$, $a=0,1$. Notice that for $\tau\ge 2$, the density of ${S}_{\tau}^0$ conditional on ${S}_{\tau-1}^0$ can be expressed as 
$f_{\varepsilon_{\tp S}}\left(s-G_0(\tau-1,{S}_{\tau-1}^0)\right)$,
and the density of $\widehat{S}_{i\tau k}^0$ is 
$\widehat{f}_{\varepsilon_{\tp S}}\left(s-\widehat{G}_0(\tau-1,\widehat{S}_{i,\tau-1, k}^0)\right)$. 
We next derive the bound of $$\left|\frac{1}{n}\sum_{i=1}^n\mathbb{E}^*\left\{{g}_a(\tau,\widehat{S}_{i\tau k}^0)-\mathbb{E}g_a(\tau,{S}_{\tau}^0)\right\}\right|,$$ for $1\le\tau\le m$.

\begin{itemize}
\item When $\tau=1$, we have $\widehat{S}_{i1k}^0={S}_{i1}$. Then $n^{-1}\sum_{i=1}^n\mathbb{E}^*{g}_a(\tau,\widehat{S}_{i\tau k}^0)-\mathbb{E} g_a(\tau,{S}_{\tau}^0)=n^{-1}\sum_{i=1}^n {g}_a(\tau,{S}_{i1})-\mathbb{E} g_a(\tau,{S}_{1})$, where ${S}_{i1}$ and ${S}_1$ are identically distributed. According to Hoeffding's inequality, the difference is 
% upper bounded by $$O(n^{-1/2}\sqrt{\log m+\log n}),$$ with probability at least $1-O(m^{-1}n^{-1})$.  
$$O_p(n^{-1/2}\sqrt{\log m+\log n}).$$

\item When $\tau=2$, by definition, we have 
\begin{eqnarray*}
    \mathbb{E} g_a(2,S_2^0)= \mathbb{E}\int_s g_a(2, s) f_{\varepsilon_{2S}}(s-G_0(1,S_1))ds,
\end{eqnarray*}
and that
\begin{eqnarray*}
    \mathbb{E}^* g_a(2, \widehat{S}_{i,2,k}^0)=\int_s g_a(2,s)\widehat{f}_{\varepsilon_{2S}}(s-\widehat{G}_0(1,S_{i,1}))ds.
\end{eqnarray*}
With some calculations, we have %$\mathbb{E}^* g_a(2, \widehat{S}_{i,2,k}^0)$ can be approximated by $\int_s g_a(2, s) f_{\varepsilon_{2S}}(s-G_0(1,S_{i,1}))ds$ with the approximation error upper bounded by %$O(\Delta_3(n,m)+L_f \Delta_1(n,m))$, 
\begin{eqnarray*}
\begin{split}
    &\left|n^{-1}\sum_{i=1}^n\mathbb{E}^*\left\{{g}_a(\tau,\widehat{S}_{i\tau k}^0)-\mathbb{E}g_a(\tau,{S}_{\tau}^0)\right\}\right| \\
    =&\mathbb{E}\int_s g_a(2, s)\left | f_{\varepsilon_{2S}}(s-G_0(1,S_1)) -  \widehat{f}_{\varepsilon_{2S}}(s-\widehat{G}_0(1,S_1))\right|ds  +O(n^{-1/2}\sqrt{\log (mn)})\\
    \le&\mathbb{E}\int_{s} g_a(2,s) |f_{\varepsilon_{2S}}(s-G_0(1,S_{i,1}))-f_{\varepsilon_{2S}}(s-\widehat{G}_0(1,S_{i,1}))|ds\\
    +&\mathbb{E}\int_{s} g_a(2,s) |\widehat{f}_{\varepsilon_{2S}}(s-\widehat{G}_0(1,S_{i,1}))-f_{\varepsilon_{2S}}(s-\widehat{G}_0(1,S_{i,1}))|ds +O(n^{-1/2}\sqrt{\log (mn)}).%\\
    %=&O(n^{-1/2}\sqrt{\log m+\log n}+\Delta_3(n,m)+L_f \Delta_1(n,m)),
\end{split}    
\end{eqnarray*}
Under the given condition, the second last line is upper bounded by $L_f \Mean |G_0(1,S_{i,1})-\widehat{G}_0(1,S_{i,1})|\le L_f \Delta_1(n,m)$, using Cauchy-Schwarz inequality. Additionally, the first term on the last line is upper bounded by $O_p(\Delta_3(n,m))$.%, with probability approaching $1$.  

% In addition, using Hoeffding's inequality, the difference between $n^{-1}\sum_{i=1}^n\int_s g_a(2, s) f_{\varepsilon_{2S}}(s-G_0(1,S_{i,1}))ds$ and $\mathbb{E} g_a(2,S_2^0)$ is upper bounded by $O(n^{-1/2}\sqrt{\log m+\log n})$, with probability at least $1-O(m^{-1}n^{-1})$. As such, $\left|n^{-1}\sum_{i=1}^n\mathbb{E}^*\left\{{g}_a(\tau,\widehat{S}_{i\tau k}^0)-\mathbb{E}g_a(\tau,{S}_{\tau}^0)\right\}\right|$ is upper bounded by $O(n^{-1/2}\sqrt{\log m+\log n}+\Delta_3(n,m)+L_f \Delta_1(n,m))$. 

\item More generally, when $\tau\ge 3$, we have 
{\footnotesize\begin{eqnarray*}
    \mathbb{E} g_a(\tp,S_\tp^0)=\mathbb{E} \int_{s_{\tp},s_{\tp-1},\cdots,s_2} g_a(\tp, s_{\tp}) f_{\varepsilon_{\tp S}}(s_{\tp}-G_0(\tp-1,s_{\tp-1}))\cdots f_{\varepsilon_{2 S}}(s_2-G_0(1,S_1)) ds_{\tp}\cdots ds_2,
\end{eqnarray*}}
and that
{\footnotesize\begin{eqnarray*}
    \mathbb{E}^* g_a(\tp, \widehat{S}_{i,\tp,k}^0)=\int_{s_{\tp},s_{\tp-1},\cdots,s_2} g_a(\tp,s_{\tp})\widehat{f}_{\varepsilon_{\tp S}}(s_{\tp}-\widehat{G}_0(\tp-1,s_{\tp-1}))\cdots\widehat{f}_{\varepsilon_{2S}}(s_2-\widehat{G}_0(1,S_{i,1}))ds_{\tp}\cdots ds_2.
\end{eqnarray*}}
Similarly, we can show that the difference $\left|n^{-1}\sum_{i=1}^n\mathbb{E}^*\left\{{g}_a(\tau,\widehat{S}_{i\tau k}^0)-\mathbb{E}g_a(\tau,{S}_{\tau}^0)\right\}\right|=\sum_{j=2}^{\tau} \zeta_{1,j}+\zeta_{2,j}+O_p(n^{-1/2}\sqrt{\log (mn)})$, 
% can be upper bounded by $\sum_{j=2}^{\tau} \zeta_{1,j}+\zeta_{2,j}+O(n^{-1/2}\sqrt{\log (mn)})$ with probability approaching $1$, 
where $\zeta_{1,j}$ is given by
{\footnotesize
\begin{eqnarray*}
    %O(n^{-1/2}\sqrt{\log (mn)})+
    \Mean\int_{s_\tau,\cdots,s_2} g_{a}(\tau,s_{\tau}) \widehat{f}_{\varepsilon_{\tp S}}(s_{\tp}-\widehat{G}_0(\tp-1,s_{\tp-1}))\cdots|\widehat{f}_{\varepsilon_{j S}}(s_j-\widehat{G}_0(j-1,s_{j-1}))-f_{\varepsilon_{j S}}(s_j-\widehat{G}_0(j-1,s_{j-1}))|\\
    \times f_{\varepsilon_{j-1 S}}(s_{j-1}-G_0(j-2,s_{j-2}))\cdots f_{\varepsilon_{2S}}(s_2-G_0(1,S_{i,1}))ds_{\tp}\cdots ds_2,
\end{eqnarray*}
}
whose absolute value can be upper bounded by
{\footnotesize
\begin{eqnarray*}
    O(1) \Mean \int_{s_j,\cdots,s_2} |\widehat{f}_{\varepsilon_{j S}}(s_j-\widehat{G}_0(j-1,s_{j-1}))-f_{\varepsilon_{j S}}(s_j-\widehat{G}_0(j-1,s_{j-1}))|\\
    \times f_{\varepsilon_{j-1 S}}(s_{j-1}-G_0(j-2,s_{j-2}))\cdots f_{\varepsilon_{2S}}(s_2-G_0(1,S_{i,1}))ds_{\tp}\cdots ds_2=O(\Delta_3(n,m)),
\end{eqnarray*}
}
%can be upper bounded by $O(\Delta_3(n,m))$, 
and $\zeta_{2,j}$ is given by
{\footnotesize
\begin{eqnarray*}
    %O(n^{-1/2}\sqrt{\log (mn)})+
    \Mean\int_{s_\tau,\cdots,s_2} g_{a}(\tau,s_{\tau}) \widehat{f}_{\varepsilon_{\tp S}}(s_{\tp}-\widehat{G}_0(\tp-1,s_{\tp-1}))\cdots|f_{\varepsilon_{j S}}(s_j-\widehat{G}_0(j-1,s_{j-1}))-f_{\varepsilon_{j S}}(s_j-G_0(j-1,s_{j-1}))|\\
    \times f_{\varepsilon_{j-1 S}}(s_{j-1}-G_0(j-2,s_{j-2}))\cdots f_{\varepsilon_{2S}}(s_2-G_0(1,S_{i,1}))ds_{\tp}\cdots ds_2,
\end{eqnarray*}
}
whose absolute value can be upper bounded by 
{\footnotesize
\begin{eqnarray*}
    O(1)L_f \Mean |\widehat{G}_0(j-1,S_{j-1}^0)-G_0(j-1,S_{j-1}^0)|\le O(1) L_f \sqrt{\Mean |\widehat{G}_0(j-1,S_{j-1}^0)-G_0(j-1,S_{j-1}^0)|^2},
\end{eqnarray*}
}
where $S_{j-1}^0$ denotes the potential state assuming the system receives the control treatment at each time. Using the change of measure theory, we obtain that $\zeta_{2,j}=O_p(L_f \sqrt{\omega} \Delta_1(n,m))$ where the big-$O_p$ term is uniform in $j$. 
% $\zeta_{2,j}=O(1)L_f \sqrt{\omega} \Delta_1(n,m)$ where the big-$O$ term is uniform in $j$ with probability approaching $1$.  %with probability at least $1-O(m^{-1}n^{-1})-o(1)$. 
\end{itemize}

To summarize, we obtain that $$\left|\frac{1}{n}\sum_{i=1}^n\mathbb{E}^*\left\{{g}_a(\tau,\widehat{S}_{i\tau k}^0)-\mathbb{E}g_a(\tau,{S}_{\tau}^0)\right\}\right|=O_p(n^{-1/2}\sqrt{\log m+\log n}+\tau \Delta_3(n,m)+L_f\sqrt{\omega}\tau \Delta_1(n,m)).$$ %the following upper bound for $\left|\frac{1}{n}\sum_{i=1}^n\mathbb{E}^*\left\{{g}_a(\tau,\widehat{S}_{i\tau k}^0)-\mathbb{E}g_a(\tau,{S}_{\tau}^0)\right\}\right|$, given by
% $O(n^{-1/2}\sqrt{\log m+\log n}+\tau \Delta_3(n,m)+L_f\sqrt{\omega}\tau \Delta_1(n,m))$.

We next bound $\left|n^{-1}\sum_{i=1}^n\mathbb{E}^*\left\{\widehat{g}_a(\tau,\widehat{S}_{i\tau k}^0)-{g}_a(\tau,\widehat{S}_{i\tau k}^0)\right\}\right|$. Notice that
{\begin{eqnarray*}
\begin{split}
    &\left|\frac{1}{n}\sum_{i=1}^n\mathbb{E}^*\left\{\widehat{g}_a(\tau,\widehat{S}_{i\tau k}^0)-{g}_a(\tau,\widehat{S}_{i\tau k}^0)\right\}\right|
    \\\le&\mathbb{E}\left|\widehat{g}_a(\tau,S_\tau^0)-g_a(\tau,S_\tau^0)\right|+\left|\frac{1}{n}\sum_{i=1}^n\mathbb{E}^*(\widehat{g}_a-g_a)(\tau,\widehat{S}_{i\tau k}^0)-\mathbb{E}(\widehat{g}_a-g_a)(\tau,{S}_{\tau }^0)\right|.
    %\\=&O(n^{-1/2}\sqrt{\log (mn)}+)+\log n}+\Delta_2(n,m)+\tau\Delta_3(n,m)+L_f\tau \Delta_1(n,m)).
\end{split}    
\end{eqnarray*}}
Similarly, we can show that the second term is $O_p(n^{-1/2}\sqrt{\log (mn)}+L_f\tau \sqrt{\omega}\Delta_1(n,m)+\tau\Delta_3(n,m))$ 
% with probability approaching $1$ 
whereas the first term can be upper bounded by $O(\sqrt{\omega}\Delta_2(n,m))$ using the change of measure theorem. 

Consequently, we have shown that
$$|\widehat{\DE}-\DE|=O_p(mn^{-1/2}\sqrt{\log (nm)}+m^2 \Delta_3(n,m)+L_f m^2 \sqrt{\omega} \Delta_1(n,m)+m\sqrt{\omega}\Delta_2(n,m)).$$
% , with probability approaching $1$, $|\widehat{\DE}-\DE|$ can be upper bounded by $O(mn^{-1/2}\sqrt{\log (nm)}+m^2 \Delta_3(n,m)+L_f m^2 \sqrt{\omega} \Delta_1(n,m)+m\sqrt{\omega}\Delta_2(n,m))$.

As for the error bound for $|\widehat{\IE}-\IE|$, it can be expressed by
{\footnotesize\begin{align*}
    \left|\widehat{\IE}-\IE\right|=&\left|\frac{1}{nM}\sum_{i=1}^n\sum_{k=1}^M\sum_{\tau=1}^m\left[\left\{\widehat{g}_1(\tau,\widehat{S}_{i\tau k}^1)-\widehat{g}_1(\tau,\widehat{S}_{i\tau k}^0)\right\}-\Mean\left\{g_1(\tau,{S}_{\tau}^1)-g_1(\tau,{S}_{\tau}^0)\right\}\right]\right|\\ 
    \le&\sum_{\tau=1}^m\left[\left|\frac{1}{nM}\sum_{i=1}^n\sum_{k=1}^M\widehat{g}_1(\tau,\widehat{S}_{i\tau k}^1)-\Mean\left\{g_1(\tau,{S}_{\tau}^1)\right\}\right|+\left|\frac{1}{nM}\sum_{i=1}^n\sum_{k=1}^M\widehat{g}_1(\tau,\widehat{S}_{i\tau k}^0)-\Mean\left\{g_1(\tau,{S}_{\tau}^0)\right\}\right|\right].
\end{align*}}
The error bound can be obtained using similar arguments in deriving the error bound of $|\widehat{\DE}-\DE|$. We omit the details to save space.}

\section{Proofs of Theorems \ref{thm:st theta asymp} and \ref{thm:bootstrap st}}
The proofs of Theorems \ref{thm:st theta asymp} and \ref{thm:bootstrap st} are very similar to those of Theorems \ref{thm:t theta asymp} and \ref{thm:bootstrap}, and we sketch an outline only. To prove the consistency of the proposed test for DE in Theorem \ref{thm:st theta asymp}, it suffices to show the joint asymptotic normality of the set of estimated varying coefficients $\{\widetilde{\theta}_{st}(\tp,\iota)\}_{\tp,\iota}$. We first notice that, the initial estimator obtained in Step 1 of Algorithm \ref{alg:DE ST} is obtained by applying Steps 1 and 2 of Algorithm \ref{alg:T DE} to each individual region. The asymptotic normality of the initial estimator can be proven using similar arguments in the proof of Theorem \ref{thm:t theta asymp}. 

Next, note that the refined estimator $(\widetilde\theta(1,\iota)^\top,\ldots,\widetilde\theta(1,\iota)^\top)^\top$ is essentially a linear transformation of the initial estimator. Using similar arguments in Section \ref{pf:de t}, we can further calculate the asymptotic bias and variance, as well as the asymptotic normality of $\widetilde\theta_{st}(\tp,\iota)$, based on the expression $\widetilde\theta_{st}(\tp,\iota)=\kappa_{\ell,h_{st}}(\iota)\widetilde\theta_{st}^0(\tp,\ell)$. %by similar techniques as \ref{pf:de t}, and derive the asymptotic normality of $\widetilde\theta_{st}(\tp,\iota)$ by Liapounov's CLT.
% Finally, applying Liapounov's CLT yields of $\{\widetilde\theta_{st}(\tp,\iota)\}_{\tp,\iota}$. 

The proof of Theorem \ref{thm:bootstrap st} is similar to that of Theorem \ref{thm:bootstrap}. The only difference lies in the dimension of parameter vector. To be specific, let $e_{i}^\beta(\tp,\iota),E_{i}^\Phi(\tp,\iota),E_{i}^\Gamma(\tp,\iota)$ be the analogs of $e_{i}^\beta(\tp)$, $E_{i}^\Phi(\tp),E_{i}^\Gamma(\tp)$ for $1\le\tp\le m$, $1\le\iota\le r$ under the spatiotemporal case.
% Let
Denote 
\begin{align}\label{eq:GARx st}
&x_{i}^{st}(\tp,\iota) = \Big(
e_{i}^\beta(\tp,\iota)^\top,
\{\vc(E_{i}^\Phi(\tp,\iota))\}^\top,
E_{i}^\Gamma(\tp,\iota)^\top
\Big)^\top\in\R^{2d(d+2)},\nonumber\\
&x_i^{st}(\iota)=\big(x_{i}(2,\iota)^\top,x_{i}(3,\iota)^\top,\ldots,x_{i}(m,\iota)^\top\big)^\top\in\R^{p_x},\quad p_x=2(m-1)dp,\nonumber\\
&x_i^{st}=\big(x_{i}^{st}(1)^\top,x_{i}^{st}(2)^\top,\ldots,x_{i}^{st}(r)^\top\big)^\top\in\R^{p_x^{st}},\quad p_x^{st}=2(m-1)dpr.
\end{align}
Define the function
{\small\begin{align*}
&F_\IE^{st}=\frac{1}{mr}\sum_{\iota=1}^{r}\sum_{\tp=2}^m\left[\left(\beta_s(\tp,\iota)+\frac{e_{\tp,\iota}^\beta}{\sqrt{n}}\right)^\top\right.\\
&\hspace{1in}\left.\cdot\sum_{j=1}^{\tp-1}\left\{\prod_{k=j+1}^{\tp-1}\left(\Phi_s(k,\iota)+\frac{E_{k,\iota}^\Phi}{\sqrt{n}}\right)\left(\Gamma_s(j,\iota)+\frac{E_{j,\iota}^\Gamma}{\sqrt{n}}\right)\right\}\right].
\end{align*}}
Similar to Theorem \ref{thm:bootstrap}, the proof of Theorem \ref{thm:bootstrap st} contains two steps. In the first step, we could employ the high-dimensional Gaussian approximation theory to bound the difference between $\widehat{\textrm{IE}}_{st}-\textrm{IE}_{st}$ and  $\widehat{\textrm{IE}}_{st}^b-\widehat{\textrm{IE}}_{st}$, assuming that  these statistics are constructed based on the oracle parameters. 
%First assume $\Mean \{x_i^{st}(x_i^{st})^\top\}$ is known, and let $y_i^{st}$ be the Gaussian analog of $x_i^{st}$ with $\Mean \{y_i^{st}(y_i^{st})^\top\}=\Mean \{x_i^{st}(x_i^{st})^\top\}$ and $Z_0^{st}$ be the statistic obtained by substituting $x_i^{st}$ in $T_0^{st}$ with $y_i^{st}$.We first derive the Gaussian approximation result to bound the Kolmogorov distance between the distributions of $T_0^{st}$ and its Gaussian analog $Z_0^{st}$. In the second step, we consider the empirical case where $\Mean \{x_i^{st}(x_i^{st})^\top\}$ is not known. 
This allows us to establish the validity of the bootstrap algorithm in the second step. As we have commented, the only difference lies in the dimension of parameters, and the results can be derived similarly using the arguments in  the proof for Theorem \ref{thm:bootstrap}.

\section{Tables and Figures}

  \begin{table}
      \caption{\label{tab:A DE t}Simulation results of DE test based on temporal model and data from city A. 
      We report the rejection probabilities of 400 replicates for different  temporal-alternating design of experiment ($TI = 1, 3, 6$), number of days ($n=8, 14, 20$), and relative improvement in percentage ($\delta = 0.00, 0.25, 0.50, 0.75, 1.00$).
      % \red{delete LME?}
      }
     \centering
     {
      \begin{tabular}{|l|c|c|r|r|r|r|r|}
      \hline
      $y$&$hour$&$n$&0.00&0.25&0.50&0.75&1.00\\
      \hline
      \multirow{9}{*}{DTI}
      & \multirow{3}{*}{1}
        &8& 6.8&24.2&47.2&64.2&76.0\\
      &&14& 6.5&34.2&65.0&82.0&91.0\\
      &&20& 5.5&38.2&74.8&90.2&96.2\\
      \cline{2-8}
      & \multirow{3}{*}{3}
        &8& 6.5&15.8&33.0&47.2&62.2\\
      &&14& 3.8&19.0&42.5&64.2&78.5\\
      &&20& 5.2&26.0&53.0&77.0&91.5\\
      \cline{2-8}
      & \multirow{3}{*}{6}
        &8& 6.8&12.5&18.2&29.8&40.8\\
      &&14& 6.8&12.0&23.5&37.8&49.5\\
      &&20& 6.8&13.0&28.8&46.0&61.8\\
      \hline
      \end{tabular}
  }
  \end{table}

  \begin{table}
      \caption{\label{tab:B DE t}Simulation results of DE test based on temporal model and data from city B. 
      We report the rejection probabilities of 400 replicates for different  temporal-alternating design of experiment ($hour = 1, 3, 6$), number of days ($n=8, 14, 20$), and relative improvement in percentage ($\delta = 0.00, 0.25, 0.50, 0.75, 1.00$).}
  \centering
  {
      \begin{tabular}{|l|c|c|r|r|r|r|r|}
      \hline
      $y$&$hour$&$n$&0.00&0.25&0.50&0.75&1.00\\
      \hline
      \multirow{9}{*}{DTI}
      &\multirow{3}{*}{1}
        &8& 4.0&14.5&29.5&49.8&64.8\\
      &&14& 4.0&21.5&50.0&79.0&93.2\\
      &&20& 3.5&22.8&62.2&86.0&97.0\\
      \cline{2-8}
      &\multirow{3}{*}{3}
        &8& 4.2& 8.5&17.0&26.8&35.8\\
      &&14& 4.2&11.5&22.8&35.2&51.0\\
      &&20& 7.2&15.3&31.0&46.8&60.5\\
      \cline{2-8}
      &\multirow{3}{*}{6}
        &8& 7.8&11.2&17.5&23.2&28.8\\
      &&14& 6.8&10.5&18.8&28.0&37.2\\
      &&20& 7.2&15.2&23.0&31.5&45.5\\
      \hline
      \end{tabular}
  }
  \end{table}

  \begin{table}
  \caption{\label{tab:A IE t}Simulation results of IE test based on temporal model and data from city A. }
  \centering
  {
  \begin{tabular}{|c|c|r|r|r|r|r|}
  \hline
  \multicolumn{1}{|p{3em}|}{TI} & \multicolumn{1}{p{3em}|}{$n$} & \multicolumn{1}{c|}{0} & \multicolumn{1}{c|}{0.25} & \multicolumn{1}{c|}{0.5} & \multicolumn{1}{c|}{0.75} & \multicolumn{1}{c|}{1} \\
  \hline
  \multirow{3}[6]{*}{1} & 8     & 4.8 & 12.0 & 46.5 & 74.8 & 87.0 \\
  \cline{2-7}          & 14    & 6.0& 25.5 & 75.2 & 89.8 & 94.5 \\
  \cline{2-7}          & 20    & 6.2 & 47.0 & 86.8 & 93.8 & 97.0 \\
  \hline
  \multirow{3}[6]{*}{3} & 8     & 4.8 & 10.0 & 21.5 & 46.8 & 64.5 \\
  \cline{2-7}          & 14    & 6.2 & 21.8 & 49.5 & 72.8 & 84.2 \\
  \cline{2-7}          & 20    & 6.0 & 23.8 & 66.0 & 83.0 & 89.2 \\
  \hline
  \multirow{3}[6]{*}{6} & 8     & 5.0 & 9.2 & 17.0 & 32.5 & 52.0 \\
  \cline{2-7}          & 14    & 5.8 & 15.5 & 37.8 & 65.2 & 77.0 \\
  \cline{2-7}          & 20    & 5.8 & 22.0 & 58.2 & 76.5 & 83.5 \\
  \hline
  \end{tabular}%
  }
  \end{table}%

  \begin{table}
  \caption{\label{tab:B IE t}Simulation results of IE test based on temporal model and data from city B.}
  \centering
  \begin{tabular}{|c|c|r|r|r|r|r|}
  \hline
  \multicolumn{1}{|p{3em}|}{TI} & \multicolumn{1}{|p{3em}|}{$n$} & \multicolumn{1}{c|}{0} & \multicolumn{1}{c|}{0.25} & \multicolumn{1}{c|}{0.5} & \multicolumn{1}{c|}{0.75} & \multicolumn{1}{c|}{1} \\
  \hline
  \multirow{3}[6]{*}{1} & 8     & 5.2 & 9.2 & 32.8 & 64.8 & 80.0 \\
  \cline{2-7}          & 14    & 5.5 & 18.2 & 66.0 & 83.5 & 91.2 \\
  \cline{2-7}          & 20    & 7.2 & 33.5 & 79.5 & 91.2 & 95.5 \\
  \hline
  \multirow{3}[6]{*}{3} & 8     & 5.0 & 8.5 & 15.5 & 30.2 & 52.8 \\
  \cline{2-7}          & 14    & 5.5 & 17.5 & 33.5 & 62.5 & 75.0 \\
  \cline{2-7}          & 20    & 5.8 & 19.5 & 52.0 & 75.0 & 85.5 \\
  \hline
  \multirow{3}[6]{*}{6} & 8     & 5.0 & 7.8 & 13.5 & 21.8 & 34.8 \\
  \cline{2-7}          & 14    & 6.5 & 13.8 & 23.5 & 50.0 & 68.0 \\
  \cline{2-7}          & 20    & 5.5 & 15.2 & 36.5 & 65.5 & 77.5 \\
  \hline
  \end{tabular}%
  \end{table}%

\begin{table}
  \caption{\label{tab:A DE st}Simulation results of DE test based on spatiotemporal model and data from city A.}%
  \centering
    \begin{tabular}{|c|r|r|r|r|r|r|r|}
    \hline
          & \multicolumn{1}{r|}{} & \multicolumn{6}{c|}{Temporal-alternating} \\
    \hline
          & DE    & \multicolumn{1}{c|}{0} & \multicolumn{2}{c|}{0.5} & \multicolumn{3}{c|}{1} \\
    \hline
    \multirow{2}[4]{*}{} & delta1 & \multicolumn{1}{r|}{0} & \multicolumn{1}{r|}{0} & \multicolumn{1}{r|}{0.5} & \multicolumn{1}{r|}{0} & \multicolumn{1}{r|}{0.5} & \multicolumn{1}{r|}{1} \\
\cline{2-8}          & delta2 & \multicolumn{1}{r|}{0} & \multicolumn{1}{r|}{0.5} & \multicolumn{1}{r|}{0} & \multicolumn{1}{r|}{1} & \multicolumn{1}{r|}{0.5} & \multicolumn{1}{r|}{0} \\
    \hline
    \multicolumn{1}{|c|}{\multirow{3}[6]{*}{TI=1}} & n=8   & 5.0 & 41.3 & 50.8 & 60.5 & 65.3 & 82.8 \\
\cline{2-8}          & n=14  & 5.3 & 55.5 & 70.3 & 74.0 & 87.3 & 94.0 \\
\cline{2-8}          & n=20  & 3.8 & 70.8 & 82.3 & 85.8 & 94.0 & 96.3 \\
    \hline
    \multicolumn{1}{|c|}{\multirow{3}[6]{*}{TI=3}} & n=8   & 4.8 & 33.0 & 36.8 & 56.8 & 59.0 & 65.5 \\
\cline{2-8}          & n=14  & 5.0 & 40.8 & 48.8 & 75.5 & 77.0 & 85.5 \\
\cline{2-8}          & n=20  & 4.0 & 57.0 & 65.8 & 80.5 & 81.3 & 90.8 \\
    \hline
    \multicolumn{1}{|c|}{\multirow{3}[6]{*}{TI=6}} & n=8   & 4.0 & 17.5 & 21.0 & 19.3 & 21.3 & 33.3 \\
\cline{2-8}          & n=14  & 3.5 & 28.3 & 34.5 & 27.5 & 43.8 & 49.5 \\
\cline{2-8}          & n=20  & 6.0 & 31.8 & 39.0 & 48.5 & 50.3 & 54.8 \\
    \hline
          & \multicolumn{1}{r|}{} & \multicolumn{6}{c|}{Spatiotempotal-alternating} \\
    \hline
          & DE    & \multicolumn{1}{c|}{0} & \multicolumn{2}{c|}{0.5} & \multicolumn{3}{c|}{1} \\
    \hline
    \multirow{2}[4]{*}{} & delta1 & \multicolumn{1}{r|}{0} & \multicolumn{1}{r|}{0} & \multicolumn{1}{r|}{0.5} & \multicolumn{1}{r|}{0} & \multicolumn{1}{r|}{0.5} & \multicolumn{1}{r|}{1} \\
\cline{2-8}          & delta2 & \multicolumn{1}{r|}{0} & \multicolumn{1}{r|}{0.5} & \multicolumn{1}{r|}{0} & \multicolumn{1}{r|}{1} & \multicolumn{1}{r|}{0.5} & \multicolumn{1}{r|}{0} \\
    \hline
    \multicolumn{1}{|c|}{\multirow{3}[6]{*}{TI=1}} & n=8   & 5.0 & 46.0 & 56.3 & 67.3 & 68.8 & 85.0 \\
\cline{2-8}          & n=14  & 6.3 & 62.3 & 75.5 & 81.0 & 91.0 & 97.3 \\
\cline{2-8}          & n=20  & 5.3 & 76.0 & 87.3 & 92.0 & 97.5 & 100.0 \\
    \hline
    \multicolumn{1}{|c|}{\multirow{3}[6]{*}{TI=3}} & n=8   & 4.3 & 38.3 & 44.0 & 62.5 & 62.5 & 68.0 \\
\cline{2-8}          & n=14  & 8.5 & 47.3 & 54.3 & 81.5 & 81.5 & 88.5 \\
\cline{2-8}          & n=20  & 6.5 & 61.8 & 71.0 & 85.3 & 85.3 & 92.8 \\
    \hline
    \multicolumn{1}{|c|}{\multirow{3}[6]{*}{TI=6}} & n=8   & 2.8 & 23.0 & 28.3 & 25.3 & 26.5 & 37.8 \\
\cline{2-8}          & n=14  & 4.5 & 34.3 & 41.3 & 34.3 & 50.3 & 55.8 \\
\cline{2-8}          & n=20  & 5.8 & 37.3 & 44.8 & 53.5 & 57.5 & 62.3 \\
    \hline
    \end{tabular}%
\end{table}

\begin{table}
  \caption{\label{tab:A IE st}Simulation results of IE test based on spatiotemporal model and data from city A.}
  \centering
    \begin{tabular}{|c|r|r|r|r|r|r|r|}
    \hline
          & \multicolumn{1}{r|}{} & \multicolumn{6}{c|}{Temporal-alternating} \\
    \hline
          & IE    & \multicolumn{1}{c|}{0} & \multicolumn{2}{c|}{0.5} & \multicolumn{3}{c|}{1} \\
    \hline
    \multirow{2}[4]{*}{} & delta1 & \multicolumn{1}{r|}{0} & \multicolumn{1}{r|}{0} & \multicolumn{1}{r|}{0.5} & \multicolumn{1}{r|}{0} & \multicolumn{1}{r|}{0.5} & \multicolumn{1}{r|}{1} \\
\cline{2-8}          & delta2 & \multicolumn{1}{r|}{0} & \multicolumn{1}{r|}{0.5} & \multicolumn{1}{r|}{0} & \multicolumn{1}{r|}{1} & \multicolumn{1}{r|}{0.5} & \multicolumn{1}{r|}{0} \\
    \hline
    \multicolumn{1}{|c|}{\multirow{3}[6]{*}{TI=1}} & n=8   & 6.0 & 57.3 & 63.8 & 83.8 & 92.8 & 94.0 \\
\cline{2-8}          & n=14  & 5.3 & 76.0 & 78.0 & 92.0 & 94.3 & 97.0 \\
\cline{2-8}          & n=20  & 4.0 & 88.8 & 90.8 & 94.3 & 96.3 & 98.3 \\
    \hline
    \multicolumn{1}{|c|}{\multirow{3}[6]{*}{TI=3}} & n=8   & 4.5 & 45.0 & 49.5 & 53.3 & 60.5 & 68.0 \\
\cline{2-8}          & n=14  & 5.3 & 60.5 & 61.8 & 64.0 & 69.5 & 84.8 \\
\cline{2-8}          & n=20  & 3.5 & 75.8 & 77.0 & 72.3 & 84.5 & 92.3 \\
    \hline
    \multicolumn{1}{|c|}{\multirow{3}[6]{*}{TI=6}} & n=8   & 6.0 & 29.8 & 32.0 & 50.8 & 61.3 & 63.8 \\
\cline{2-8}          & n=14  & 4.8 & 50.5 & 51.0 & 59.0 & 68.0 & 82.5 \\
\cline{2-8}          & n=20  & 4.8 & 59.5 & 61.5 & 77.5 & 83.5 & 88.3 \\
    \hline
          & \multicolumn{1}{r|}{} & \multicolumn{6}{c|}{Spatiotempotal-alternating} \\
    \hline
          & IE    & \multicolumn{1}{c|}{0} & \multicolumn{2}{c|}{0.5} & \multicolumn{3}{c|}{1} \\
    \hline
    \multirow{2}[4]{*}{} & delta1 & \multicolumn{1}{r|}{0} & \multicolumn{1}{r|}{0} & \multicolumn{1}{r|}{0.5} & \multicolumn{1}{r|}{0} & \multicolumn{1}{r|}{0.5} & \multicolumn{1}{r|}{1} \\
\cline{2-8}          & delta2 & \multicolumn{1}{r|}{0} & \multicolumn{1}{r|}{0.5} & \multicolumn{1}{r|}{0} & \multicolumn{1}{r|}{1} & \multicolumn{1}{r|}{0.5} & \multicolumn{1}{r|}{0} \\
    \hline
    \multicolumn{1}{|c|}{\multirow{3}[6]{*}{TI=1}} & n=8   & 4.3 & 59.3 & 66.0 & 85.8 & 94.3 & 96.0 \\
\cline{2-8}          & n=14  & 6.3 & 78.5 & 80.3 & 93.0 & 96.0 & 98.0 \\
\cline{2-8}          & n=20  & 6.5 & 90.0 & 92.0 & 95.8 & 97.5 & 99.8 \\
    \hline
    \multicolumn{1}{|c|}{\multirow{3}[6]{*}{TI=3}} & n=8   & 5.0 & 47.0 & 51.5 & 55.0 & 62.0 & 70.0 \\
\cline{2-8}          & n=14  & 5.5 & 62.0 & 63.8 & 65.8 & 71.5 & 85.8 \\
\cline{2-8}          & n=20  & 5.3 & 77.0 & 78.8 & 73.3 & 86.3 & 93.5 \\
    \hline
    \multicolumn{1}{|c|}{\multirow{3}[6]{*}{TI=6}} & n=8   & 6.0 & 31.3 & 34.0 & 51.8 & 62.3 & 64.8 \\
\cline{2-8}          & n=14  & 4.8 & 52.0 & 53.3 & 61.3 & 70.5 & 84.3 \\
\cline{2-8}          & n=20  & 4.8 & 62.0 & 63.0 & 79.8 & 86.0 & 90.3 \\
    \hline
    \end{tabular}%
\end{table}%

  \begin{figure}[H]
  \begin{minipage}[t]{1\textwidth}
  \centering
  \includegraphics[width=1\textwidth,height=1.7in]{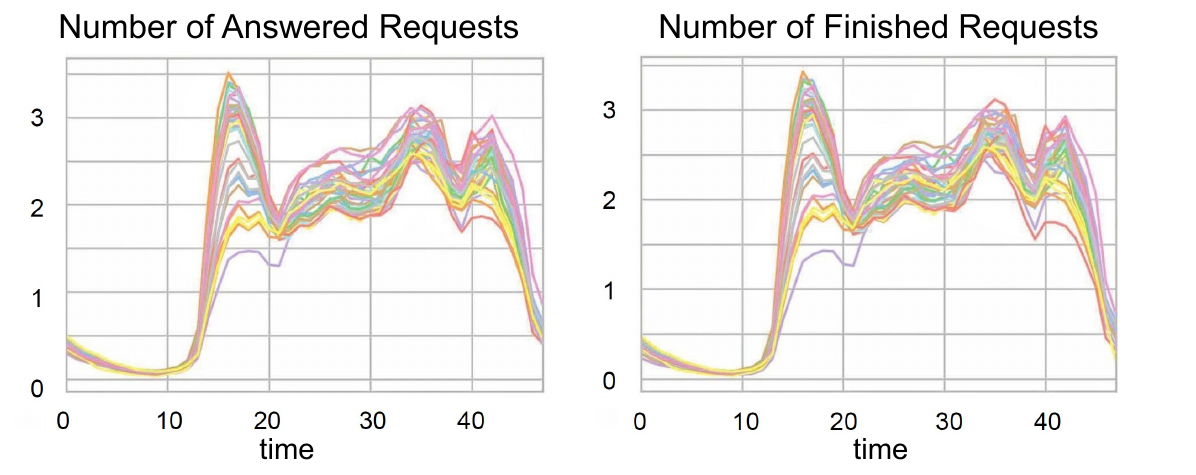}
  \end{minipage}%

  \begin{minipage}[t]{1\textwidth}
  \centering
  \includegraphics[width=1\textwidth,height=1.7in]{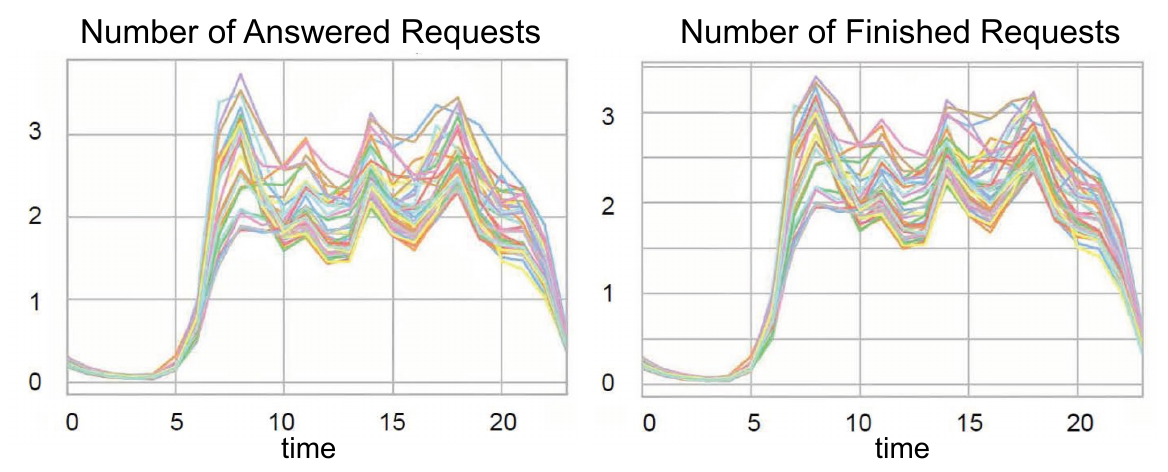}
  \end{minipage}%
  \caption{Scaled numbers of answered and finished requests from City A (the first row) and City B (the second row) across 40 days .}
  \label{fig:cityAB_t_pattern2}
  \end{figure}

  \begin{figure}[H]
  \centering
  \includegraphics[width=\textwidth]{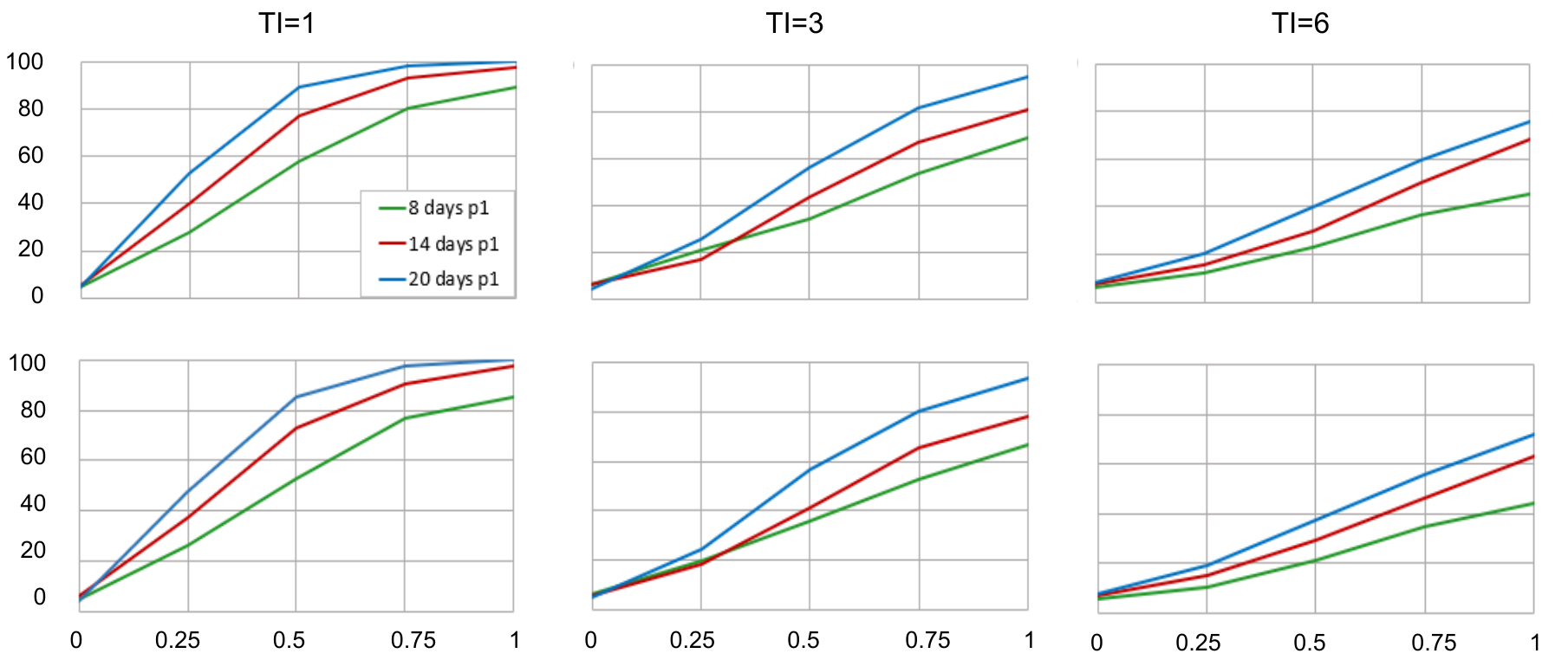}
  \caption{Empirical rejection rates of the proposed test for DE, with different combinations of $n,\delta,\textrm{TI}$ and outcomes based on the real dataset from city A (the number of answered requests in the first row and the number of finished requests in the second row). }
  \label{fig:DE2_t_cityA}
  \end{figure}

  \begin{figure}[H]
  \centering
  \includegraphics[width=\textwidth]{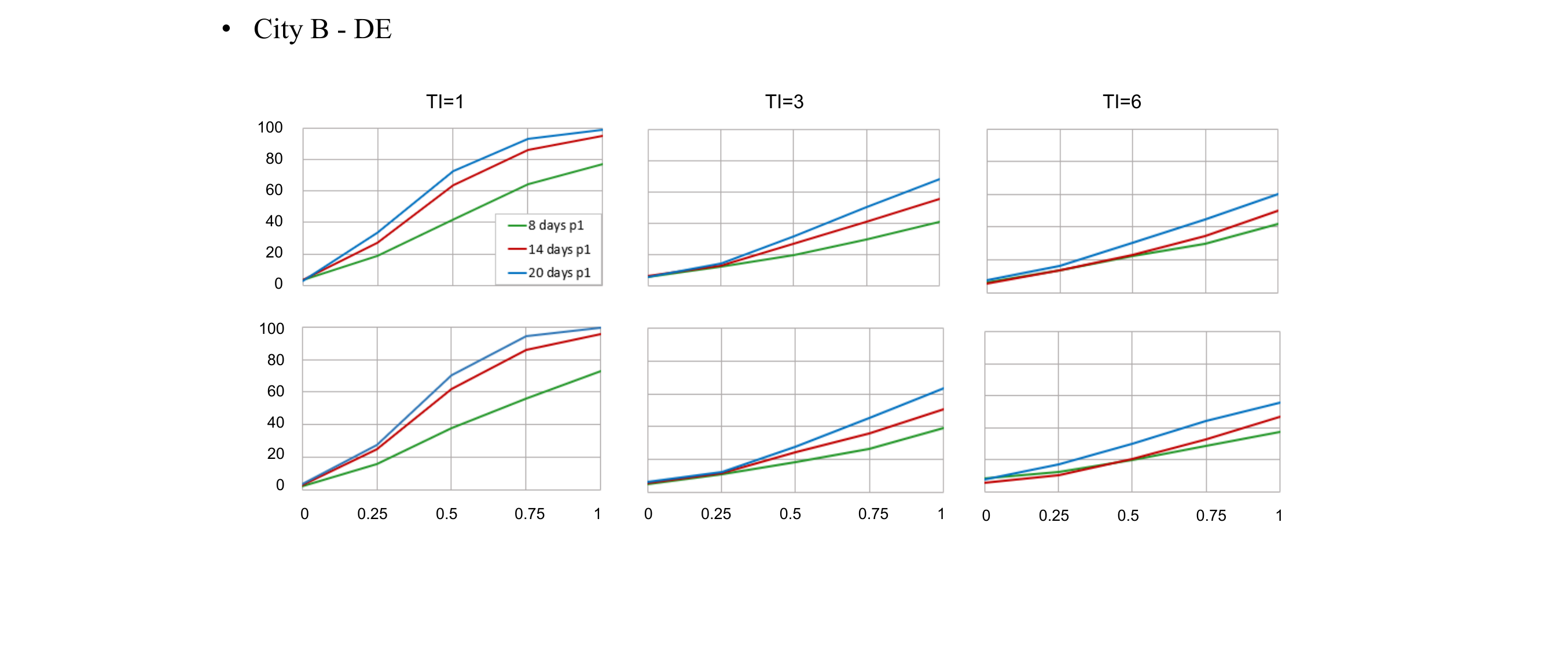}
  \caption{Empirical rejection rates of the proposed test for DE, with different combinations of $n,\delta,\textrm{TI}$ and outcomes based on the real dataset from city B (the number of answered requests in the first row and the number of finished requests in the second row). }
  \label{fig:DE2_t_cityB}
  \end{figure}

  \begin{figure}[H]
  \begin{minipage}[t]{0.5\textwidth}
  \centering
  \includegraphics[width=3.2in]{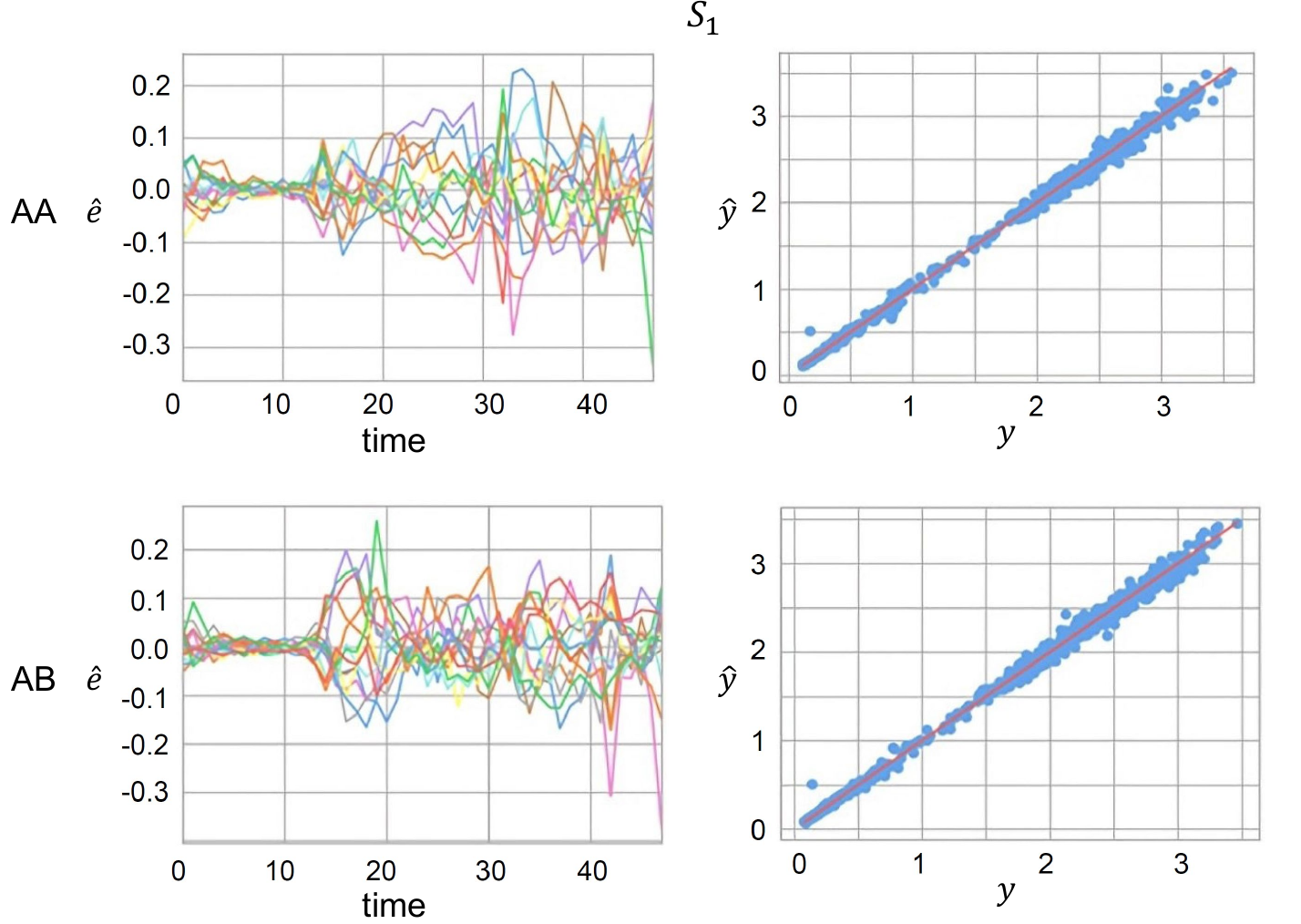}
  % \caption{fig1}
  % \label{fig:side:a}
  \end{minipage}%
  \begin{minipage}[t]{0.5\textwidth}
  \centering
  \includegraphics[width=3.2in]{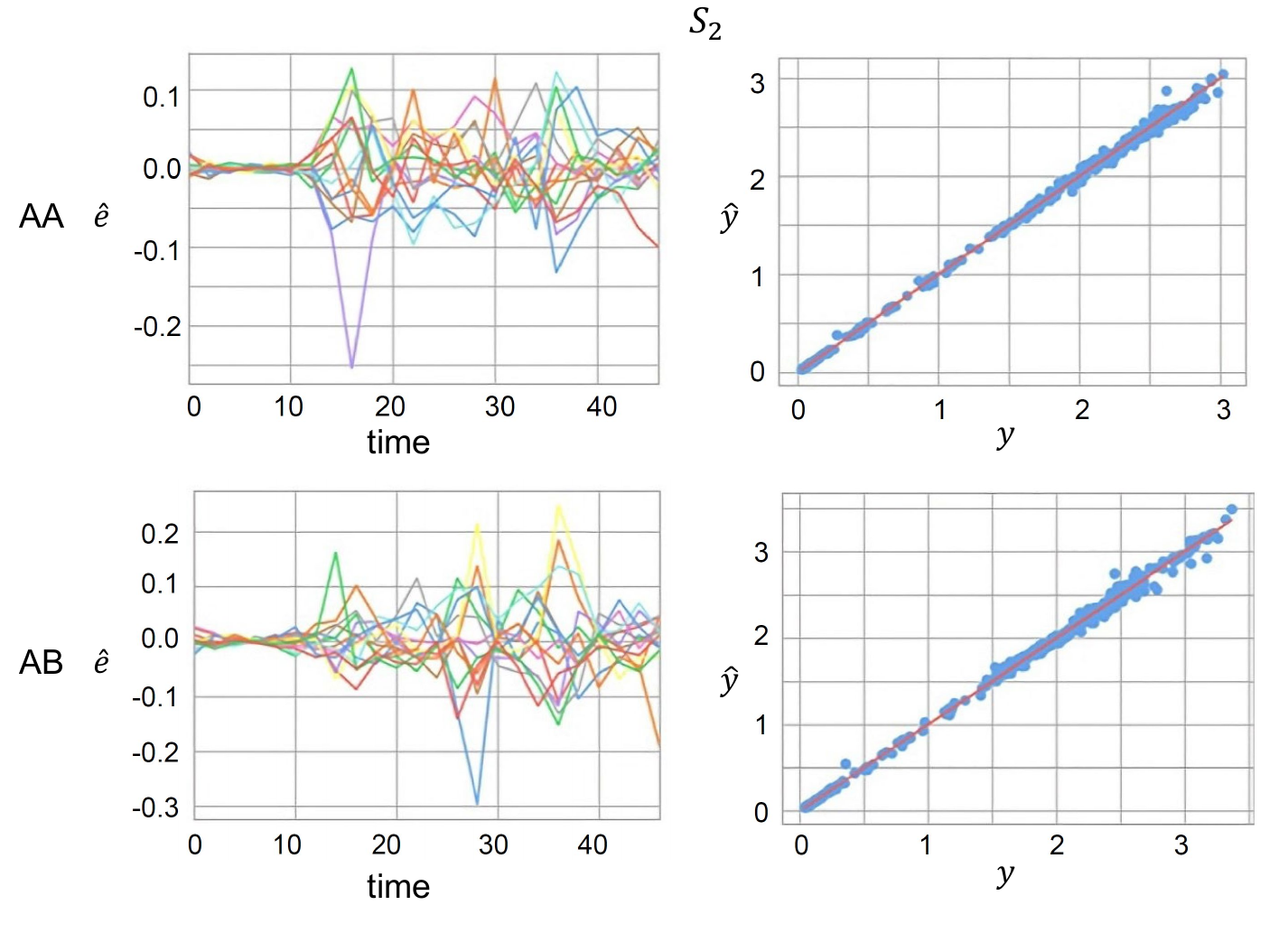}
  % \caption{fig2}
  % \label{fig:real_DE_t}
  \end{minipage}
  \begin{minipage}[t]{0.5\textwidth}
  \centering
  \includegraphics[width=3.2in]{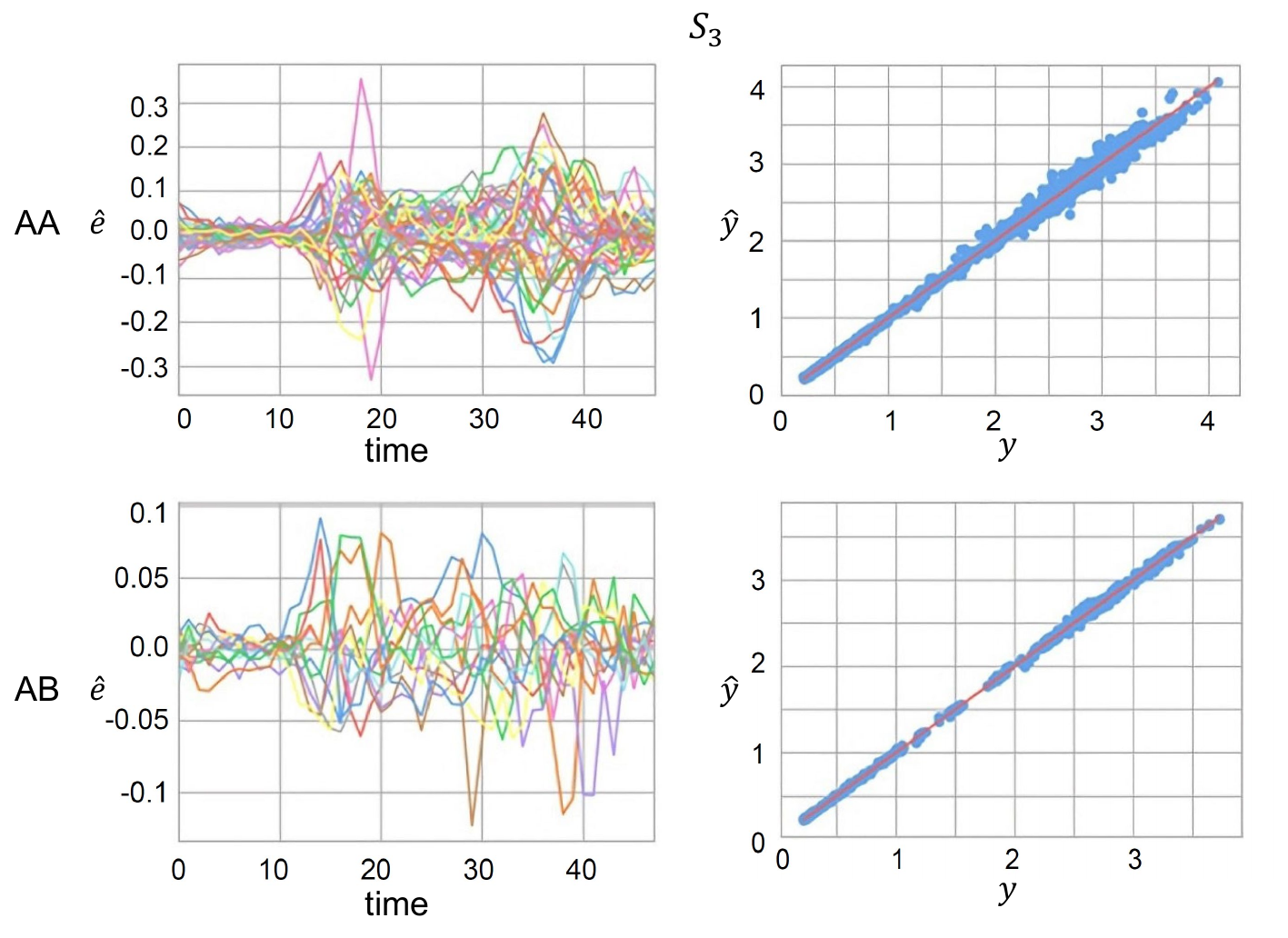}
  % \caption{fig3}
  % \label{fig:side:a}
  \end{minipage}%
  \begin{minipage}[t]{0.5\textwidth}
  \centering
  \includegraphics[width=3.2in]{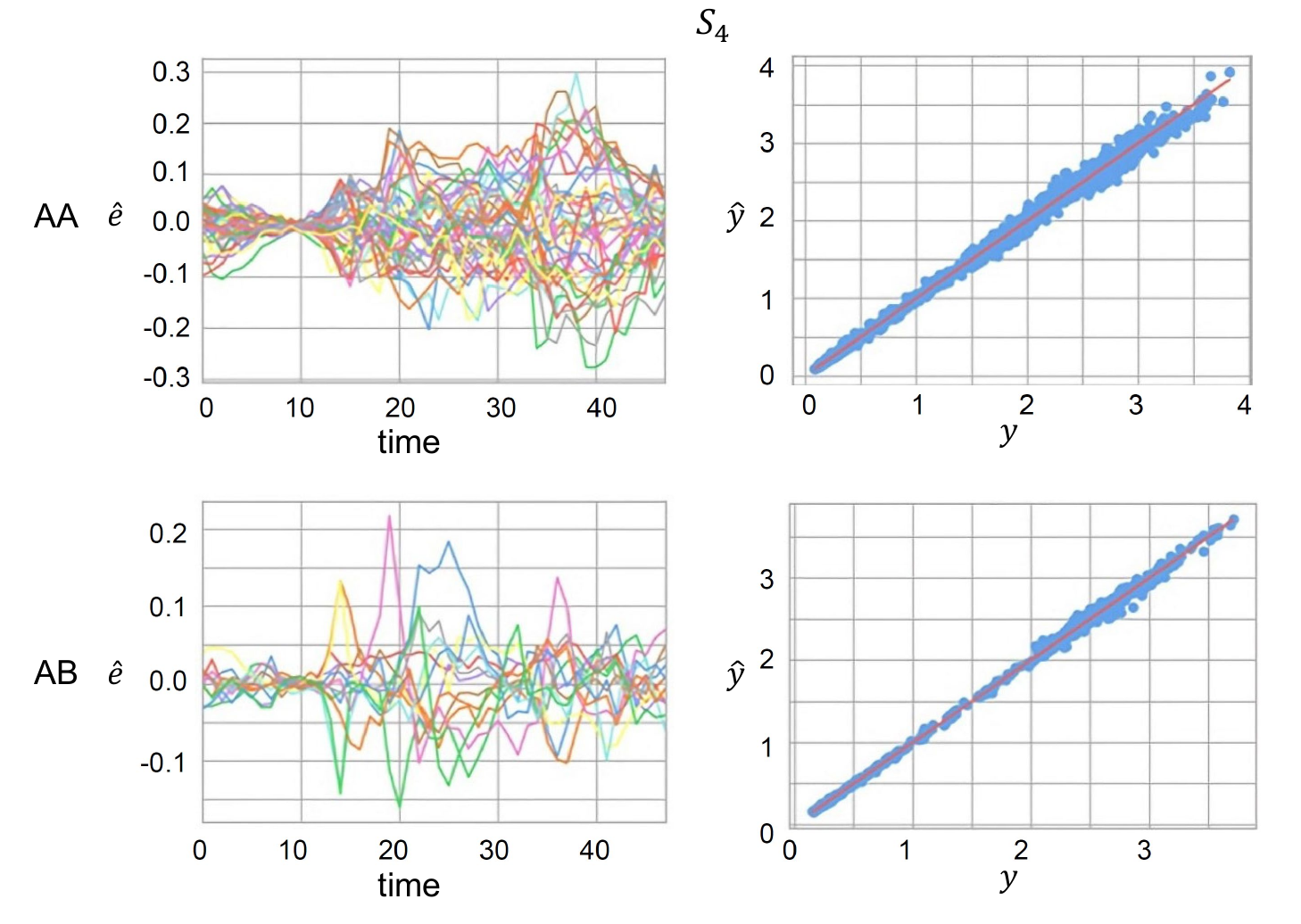}
  \end{minipage}
  \caption{Plots of the fitted drivers' total income against the observed values as well as the corresponding residuals. Data are 
  %Residuals plots of the fitted drivers' total income over true drivers' total income plots for both 
  collected from an A/A or A/B experiment under the temporal alternation design.}
  \label{fig:real_DE_t2}
  \end{figure}

  \begin{figure}[H]
  \begin{minipage}[t]{0.5\textwidth}
  \centering
  \includegraphics[width=3.2in]{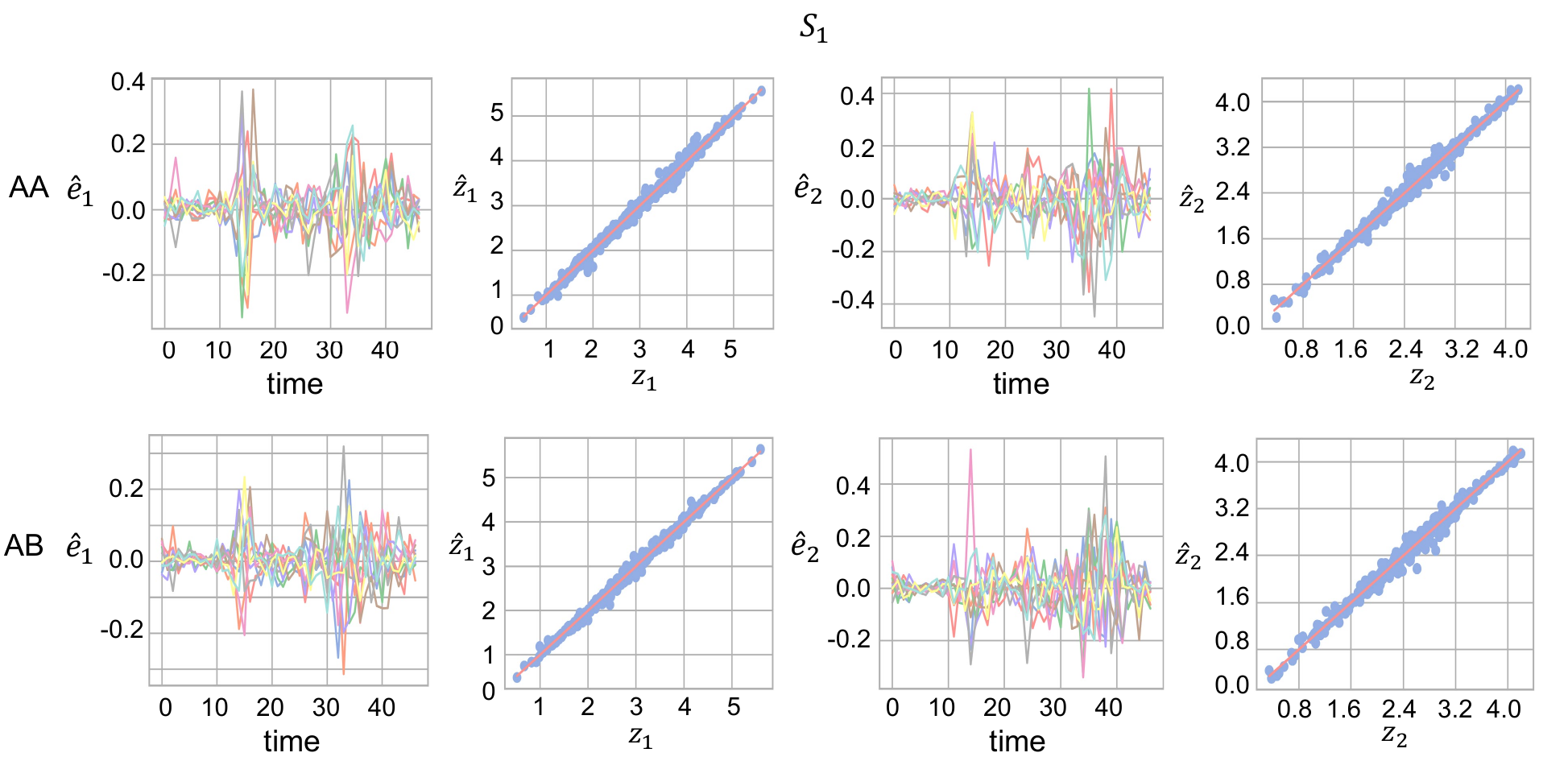}
  % \caption{fig1}
  % \label{fig:side:a}
  \end{minipage}%
  \begin{minipage}[t]{0.5\textwidth}
  \centering
  \includegraphics[width=3.2in]{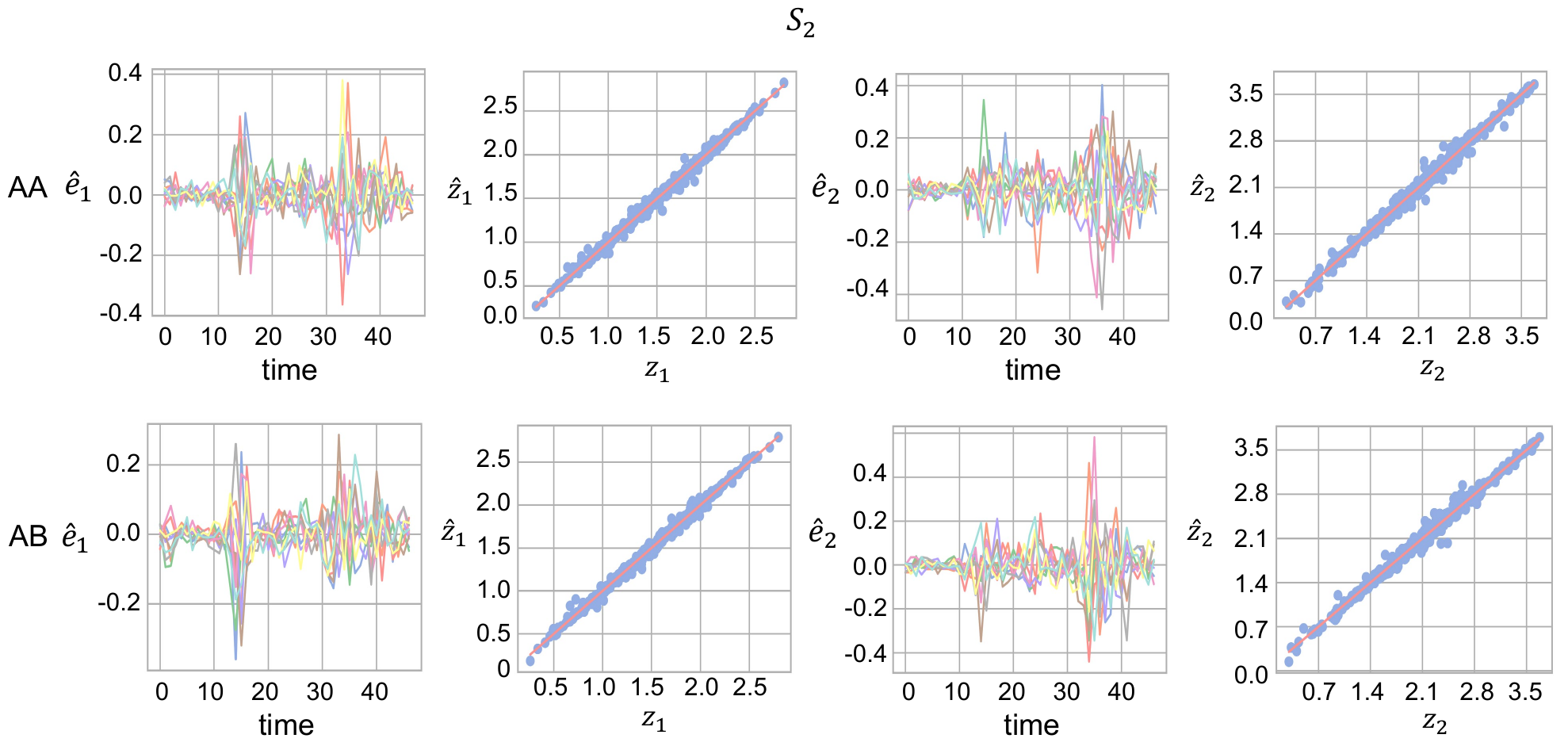}
  % \caption{fig2}
  % \label{fig:real_DE_t}
  \end{minipage}
  \begin{minipage}[t]{0.5\textwidth}
  \centering
  \includegraphics[width=3.2in]{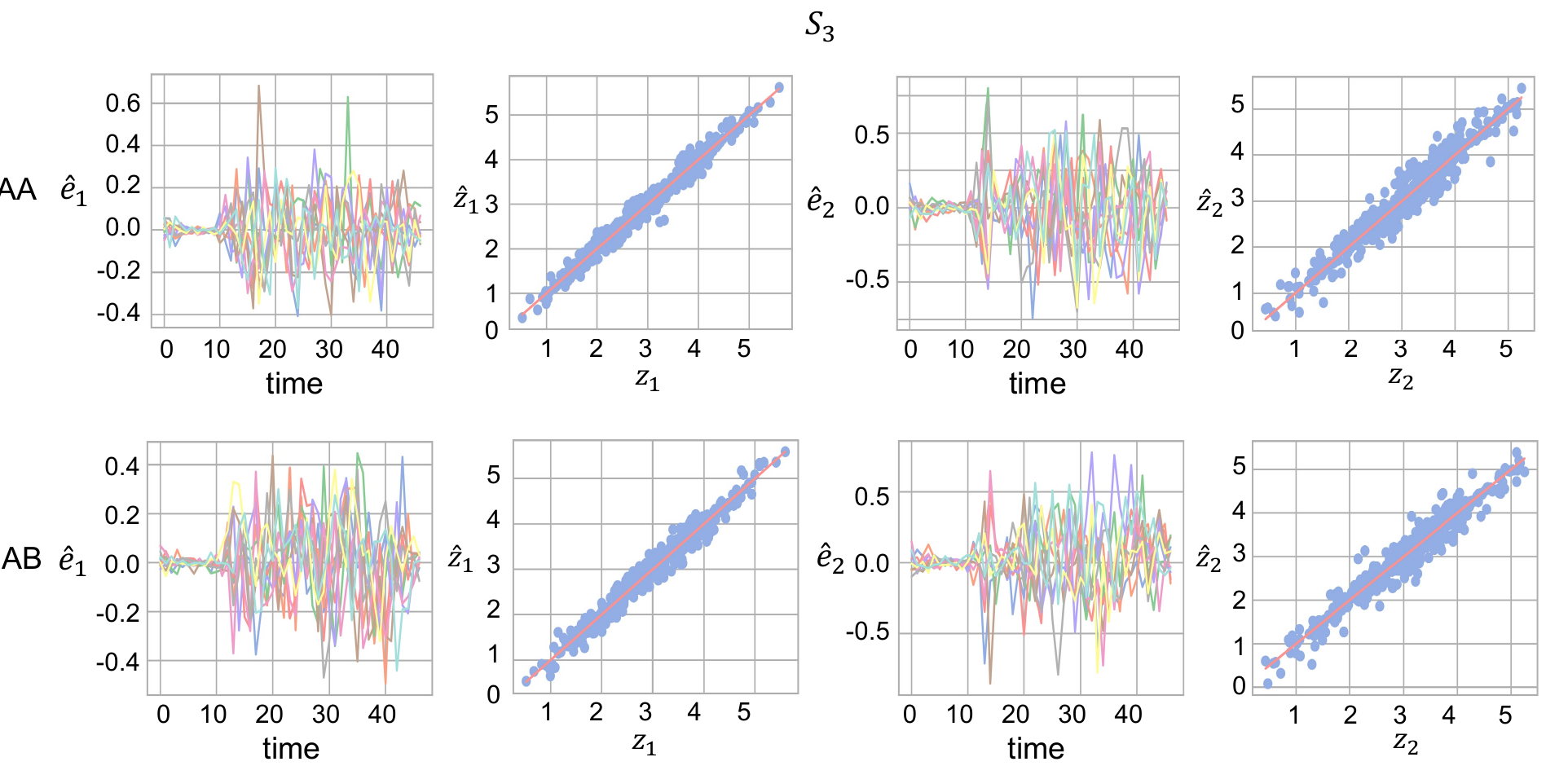}
  % \caption{fig3}
  % \label{fig:side:a}
  \end{minipage}%
  \begin{minipage}[t]{0.5\textwidth}
  \centering
  \includegraphics[width=3.2in]{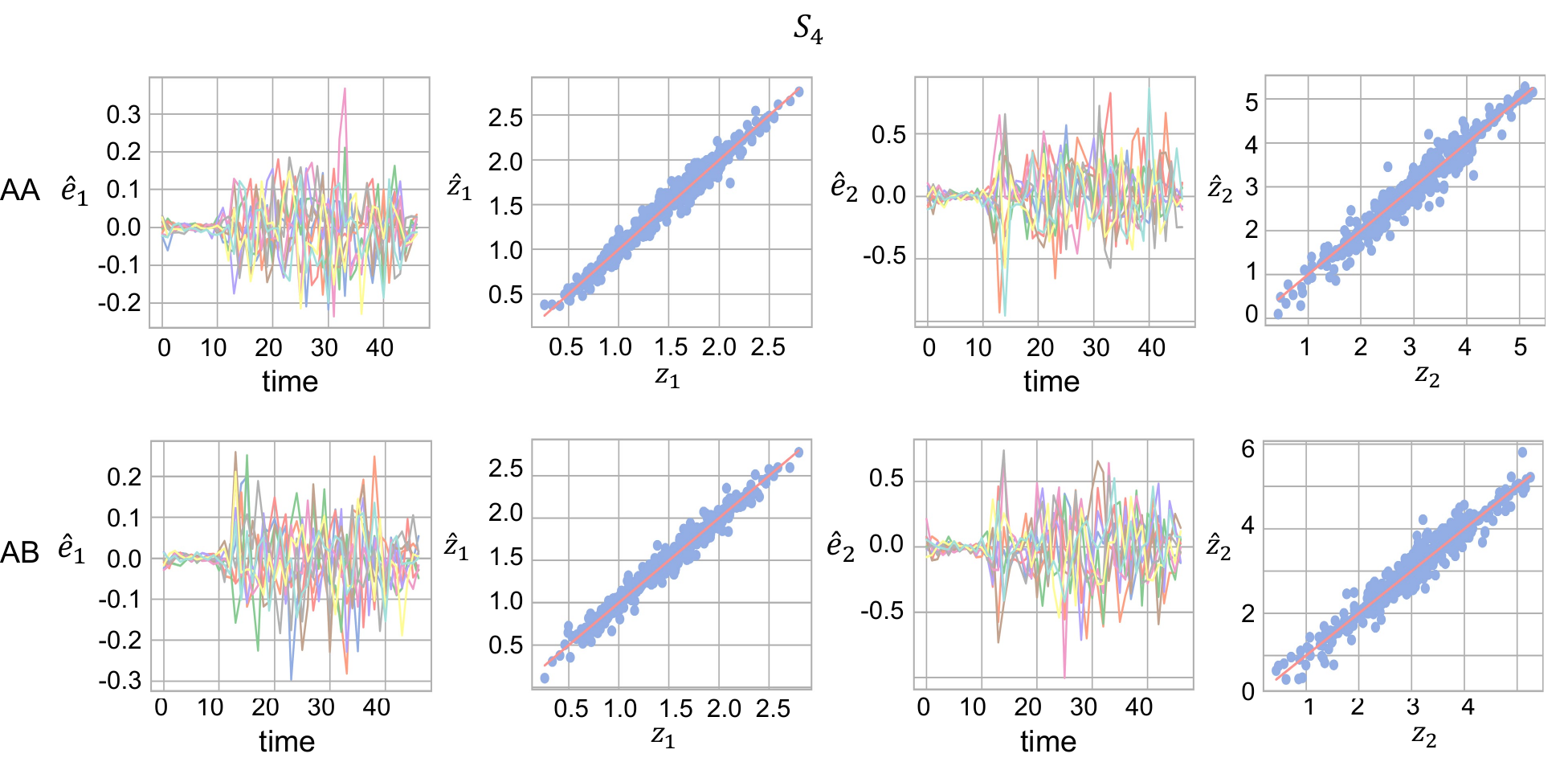}
  \end{minipage}
  \caption{\label{fig:real_IE_t}Plots of the fitted number of orders ($\widehat{e}_1$) and drivers' online time ($\widehat{e}_2$) against their observed values, as well as the corresponding residuals. Data are collected from an A/A or A/B experiment under the temporal alternation design.}%experiments of the four datasets based on one-region temporal experiment design.}
  \label{fig:sideaaa}
  \end{figure}

\section{Codes}

\subsection{Code for cross validation}

\begin{python}
import numpy as np
import pandas as pd
import statsmodels.api as sm
import statsmodels.formula.api as smf
from itertools import product
import multiprocessing as mp
import os
import warnings
warnings.filterwarnings("ignore")
import sys
path='.../temporal/src'  
if path not in sys.path:
    sys.path.append(path)
from sklearn.model_selection import KFold
from model_new import VCM

### simulation settings ###
file = 'V2_hangzhou_serial_order_dispatch_AA.csv'
ycol = 'gmv'
xcols = ['cnt_call','sum_online_time']
scols = ['cnt_call_1','sum_online_time_1']
acol = 'is_exp'
regcols = ['const'] + xcols

df = pd.read_csv('C:/Users/annie/OneDrive - pku.edu.cn/projects/3. Finished/stvcm/Code+Data20210825/temporal/data/'+file)
df['const'] = 1
xycols = [ycol] + regcols +['date', 'time']
df = df[xycols]  

NN = 40
idx = [i+1 for i in range(NN)]

kf = KFold(n_splits=5, shuffle=True)

param_grid = [0.05*i for i in range(20)] * NN ** (-1/3)

K = 3; M =48

res = [] 

for train_index, test_index in kf.split(idx):
    df_train = df.loc[df['date'].isin(train_index)].set_index(['date','time'])
    df_test = df.loc[df['date'].isin(test_index)].set_index(['date','time'])
    for hc in param_grid:
        Amat = df_train.groupby('date').apply(lambda dt: np.dot(dt[regcols].T.values, dt[regcols].values)).sum()
        bvec = df_train.groupby('date').apply(lambda dt: np.dot(dt[regcols].T, dt[ycol])).sum()
        eps_diag = np.eye(Amat.shape[0])*1e-3
        theta = np.linalg.solve(Amat+eps_diag, bvec)
        theta = pd.DataFrame(theta.reshape((M, K)), columns=regcols)
        tmat = np.mat(np.reshape(np.repeat(np.arange(M)/(M-1), M), (M,M)))
        theta = smooth(theta.T, ker_mat((tmat.T-tmat),hc)).T
        df_test['fitted'] = df_test[regcols].dot(theta_DE.values.flatten())
        df_test['resid'] = df_test[ycol] - df_test['fitted']
        res.append(sum((df_test['resid'])**2))

res = np.array(res).reshape(5,20)
res = res.sum(axis=0)

np.array(param_grid)[np.where(np.min(res))]
\end{python}

\subsection{Main code}

\begin{python}
import numpy as np
import pandas as pd
import statsmodels.api as sm
import statsmodels.formula.api as smf
from itertools import product
import multiprocessing as mp
from numpy import kron
import os
import warnings
warnings.filterwarnings("ignore")
import sys
path='.../Spatio-temporal/src'
if path not in sys.path:
    sys.path.append(path)
from model_st_new import VCM

### simulation settings ###
ycol = 'gmv'#,'cnt_grab', 'cnt_finish']
xcol = 'cnt_call'#,'sum_online_time']
scol = 'cnt_call_1'#the lag term
acol = 'is_exp'
acol_n = 'is_exp_n'
regcols = ['const'] + [xcol]
adj_mat = np.array([[0,1,1,1,0,0,0,0,0,0],
          [1,0,0,1,1,0,0,0,0,0],
          [1,0,0,1,0,1,0,0,0,0],
          [1,1,1,0,1,1,1,0,0,0],
          [0,1,0,1,0,0,1,1,0,0],
          [0,0,1,1,0,0,1,0,1,0],
          [0,0,0,1,1,1,0,1,1,1],
          [0,0,0,0,1,0,1,0,0,1],
          [0,0,0,0,0,1,1,0,0,1],
          [0,0,0,0,0,0,1,1,1,0]])
G = 10
adj_mat = adj_mat/np.repeat(adj_mat.sum(axis=0),G).reshape(G,G)
nsim = 400

two_sided = False
wild_bootstrap = False
interaction = False

DDS = [0.00, 0.005, 0.01]
IIS = [0.00, 0.005, 0.01]
IIS_n = [0.00, 0.005, 0.01]
NNs = [8,14,20]
TIs = [1,3,6]
designs = ['st','t']

wbi = 1 if wild_bootstrap else 0
tsi = 1 if two_sided else 0
ini = 1 if interaction else 0
hc = 0.01
hc_b = 0.01
IE = True

DD = 0
for (II, II_n, TI, design, NN) in product(IIS, IIS_n, TIs, designs, NNs):
    
    file = 'V1_hangzhou_pool.csv'
    df = pd.read_csv('../data/'+file, index_col=['grid_id','date','time'])
    path = '../res/IE_{}_{}_{}_{}_{}.npy'.format(design, file, NN, TI, DDS)
    if os.path.exists(path):
        continue

    df['const'] = 1
    M = len(df.index.get_level_values(2).unique())
    N = len(df.index.get_level_values(1).unique())
    NM = M*N
    if IE:
        df[scol] = np.append(np.delete(df[xcol].values
        *(df.index.get_level_values(2)>0),0),0)
        df[scol][df[scol]==0] = np.nan
        xyscols = [ycol] + regcols + [scol]
        df = df[xyscols]
    else:
        xycols = [ycol] + regcols
        df = df[xycols]
    df[acol] = -1

    model0 = VCM(df, ycol, xcol, acol, scol,IE,
                         interaction=interaction,
                         two_sided=two_sided, 
                         wild_bootstrap=wild_bootstrap, 
                        center_x=True, scale_x=True,hc=hc)
    model0.estimate(null = True)
    df['fitted_DE'] = model0.holder['fitted_DE'].values
    df['eta_DE'] = model0.holder['eta_DE'].values
    df['eps_DE'] = model0.holder['eps_DE'].values 
    df['fitted_IE'] = model0.holder['fitted_IE'].values
    df['eta_IE'] = model0.holder['eta_IE'].value
    df['eps_IE'] = model0.holder['eps_IE'].values

    def generate(df, N, ycol, regcols, acol, ti=1, delta=0, delta_s=0, delta_s_n=0):
        grids = (df.index.get_level_values(0).unique())
        G = len(grids)
        dates = (df.index.get_level_values(1).unique())
        number_of_days = len(dates)
        M = len(df)// G // number_of_days

        dates_ = np.random.choice(dates, size=(N,), replace=True)
        df_ = df.loc[[(x,y,z) for x in grids for y in dates_ for z in range(M)],:]
        df_ = df_.reset_index()
        df_['date'] = np.tile(np.repeat(np.arange(N),M), G)
        df_.set_index(['grid_id','date','time'], inplace=True)

        mt = int(24/ti)
        if ti < 24: # intra-day time interval
            abv = np.tile(np.repeat([-1,1], M//mt), mt//2)
            bav = np.tile(np.repeat([1,-1], M//mt), mt//2)
            vec = np.hstack([abv, bav])
        elif ti == 24: # inter-day time interval
            av = -np.ones(M)
            bv = np.ones(M)
            vec = np.hstack([av, bv])
        gvs = np.array([])
        gv = np.tile(vec, N//2)
        if design == 'st':
            for i in range(G):
                gvs = np.append(gvs, np.random.choice([-1,1])*gv)
        else:
            for i in range(G):
                gvs = np.append(gvs, gv)
        df_[acol] = gvs
        df[acol_n] = np.dot(adj_mat, ((df[acol].values+1)/2).reshape(G,M*N)).ravel()

        if IE:
            idx1 = np.arange(df_.shape[0])[df_.index.get_level_values(2)>0]
            a=(df_['fitted_IE'] + \
                       df_['eps_IE'] * np.repeat(np.random.randn(N*G), M) + \
                                df_['eta_IE'] * np.repeat(np.random.randn(N*G), M)).values
            df_[xcol].iloc[idx1]=a[~np.isnan(a)]
            df_[xcol] *= (1+delta_s_n)
            df_.loc[df_[acol]==1, xcol] *= (1+delta_s)
            df_[scol] = np.append(np.delete(df_[xcol].values
            *(df_.index.get_level_values(2)>0),0),0)
            df_[scol][df_[scol]==0] = np.nan
        df_[ycol] = (df_['fitted_DE'] + \
                        df_['eps_DE'] * np.repeat(np.random.randn(N*G), M) + \
                        df_['eta_DE'] * np.repeat(np.random.randn(N*G), M)).values
        df_[ycol] *= (1+delta_s_n)
        df_.loc[df_[acol]==1, ycol] *= (1+delta+delta_s)

        return df_    

    def one_step(seed):
        
        np.random.seed(seed)
        ret = []
        
        df_ = generate(df, NN, ycol, regcols, acol, TI, DD, II, II_n) 
        model = VCM(df_, ycol, xcol, acol, acol_n, scol,IE,
                     interaction=interaction,
                     two_sided=two_sided, 
                     wild_bootstrap=wild_bootstrap, 
                    center_x=True, scale_x=True,hc=hc)
        if IE==0:
            model.inference()
            ret.append([model.holder['test_stats_wb'], model.holder['test_stat'],
                   model.holder['pvalue1'], model.holder['pvalue2']])
        else:
            model.estimate()
            ret.append(model.holder['test_stat_IE'])
            
        return ret

    pool = mp.Pool(20)
    rets = pool.map(one_step, range(nsim))
    rets = np.array(rets)
    pool.close()
    
    path = '../res/IE_{}_{}_{}_{}_{}.npy'.format(design, file, NN, TI, DDS)
    
    np.save(path, rets)
\end{python}

\section{Further Discussions and Extensions}

\subsection{Endogeneity bias}\label{sec:endogeneity}
   {In this subsection, we discuss how to remove the endogeneity bias when the random effects appear in the state regression model as well. Specifically, Model 1 becomes
  \begin{align*}
  &Y_{i,\tp} = \beta_0(\tp) + S_{i,\tp}^\top\beta(\tp) + A_{i,\tp}\gamma(\tp) + e_{i,\tp}
  = Z_{i,\tp}^\top \theta(\tp) + e_{i,\tp}, \\
  & S_{i,\tp+1} =\phi_0(\tp)+\Phi(\tp) S_{i,\tau}+A_{i,\tp} \Gamma(\tp) + e_{i,\tp S}
  =  \Theta(\tp)Z_{i,\tp} + e_{i,\tp S},
  \end{align*}
  where 
  % $\theta(\tp) = (\beta_0(\tp), \beta(\tp)^\top,  \gamma(\tp))^\top$ is the vector of time-varying coefficients and $\Theta(\tp) = [\phi_0(\tp) ~~ \Phi(\tp) ~~ \Gamma(\tp)]$ is a $d\times (d+2)$ coefficient matrix, 
  $e_{i,\tp S}=\eta_{i,\tp S} + \varepsilon_{i,\tp S}$, $\eta_{i,\tp S}$ characterizes the day-specific temporal variation across different days and $\varepsilon_{i,\tp S}$ is the measurement error. We assume that $\eta_{i,\tp S},\varepsilon_{i,\tp S}$ are mutually independent;  $\{\varepsilon_{i,\tp S}\}_{i,\tp}$ are independent measurement errors with zero means and $\mbox{Cov}(\varepsilon_{i,\tp S})=\Sigma_{\varepsilon,\tp S}$; and $\{\eta_{i,\tp S}\}_{i,\tp}$ are identical copies of a mean-zero stochastic process with covariance function and $\{\Sigma_{\eta_S}(\tp_1,\tp_2)\}_{\tp_1,\tp_2}$.

  \change{Due to the potential dependencies between these random effects, past and future features are no longer conditionally independent.
  % , leading to the violation of the Markov assumption.  As such, the proposed model is no longer MDP and corresponds to a special case of partially observable MDP \citep[POMDP, see e.g.,][]{sutton2018reinforcement} where the random effects are unobserved. 
  Directly applying existing OPE methods or our proposal developed in Section \ref{sec:po_mdp} would yield biased policy value estimators.  Note that the predictor $S_{i,\tau}=\Theta(\tp-1)Z_{i,\tp-1} + e_{i,\tp-1, S}$ at time $\tau$ is dependent upon the $e_{i,\tau}$ due to the existence of the random effects in these residuals, resulting in endogeneity in the state regression model. As a result, the resulting OLS estimator is biased, leading to inconsistent estimation of IE. }
  
  We next outline two approaches to remove the endogeneity bias. The first approach relies on the use of historical data in which the actions were the set to baseline policy. According to the state regression model, $\{S_{t}\}_t$ in the historical data satisfies
  $$S_{t+1}=\phi_0^*(t)+\Phi^*(k)S_1+e_{tS}^*,$$
  where $\phi_0^*(t)=\sum_{k=1}^t\phi_0(k)\prod_{\ell=k+1}^t\Phi(\ell)$, $\Phi^*(k)=\prod_{k=1}^t\Phi(k)$ and the error $e_{tS}^*$ is independent of $S_1$. As such, the OLS estimator $\widehat\Phi^*(k)$ is consistent. When $\{\Phi(k)\}_k$ are nonzero, it allows us to consistently estimate these regression coefficients. On the other hand, when the actions are independent of the states, the regression coefficients $\{\Gamma(\tp)\}_{\tp}$ can be consistently estimated using data collected from online experiments. This allows us to consistently estimate IE based on \eqref{eq:DE IE T}. 
  
  The second approach requires the random effects to satisfy certain covariance structures. In particular, we require the correlation between $\eta_{i,\tau_1S}$ and $\eta_{i,\tau_2S}$ to decay to zero as $|\tau_{1S}-\tau_{2S}|$ approaches infinity. For a given sufficiently large $m_1$, the residual error $e_{tS}$ and the past state $S_{t-m_1}$ become asymptotically uncorrelated. According to the state regression model, we obtain that
  $$S_{t}=\phi(0)+\Phi(t)S_{t-m_1}+\sum_{k=t-m_1}^{t-1}\Gamma_t(k)A_k+e_{tS},$$
  where $\Gamma_t(k)=(\Phi(t-1) \Phi(t-2) \ldots \Phi(k+1)) \Gamma(k)$ and can be consistently estimated via OLS. As such, IE can be consistently estimated as well by noting that
  $$\IE=\sum_{t=2}^m \beta(t)^\top \left\{ \sum_{k=1}^{t-1}  \Phi(k) \left(\sum_{\ell=k-m_1}^{k-1}\Gamma_k(\ell)\right) \right\}.$$}

\subsection{High-dimensional models}\label{subsec:highd}
{We extend the proposed method to settings with high-dimensional state information in this section. For simplicity, we focus on the linear temporal varying coefficient model example. In the high-dimensional setting, we assume most elements in the regression coefficients $\beta(\tp)$ and $\Phi(\tau)$ are equal to zero. Hypothesis testing is challenging since many penalized estimators such as the Lasso \citep{tibshirani1996regression} or the Dantzig selector \citep{candes2007dantzig} does not have a traceable limiting distribution. 

One solution is to employ regularization methods with folded-concave penalty functions such as the smoothly clipped absolute deviation \citep[SCAD,][]{fan2001variable}, adaptive Lasso \citep{zou2006adaptive} or minimal concave penalty \citep[MCP,][]{zhang2010nearly} in Step 1 of Algorithms 1 and 2 to obtain sparse estimators. Under certain minimal-signal-strength assumptions, the resulting estimators possess the ``oracle" property in that they are selection consistent and asymptotically equivalent to the oracle OLS estimators computed as if the supports were known in advance \citep{fan2011nonconcave}. As such, the proposed Wald-type test statistics for DE remain valid. The bootstrap procedure is equally applicable even when the number of parameters is much larger than the sample size \citep{dezeure2017high,zhang2017simultaneous}. We may also apply sample splitting \citep{dezeure2015high} or the recursive online-score estimation (ROSE) algorithm \citep{shi2021statistical} to account for model selection uncertainty. 

Another solution is to employ the debiasing method \citep{javanmard2014confidence,van2014asymptotically,zhang2014confidence,ning2017general} to allow for valid inference without the minimal-signal-strength assumption. Specifically, we first apply penalized regression with LASSO, SCAD or MCP to obtain the initial regression estimators. We next debias these initial estimators using decorrelated estimation \citep[see e.g.][Equation 14]{shi2021testing}. This strategy guarantees each entry of the final estimator is asymptotically normal, regardless of whether the minimal-signal-strength assumption holds or not. These final estimators can be subsequently used for testing DE and IE. }

%\subsection{Test Procedures based on the Unsmoothed Estimator}\label{subsec:unsmooth}

\subsection{Test Procedures based on the Unsmoothed Estimator}\label{subsec:unsmooth}

  %Test procedures for DE and IE based on the unsmoothed estimator can be obtained by slightly modifying the Algorithm 1 and 2. 
 As commented in the main text, we can also use the unsmoothed estimators to test DE and IE. The resulting tests require weaker conditions on $m$ compared to those built upon the smoothed estimators. Specifically, $m$ is allowed to be either fixed, or to diverge to infinity. To the contrary, tests based on smoothed estimators require $m$ to diverge with $n$ at certain rate. 
  
  Test statistics based on the unsmoothed estimators are given by
  \begin{eqnarray*}
  \widetilde{\DE}=\sum_{\tau=1}^m\widehat{\gamma}(\tau),\quad\widetilde{\IE}=\sum_{\tp=2}^m \widehat{\beta}(\tp)^\top \left\{ \sum_{k=1}^{\tp-1} \left(\prod_{l=k+1}^{\tp-1} \widehat{\Phi}(l)\right) \widehat{\Gamma}(k) \right\}.
  \end{eqnarray*}
  The standard error of $\widetilde{\DE}$ is computed based on $\widehat{\bm{V}}_\theta$ which we denote by $\widehat{se}(\widetilde{\DE})$. The residuals and pseudo-outcomes for computing bootstrap samples are also constructed based on the OLS estimators $\widehat\theta(\tau)$ and $\widehat\Theta(\tau)$. %And then we have the following results immediately 
  The following results follow immediately from Theorem 1(i).

  \setcounter{prop}{2}
  \begin{prop}
  Suppose the assumptions in Theorem 1 hold. %, for the hypotheses \eqref{hypo:de t}, we have the following conclusions:
  Then under $H_0^{DE}$, we have  $\prob(\widetilde{\DE}/\widehat{se}(\widetilde{\DE})>z_{\alpha})=\alpha+o(1)$; under $H_1^{DE}$, we have $\prob(\widetilde{\DE}/\widehat{se}(\widetilde{\DE})>z_{\alpha})\to 1$. %where $z_{\alpha}$ denotes the upper $\alpha$th quantile of a standard normal distribution.  
  \end{prop}  

  %For IE, when $m$ is fixed or small, the validity of the bootstrap procedure based on the unsmoothed estimators can be guaranteed by the delta method whose convergence rate is $n^{-1/2}$. When $m$ diverges, the proof is similar to Theorem 2 except that there is no bias. In all, we have the following result.
  Similar to Theorem 2, we can show that the bootstrap procedure based on the unsmoothed estimators is valid to infer IE as well.

  \begin{prop}
  Suppose that there exist some constants $0<c_1\le 1, 0\le c_2<3/2$ such that $c_1\leq \Mean \|\varepsilon_{\tp,S}\|^2$, $\Mean e_\tp^2\leq c_1^{-1}$ for all $1\leq \tp\leq m$ and that $m=O(n^{c_2})$. Suppose the assumptions in Theorem \ref{thm:t theta asymp} as well as Assumptions \ref{asmp:st1} holds.  Then, with probability approaching 1, 
  \begin{equation*}%\label{result2}
  \sup_{z}|\prob(\widetilde{\IE}-\IE\leq z)-\prob(\widetilde{\IE}^b-\widetilde{\IE} \leq z |\textrm{Data}) \vert\leq Cn^{-1/8},
  \end{equation*}
  for some positive constant $C>0$.
  \end{prop}

\subsection{Advantage of the decomposition of DE and IE}
\label{sec:decomp}

\change{
Recall that the DE represents the sum of the short-term treatment effects on the immediate outcome over time assuming that the baseline policy is being employed in the past. In contrast, IE characterizes the carryover effects of past policies through their impact on the state variables (e.g., the demand and supply in the ridesharing platform).

Gaining insights into both DE and IE is instrumental in understanding the mechanisms through which the new policy surpasses the existing one, thereby paving the way for the creation of even more effective strategies. For instance, if the new policy's DE exceeds that of the current policy, but its IE is smaller, then adopting either policy in isolation would yield similar results on average. However, studying this decomposition enables us to derive a hybrid strategy that employs the existing policy during the first half of the day and switches to the new policy for the latter half. Given that DE characterizes short-term effects and IE measures delayed effects, it is reasonable to expect this hybrid approach to outperform both original policies.
To see this, we use the temporal case as an instance and denote 
\begin{align*}
    &\DE_\tp=\Mean \{R_{\tp}(1,S_\tp^*(\bm{0}_{\tp-1}),0, S_{\tp-1}^*(\bm{0}_{\tp-2}),\ldots,S_1)-R_\tp(0,S_\tp^*(\bm{0}_{\tp-1}),0, S_{\tp-1}^*(\bm{0}_{\tp-2}),\ldots,S_1)\},\\ 
    &\IE_\tp=\Mean  \{R_{\tp}(1,S_\tp^*(\bm{1}_{\tp-1}),1, S_{\tp-1}^*(\bm{1}_{\tp-2}),\ldots,S_1)-R_\tp(1,S_\tp^*(\bm{0}_{\tp-1}),0, S_{\tp-1}^*(\bm{0}_{\tp-2}),\ldots,S_1)\}.
\end{align*}
We remark that $\DE_\tp$ represents the direct effect on $R_\tp$ of applying the new policy only during time interval $\tp$  and $\IE_\tp$ represents the indirect effect on $R_\tp$ of applying the new policy from time interval 1 to $(\tp-1)$. Denote $\IE_1=0$, $\overline{\DE}=(\DE_1,\DE_2,\ldots,\DE_m)^\top$ and $\overline{\IE}=(\IE_1,\IE_2,\ldots,\IE_m)^\top$. Then we have
$$\DE=\mathbf{1}_m^\top\overline{\DE}\ \text{ and }\ \IE=\mathbf{1}_m^\top\overline{\IE}$$
where $\mathbf{1}=(1,1,\ldots,1)^\top\in\mathbb{R}^m$.
Suppose that we are interested in the policy effects of a specific time period $1\le m_1\le\tp\le m_2\le m$ and denote the corresponding DE and IE by $\DE_{m_1,m_2}$ and $\IE_{m_1,m_2}$. Let $\mathbf{1}_{m_1,m_2}$ be the $m$-dimensional vector whose $(m_1,m_1+1,\ldots,m_2)$th elements are 1 and the other elements are 0. Then 
$$\DE_{m_1,m_2}=\mathbf{1}_{m_1,m_2}^\top\overline{\DE}\ \text{ and }\ \IE_{m_1,m_2}=\mathbf{1}_{m_1,m_2}^\top\overline{\IE}.$$
Using the same technique of inferring DE and IE, we can test 
\begin{eqnarray*}
    &H_0(\DE_{m_1,m_2}):\DE_{m_1,m_2}\le 0 \ \ \text{versus}\ \ H_1(\DE_{m_1,m_2}):\DE_{m_1,m_2}>0;\\
    &H_0(\IE_{m_1,m_2}):\IE_{m_1,m_2}\le 0 \ \ \text{versus}\ \   H_1(\IE_{m_1,m_2}):\IE_{m_1,m_2}>0;
\end{eqnarray*}
which can guide the strategy design.
For instance, if $H_0(\DE_{m_1,m_2})$ is rejected and $H_0(\IE_{m_1,m_2})$ is not, we can apply the new policy from time $m_1$ and $m_2$ and keep the old policy from time 1 to $m_1-1$ to save the cost.
}

\end{document}